\documentclass[preprint]{aastex}
\usepackage[caption=false]{subfig}

\shorttitle{Remnant Disk in the GC}
\shortauthors{Yelda, Ghez, Lu, Do, Meyer, Morris, Matthews}

\newcommand{\msun}{M$_{\odot}$\,}

\def\lesssim{\mathrel{\hbox{\rlap{\hbox{\lower4pt\hbox{$\sim$}}}\hbox{$<$}}}}
\def\gtrsim{\mathrel{\hbox{\rlap{\hbox{\lower4pt\hbox{$\sim$}}}\hbox{$>$}}}}

\def\arcsec{\hbox{$^{\prime\prime}$}}

\def\farcm{\hbox{.\kern -0.7ex\raisebox{.9ex}{\scriptsize$\prime$}}}
\def\farcs{\hbox{\kern 0.13ex.\kern -0.95ex%
\raisebox{.9ex}{\scriptsize$\prime\prime$}\kern -0.1ex}}

\def\apj{ApJ}
\def\apjl{ApJL}

\def\nat{Nature}
\def\araa{ARAA}
\def\aaps{A\&AS}

\def\it{}
\def\mnras{MNRAS}
\def\aap{AAP}
\def\na{New Astronomy}
\def\nar{New Astron. Rev.}
\def\pasp{PASP}

\def\lesssim{\mathrel{\hbox{\rlap{\hbox{\lower4pt\hbox{$\sim$}}}\hbox{$<$}}}}
\def\gtrsim{\mathrel{\hbox{\rlap{\hbox{\lower4pt\hbox{$\sim$}}}\hbox{$>$}}}}

\def\arcsec{\hbox{$^{\prime\prime}$}}

\def\farcm{\hbox{.\kern -0.7ex\raisebox{.9ex}{\scriptsize$\prime$}}}
\def\farcs{\hbox{\kern 0.13ex.\kern -0.95ex%
\raisebox{.9ex}{\scriptsize$\prime\prime$}\kern -0.1ex}}

\def\arcsec{''}

\def\deg{\hbox{$^\circ$}}
\def\la{\mathrel{\hbox{\rlap{\hbox{\lower4pt\hbox{$\sim$}}}\hbox{$<$}}}}
\def\ga{\mathrel{\hbox{\rlap{\hbox{\lower4pt\hbox{$\sim$}}}\hbox{$>$}}}}

\def\arcsec{\hbox{$^{\prime\prime}$}}

\def\farcm{\hbox{$.\mkern-4mu^\prime$}}
\def\farcs{\hbox{$.\!\!^{\prime\prime}$}}

\def\fnum@figure{{\rmfamily Fig.\space\thefigure.---}}%
\def\fnum@table{{\rmfamily Table \thetable:}}%
\def\fnum@plate{{\bfseries Plate \theplate.}}%
\def\fps@figure{bp}%
\def\fps@table{bp}%
\def\fps@plate{bp}%
\def\eps@scaling{1.0}%

\begin{document}

\title{Properties of the Remnant Clockwise Disk of Young Stars in the Galactic Center}
\author{
S. Yelda\altaffilmark{1}, 
A. M. Ghez\altaffilmark{1},
J. R. Lu\altaffilmark{2}, 
T. Do\altaffilmark{3},
L. Meyer\altaffilmark{1},
M. R. Morris\altaffilmark{1},
K. Matthews\altaffilmark{4}}
\altaffiltext{1}{UCLA Department of Physics and Astronomy, Los Angeles, CA 90095
-1547; syelda, ghez, leo, morris@astro.ucla.edu}
\altaffiltext{2}{Institute for Astronomy, University of Hawaii, Honolulu, HI 
96822; jlu@ifa.hawaii.edu}
\altaffiltext{3}{Dunlap Institute for Astronomy and Astrophysics, University of
Toronto, 50 St. George Street, Toronto M5S 3H4, ON, Canada; do@di.utoronto.ca}
\altaffiltext{4}{Astrophysics, California Institute of Technology, MC 249-17, 
Pasadena, CA 91125; kym@caltech.edu}

\begin{abstract}
We present new kinematic measurements and modeling of a sample of 116 young
stars in the central parsec of the Galaxy in order to investigate the properties of
the young stellar disk. The measurements were derived from a combination 
of speckle and laser guide star adaptive optics imaging and integral 
field spectroscopy from the Keck telescopes. Compared to
earlier disk studies, the most important kinematic measurement improvement is
in the precision of the accelerations in the plane of the sky, which have a
factor of six smaller uncertainties ($\sigma\sim$10 $\mu$as yr$^{-2}$). 
We have also added the first radial velocity 
measurements for 8 young stars, increasing the sample at the largest radii 
(6$\arcsec$-12$\arcsec$) by 25\%. We derive the ensemble properties 
of the observed stars using Monte-Carlo simulations of 
mock data. There is one highly significant kinematic feature
($\sim$20$\sigma$), corresponding to the well-known clockwise disk, and 
no significant feature is detected at the location of the previously 
claimed counterclockwise disk. The true disk fraction is estimated to be $\sim$20\%, 
a factor of $\sim$2.5 lower than previous claims, suggesting that we may be 
observing the remnant of what used to be a more densely populated stellar disk.
The similarity in the kinematic properties of the B stars and the O/WR stars suggests 
a common star formation event. The intrinsic eccentricity distribution of the disk 
stars is unimodal, with an average value of $\langle e \rangle$=0.27$\pm$0.07,
which we show can be achieved through dynamical relaxation in an initially circular
disk with a moderately top-heavy mass function. 
\end{abstract}

\section{Introduction}
\label{sec:intro}
Spectroscopic observations of the Galaxy's central 
parsec have revealed an enigmatic population of nearly 200 hot, early-type stars, 
including Wolf-Rayet (WR) stars and O and B type main sequence stars, giants, and 
supergiants \citep{allen90,krabbe91,krabbe95,blum95, tamblyn96,najarro97,
ghez03,paumard06,bartko10,do13}. Their location in the Galactic center (GC) 
raises the question of how stars can form in such a hostile environment, as the 
tidal forces from the 4$\times$10$^6$ \msun supermassive black hole 
\citep[SMBH;][]{ghez08,gillessen09} would prevent the collapse of 
typical molecular clouds within its radius of influence 
\citep[$r_{infl}\sim$2 pc;][]{sanders92,morris93}.  

Clues to the origin of these stars can be gained through the detailed study 
of their orbital dynamics, as the age of the population 
\citep[$\sim$3-8 Myr;][]{paumard06,lu13} is much less than the
relaxation timescale in the Galactic center \citep[$\sim$1 Gyr;][]{hopman06}.
A particularly prominent feature that has been observed is a stellar disk
containing a large fraction of the O and WR stars orbiting the black hole in a 
clockwise (CW) sense, with an inner edge at a projected radius of 
$R$ = 0$\farcs$8 \citep{levin03,genzel03,paumard06,lu09,bartko09}.  
At smaller radii, dynamical effects such as vector resonant relaxation
\citep{rauch96,hopman06,alexanderTal07} will randomize the orbital planes within
the lifetimes of the B stars, which is in agreement with observations
\citep{schodel03,ghez05,eisenhauer05,gillessen09}.
The coherent motion of the disk stars may be indicative of {\em in situ} formation 
in a massive, gas disk around the SMBH \citep{levin03}. In standard models of accretion 
disks around central black holes, the disks are expected to fragment under their own 
self-gravity and lead to the formation of stars \citep{kolykhalov80,shlosman87,goodman03,
nayakshin06a,nayakshin07}. In such models, the steady build-up of the gas disk leads to 
stars on circular orbits, as the gas will have circularized prior to star formation.
However, there is growing evidence that the young stars are on more eccentric
orbits \citep{paumard06,beloborodov06,lu09,bartko09,gillessen09}. 
Several theories have invoked the infall of giant molecular clouds or the 
collision of two clouds to produce initially eccentric 
stellar disks \citep{mapelli08,yusef08,wardle08,bonnell08}. In any case, the surface 
density predicted by {\em in situ} formation scenarios falls off 
like $r^{-2}$ \citep{lin87,levin07} and agrees well with observations of the disk
\citep{paumard06,lu09,bartko09}.  

A stellar disk may also result from the inward migration of a massive cluster 
whose stars are tidally stripped as it spirals inward under dynamical friction
\citep{gerhard01}. However, this theory has been
difficult to reconcile with observations, most notably the surface density profile.
During the infall, the cluster will deposit stars throughout 
the GC with a radial profile of $r^{-0.75}$, much shallower than that observed 
\citep{berukoff06}.  Furthermore, in order for the cluster to reach the small
galactocentric radii that the young stars occupy, unrealistic cluster properties
are required, such as an initial cluster mass of $>$10$^5$ \msun or the presence of 
an intermediate mass black hole \citep[IMBH;][]{hansen03,gurkan05,berukoff06}
containing too large a fraction of the total cluster mass \citep{kim04}. 

While there is consensus in the literature regarding the existence of the clockwise 
disk and its surface density profile, many of its properties have yet to be 
well characterized, in part because interpretations of kinematic studies rely on the 
ability to assign disk membership. For example, \citet{bartko09} 
reported a bimodal eccentricity distribution for the disk, 
which is difficult to explain dynamically.  The authors could not rule out that 
contamination by non-members of the disk led to the second peak seen at $e$ = 0.9 - 1.0.
Contamination may also affect the interpretation of the geometric structure
of the disk, which was recently claimed to be highly warped \citep{bartko09}.

Further controversy exists regarding the kinematic properties of the stars that 
are not on the clockwise disk. Claims of a second, counterclockwise (CCW) disk have
been made \citep{genzel03,paumard06}, although this structure was not detected 
by \citet{lu09} and was later reinterpreted as a possible streamer or dissolving disk by 
\citet{bartko09}. Precise orbital parameter estimates are necessary 
for resolving this issue, as the presence of a second structure has implications for
both star formation and stellar dynamical evolution in the Galactic center.

We have carried out a detailed kinematic analysis on the Galactic center's 
young star population using high precision astrometric measurements over a 
16-year baseline.  Both the size and radial extent of our sample have
increased by a factor of $\sim$3-4 over our previous efforts in \citet{lu09}.
The data sets and sample are presented in \S \ref{sec:data}.
The data analysis, including image processing and astrometric and orbital
analysis techniques, is detailed in \S \ref{sec:dataAnalysis}.
To explore the impacts of measurement error and the assumptions used in our analysis, 
simulations are run on mock data sets, which are presented in parallel
with the observed results in \S \ref{sec:results}. We discuss our findings in
\S\ref{sec:disc} and conclude in \S\ref{sec:conc}.

\section{Sample and Data Sets}
\label{sec:data}

\subsection{Sample}
\label{sec:sample}
\begin{figure}
\epsscale{0.8}
\plotone{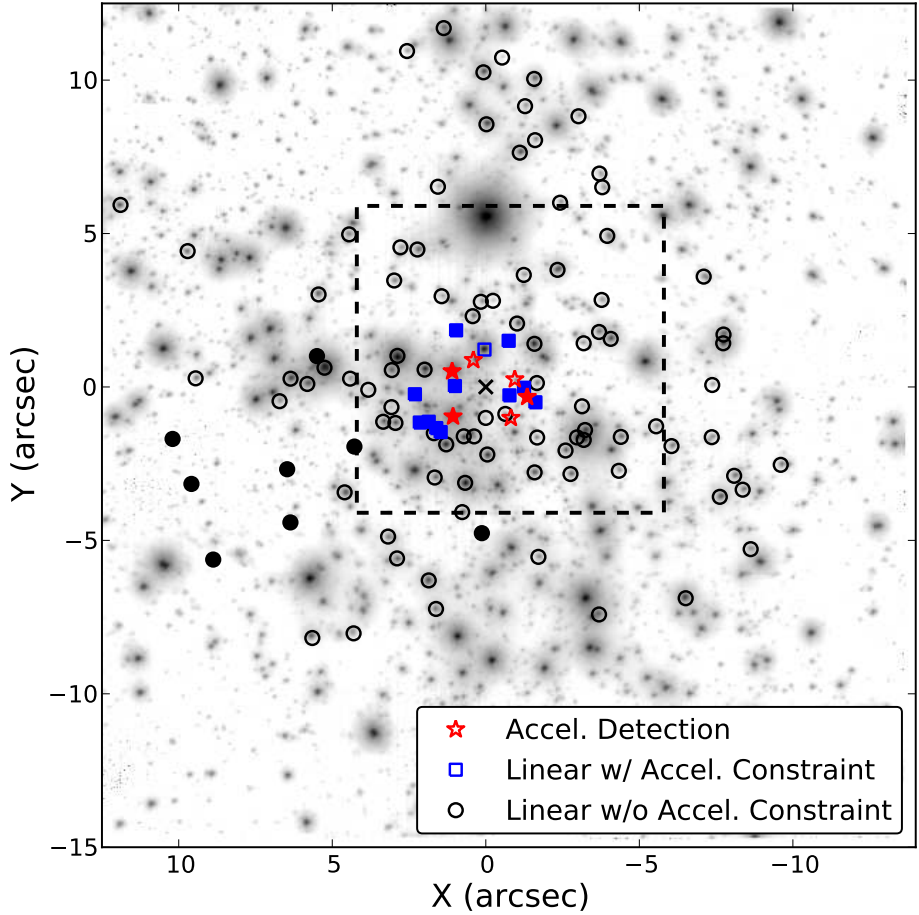}
\figcaption{The location of the 116 young stars with RV and astrometric measurements 
that comprise the sample for this study. Sources are identified based on their 
astrometric properties: acceleration detections ({\em red stars}), linearly-moving
with acceleration constraints ({\em blue squares}), and linearly-moving without 
acceleration constraints ({\em black circles}). {\em Filled stars} and 
{\em filled squares} mark sources with new acceleration detections and 
acceleration constraints, while the {\em filled circles} mark the sources with
new radial velocity measurements from GCOWS. In the background is the wide-field
adaptive optics mosaic image from 2008 May covering the central $\sim$1 pc of the 
Galaxy. The dashed black box denotes the central 10$\arcsec$ field of view where
the highest astrometric precision is achieved. 
}
\label{fig:sample}
\end{figure}

There are 116 stars that form the sample of this study (see 
Figure \ref{fig:sample}). These stars are selected based on the following
two criteria:
\begin{enumerate}
\item {\em Spectroscopic selection criteria:} stars that are spectroscopically 
identified as young and have spectral line measurements with sufficient 
signal-to-noise ratio to measure a radial velocity (RV).  We use both existing
RV measurements \citep{paumard06,bartko09} and new measurements based on 
spectroscopic data reported in \citet{do09} and \citet{do13}.
\item {\em Location selection criteria:} located outside a projected radius
of $R$ = 0$\farcs$8, which has 
been previously identified as the inner edge of the clockwise disk \citep{paumard06}
since stars interior to this radius appear to be randomly oriented 
\citep{schodel03,ghez05,gillessen09},
and within a 27$\arcsec\times$27$\arcsec$ region that is centered roughly on Sgr A*
and that is defined by our widest imaging field of view (see next section).
\end{enumerate}
We assume initially that all of the young stars meeting the above criteria belong
to the same population since the estimated age of the O/WR stars is $\sim$3-8 Myr 
\citep{paumard06,lu13} and the B stars have main sequence lifetimes of 
up to $\sim$30 Myr for the faintest stars in our sample ($K$ = 15.9). 
%This is therefore a reasonable assumption to make, however, 
We explicitly 
test this assumption in \S\ref{sec:bstars} by separating the 
sample into two subsets based on K magnitude.

\subsection{Imaging Observations}
The astrometric measurements in this study are based on three types of high-angular
resolution 2 micron imaging observations (speckle imaging, narrow-field adaptive optics 
imaging, wide-field mosaic AO imaging), which have been obtained at the W. M. Keck
observatory over a 16-year time period. As has been reported in previous 
publications \citep{ghez98,ghez00,ghez05,ghez08,lu09}, the earliest data sets were
obtained with K-band (2.2 micron) speckle imaging between 1995 and 2005
using the Near Infrared Camera \citep[NIRC;][]{matthews94,matthews96}, 
which has a $\sim$5$\arcsec\times$5$\arcsec$ field of view.
From the 27 epochs of available speckle data, we use those epochs with more than
900 frames to insure robust coordinate transformations (see \S \ref{sec:align}). 
This excludes only 2000 April (805 frames), resulting in 26 speckle epochs with a 
time baseline of 10 years (see Table \ref{tab:speckle_obs}).

Since 2004, we have utilized the Keck II adaptive optics (AO) system in conjunction 
with the facility near infrared camera NIRC2 (PI: K. Matthews) in its narrow-field 
mode, which has a plate scale of 9.952 mas pix$^{-1}$ 
\citep{yelda10} and a 10$\arcsec$ field of view \citep[$\sim$0.4 pc at the 8 kpc 
distance to the GC;][]{ghez08}. Here we include all existing Keck AO observations
through 2011, which includes 19 epochs and a time baseline of seven years
\citep{ghez05lgs,ghez08,lu09,meyer12}. 
As compared to our previous work on the young stars in \citet{lu09}, which included
only two years of deep, narrow-field AO imaging observations, we tripled
the time baseline for this type of observation. The observational setup 
was the same as the 2006-2007 laser guide star adaptive optics (LGSAO) 
observations reported in \citet{ghez08}. Specifically, a 20-point pseudo-random 
0.7$\arcsec\times$0.7$\arcsec$ dither pattern was used, with an initial position
that placed IRS16NE at pixel (229, 720). The images were taken at a position 
angle (PA) of 0$\deg$, and each frame consisted of 10 co-added 2.8 s integrations. 
At least three exposures were taken at each dither position.
The star USNO 0600-28577051 ($R$=13.7 mag and $\Delta$r$_{SgrA^*}$=19$\arcsec$) was
used to correct for tip and tilt in the LGSAO observations and served as the
natural guide star in the NGSAO observation of the Galactic center.
Table \ref{tab:narrowAO_obs} summarizes the narrow-field AO imaging observations
used in this study.

To measure the proper motions of the young stars at larger radii from Sgr A*
($R \gtrsim$ 7$\arcsec$), we obtained three epochs of 
$K'$-band LGSAO mosaics with the NIRC2 narrow camera that cover 
27$\arcsec\times$27$\arcsec$ ($\sim$1.1 pc $\times$ 1.1 pc).
These observations were taken on 2006 May 3, 2008 May 20, and 2010 June 5.
The tip-tilt star, PA, filter, exposure time per frame, and initial position were 
the same 
as those used for the deep narrow-field AO imaging data set, which covered the 
central 10$\arcsec\times$ 10$\arcsec$. In order to obtain the large 
field of view, we used a 9-position box pattern with a 8.5$\arcsec$ dither offset
and obtained 3-7 frames at each dither position. For the first two epochs,
we also obtained a 4-position box pattern with 4$\arcsec$ dithers, providing large
overlaps between all tiles in the mosaic.  At least three exposures
were taken at each dither position.  We refer to these wide-field data as 
``mosaics'' and the details of the observations can be found in Table 
\ref{tab:mosaicObs}.

\begin{deluxetable}{lcccccccc}
\tabletypesize{\scriptsize}
\tablewidth{0pt}
\tablecaption{Summary of Speckle Imaging Observations}
\tablehead{
  \colhead{Date} & 
  \colhead{Frames} & 
  \colhead{Frames} & 
  \colhead{FWHM} & 
  \colhead{Strehl} & 
  \colhead{N$_{stars}$} & 
  \colhead{K$_{lim}\tablenotemark{a}$} & 
  \colhead{$\sigma_{pos}\tablenotemark{b}$} & 
  \colhead{Data Source\tablenotemark{c}} \\ 
  \colhead{(UT)} & 
  \colhead{Obtained} & 
  \colhead{Used} & 
  \colhead{(mas)} & 
  \colhead{} & 
  \colhead{} & 
  \colhead{(mag)} & 
  \colhead{mas} & 
  \colhead{} 
}
\startdata
      1995 June 9-12  & 15114 & 1800 & 57 & 0.06 &  151 & 15.4 & 1.06 &             ref. 1\\ 
     1996 June 26-27  & 9261 &  865 & 60 & 0.03 &   77 & 14.1 & 1.76 &             ref. 1\\ 
         1997 May 14  & 3811 & 1837 & 61 & 0.04 &  139 & 15.4 & 1.28 &             ref. 1\\ 
      1998 April 2-3  & 9751 & 1639 & 62 & 0.04 &   83 & 14.6 & 1.52 &             ref. 2\\ 
      1998 May 14-15  & 16531 & 2102 & 69 & 0.04 &  126 & 15.4 & 1.32 &             ref. 2\\ 
       1998 July 3-5  & 9751 &  933 & 61 & 0.06 &  127 & 15.3 & 1.24 &             ref. 2\\ 
        1998 Aug 4-6  & 20375 & 1933 & 61 & 0.06 &  172 & 15.6 & 0.84 &             ref. 2\\ 
          1998 Oct 9  & 4776 & 1082 & 55 & 0.07 &  120 & 15.3 & 1.49 &             ref. 2\\ 
        1999 May 2-4  & 19512 & 1857 & 70 & 0.07 &  183 & 15.7 & 1.06 &             ref. 2\\ 
     1999 July 24-25  & 19307 & 2108 & 55 & 0.09 &  232 & 15.8 & 0.75 &             ref. 2\\ 
      2000 May 19-20  & 21492 & 2492 & 55 & 0.08 &  242 & 15.8 & 0.67 &             ref. 3\\ 
     2000 July 19-20  & 15124 & 1581 & 61 & 0.07 &  194 & 15.6 & 1.11 &             ref. 3\\ 
         2000 Oct 18  & 2587 & 1517 & 59 & 0.04 &   77 & 14.4 & 1.34 &             ref. 3\\ 
        2001 May 7-9  & 11343 & 1994 & 54 & 0.07 &  175 & 15.5 & 1.03 &             ref. 3\\ 
     2001 July 28-29  & 15920 & 1695 & 54 & 0.11 &  239 & 16.0 & 0.79 &             ref. 3\\ 
    2002 April 23-24  & 16130 & 1958 & 66 & 0.05 &  183 & 15.7 & 1.15 &             ref. 3\\ 
      2002 May 23-24  & 18338 & 1443 & 58 & 0.08 &  252 & 15.8 & 0.85 &             ref. 3\\ 
     2002 July 19-20  & 8878 & 1118 & 61 & 0.06 &  125 & 15.3 & 1.40 &             ref. 3\\ 
    2003 April 21-22  & 14475 & 1841 & 61 & 0.04 &  121 & 15.3 & 1.06 &             ref. 3\\ 
     2003 July 22-23  & 6948 & 1703 & 64 & 0.07 &  180 & 15.7 & 1.17 &             ref. 3\\ 
       2003 Sept 7-8  & 9799 & 1723 & 63 & 0.07 &  182 & 15.7 & 1.22 &             ref. 3\\ 
    2004 April 29-30  & 20140 & 1423 & 62 & 0.08 &  185 & 15.7 & 0.75 &             ref. 4\\ 
     2004 July 25-26  & 14440 & 2161 & 59 & 0.08 &  200 & 15.7 & 0.86 &             ref. 4\\ 
         2004 Aug 29  & 3040 & 1301 & 57 & 0.08 &  167 & 15.6 & 1.25 &             ref. 4\\ 
    2005 April 24-25  & 15770 & 1679 & 59 & 0.06 &  162 & 15.6 & 0.99 &             ref. 5\\ 
     2005 July 26-27  & 14820 & 1331 & 60 & 0.05 &  111 & 15.2 & 1.19 &             ref. 5\\ 
\enddata 
\label{tab:speckle_obs}

\tablenotetext{a}{K$_{lim}$ is the magnitude at which the cumulative distribution function of the observed K magnitudes reaches 90\% of the total sample size.}
\tablenotetext{b}{Positional error taken as error on the mean from the three sub-images in each epoch and includes stars with $K <$ 15.}
\tablenotetext{c}{Data originally reported in (1) \citet{ghez98}, (2) \citet{ghez00}, (3) \citet{ghez05}, (4) \citet{lu05}, and (5) \citet{rafelski07}.}

\end{deluxetable}
\clearpage

\begin{deluxetable}{lcccccccc}
\tabletypesize{\scriptsize}
\tablewidth{0pt}
\tablecaption{Summary of AO Imaging Observations}
\tablehead{
  \colhead{Date} & 
  \colhead{Frames} & 
  \colhead{Frames} & 
  \colhead{FWHM} & 
  \colhead{Strehl} & 
  \colhead{N$_{stars}$} & 
  \colhead{K$_{lim}\tablenotemark{a}$} & 
  \colhead{$\sigma_{pos}\tablenotemark{b}$} & 
  \colhead{Data Source\tablenotemark{c}} \\ 

  \colhead{(UT)} & 
  \colhead{Obtained} & 
  \colhead{Used} & 
  \colhead{(mas)} & 
  \colhead{} & 
  \colhead{} & 
  \colhead{(mag)} & 
  \colhead{mas} & 
  \colhead{} 
}
\startdata
        2004 July 26  &   10 &   10 & 60 & 0.28 &  598 & 15.9 & 0.30 &      LGSAO; ref. 6\\ 
        2005 June 30  &   10 &   10 & 61 & 0.26 &  929 & 16.3 & 0.32 &      LGSAO; ref. 8\\ 
        2005 July 31  &   59 &   31 & 57 & 0.18 & 1865 & 19.0 & 0.10 &      LGSAO; ref. 7\\ 
        2006 May 2-3  &  153 &  107 & 58 & 0.24 & 1952 & 19.1 & 0.05 &      LGSAO; ref. 7\\ 
     2006 June 19-20  &  289 &  156 & 57 & 0.30 & 2460 & 19.5 & 0.08 &      LGSAO; ref. 7\\ 
        2006 July 16  &   70 &   64 & 58 & 0.28 & 2179 & 19.3 & 0.09 &      LGSAO; ref. 7\\ 
         2007 May 17  &  101 &   76 & 58 & 0.28 & 2514 & 19.4 & 0.09 &      LGSAO; ref. 7\\ 
      2007 Aug 11-12  &  139 &   78 & 57 & 0.24 & 1879 & 19.0 & 0.08 &      LGSAO; ref. 7\\ 
         2008 May 15  &  138 &  134 & 54 & 0.25 & 2089 & 19.4 & 0.06 &      LGSAO; ref. 9\\ 
        2008 July 24  &  179 &  104 & 58 & 0.27 & 2189 & 19.3 & 0.04 &      LGSAO; ref. 9\\ 
          2009 May 4  &  311 &  149 & 57 & 0.27 & 2316 & 19.2 & 0.08 &      LGSAO; ref. 9\\ 
        2009 July 24  &  146 &   75 & 62 & 0.21 & 1701 & 18.9 & 0.09 &      LGSAO; ref. 9\\ 
         2009 Sept 9  &   55 &   43 & 61 & 0.25 & 1921 & 18.9 & 0.11 &      LGSAO; ref. 9\\ 
          2010 May 5  &  219 &  158 & 63 & 0.23 & 2037 & 19.1 & 0.06 &      LGSAO; ref. 9\\ 
         2010 July 6  &  136 &  117 & 61 & 0.23 & 1956 & 18.9 & 0.08 &      LGSAO; ref. 9\\ 
         2010 Aug 15  &  143 &  127 & 60 & 0.21 & 1826 & 19.0 & 0.07 &      LGSAO; ref. 9\\ 
         2011 May 27  &  164 &  114 & 66 & 0.19 & 1563 & 18.8 & 0.13 &      LGSAO; ref. 9\\ 
        2011 July 18  &  212 &  167 & 59 & 0.21 & 2031 & 19.2 & 0.08 &      NGSAO; ref. 9\\ 
         2011 Aug 23  &  218 &  196 & 59 & 0.27 & 2372 & 19.4 & 0.05 &      LGSAO; ref. 9\\ 
\enddata 
\label{tab:narrowAO_obs}

\tablenotetext{a}{K$_{lim}$ is the magnitude at which the cumulative distribution function of the observed K magnitudes reaches 90\% of the total sample size.}
\tablenotetext{b}{Positional error taken as error on the mean from the three sub-images in each epoch and includes stars with $K <$ 15.}
\tablenotetext{c}{Data originally reported in (6) \citet{ghez05lgs}, (7) \citet{ghez08}, (8) \citet{lu09}, and (9) \citet{yelda12} and this work.}

\end{deluxetable}
\clearpage

\begin{deluxetable}{lcccccccccc}
\tabletypesize{\scriptsize}
\tablewidth{0pt}
\tablecaption{Summary of Wide-field Mosaic Observations}
\tablehead{
  \colhead{Date} & 
  \colhead{N Dither} & 
  \colhead{$<N_{frmObtained}>$} & 
  \colhead{$<N_{frmUsed}>$} & 
  \colhead{FWHM} & 
  \colhead{Strehl} & 
  \colhead{N$_{stars}$} & 
  \colhead{K$_{lim}\tablenotemark{a}$} & 
  \colhead{$\sigma_{pos}\tablenotemark{b}$} & 
  \colhead{Data Source\tablenotemark{c}} \\ 

  \colhead{(UT)} & 
  \colhead{Positions} & 
  \colhead{per Position} & 
  \colhead{per Position} & 
  \colhead{(mas)} & 
  \colhead{} & 
  \colhead{} & 
  \colhead{(mag)} & 
  \colhead{mas} & 
  \colhead{} 
}
\startdata
   2006 May 3 &   13 & 3.4 & 3.4 & 63 & 0.20 & 6583 & 18.1 & 1.63&     LGSAO; ref. 10\\ 
  2008 May 20 &   13 & 3.4 & 2.9 & 78 & 0.11 & 4494 & 17.1 & 1.88&     LGSAO; ref. 10\\ 
  2010 June 5 &    9 & 7.2 & 5.4 & 76 & 0.12 & 5189 & 17.6 & 1.71&     LGSAO; ref. 10\\ 
\enddata 
\label{tab:mosaicObs}

\tablenotetext{a}{K$_{lim}$ is the magnitude at which the cumulative distribution function of the observed K magnitudes reaches 90\% of the total sample size.}
\tablenotetext{b}{Positional errors include distortion error (see text).}
\tablenotetext{c}{Data originally reported in (10) \citet{do13}.}

\end{deluxetable}
\clearpage

\subsection{Spectroscopic Observations}
To spectroscopically identify young stars and measure their line-of-sight
motions, high angular resolution spectroscopic observations were obtained with the 
integral field spectrograph OSIRIS in conjunction with the LGSAO system on Keck II 
\citep{larkin06}.  The central 4$\arcsec$ have been observed since 2006 with the 
Kn3 narrowband filter centered on the Br$\gamma$ line ($\lambda =$ 2.1661 $\mu$m)
and using the 35 mas plate scale.  In 2010, we began the Galactic Center OSIRIS
Wide-field Survey (GCOWS), in which observations were taken along the eastern 
portion of the clockwise disk in order to maximize the number of young star 
identifications \citep{do13}. These observations reached a radial extent
of $R \sim$14$\arcsec$ east of Sgr A* and used the 50 mas plate scale.
The details of our OSIRIS observations are presented in \citet{ghez08},
\citet{do09}, and \citet{do13}.  While the spectroscopic identification
of young stars using OSIRIS has been reported elsewhere 
\citep{do09,do13}, we report the radial velocities for 38 stars
from this instrument for the first time here.  For eight
of these stars, this is the first report of an RV measurement in the
literature.

% Generated by python code:
% syelda_yngstars.plot_sample()

\section{Data Analysis}
\label{sec:dataAnalysis}
\subsection{Astrometry}
\label{sec:astrometry}
\subsubsection{Image Processing}
All data sets were reduced using standard data processing techniques,
including sky subtraction, flat-fielding, and bad-pixel and cosmic-ray rejection.
The AO data were corrected for both optical distortion using the latest solution for 
the NIRC2 narrow camera and achromatic differential atmospheric refraction 
\citep{yelda10}. Based on this distortion solution,
we derive an improved solution for the NIRC speckle camera 
(Appendix \ref{app:nircDist}) using an approach similar to that of \citet{lu09}. The
updated NIRC distortion coefficients are presented in Table \ref{tab:nircCoef}.

For each observing run, individual frames are combined to make an average map. The
details of this process depend on the observing technique used.
The speckle data are combined to create an average image for each epoch using a
weighted shift-and-add technique as described in \citet{hornsteinPhd07}.
The final speckle images cover a field of view of 
$\sim$6$\arcsec\times$6$\arcsec$, centered approximately on Sgr A*.
For the adaptive optics narrow-field data, frames are selected based on the image
quality, as measured by the full width at half-maximum (FWHM) of the point spread
function. We choose to keep only those frames whose FWHM is within 125\%
of the minimum observed FWHM measured in a given epoch. These images 
are then combined with a weighted average, where the weights are set equal to the
Strehl ratio of each image. For each epoch of mosaic data, we create an average 
image at each dither position (i.e., 13 for each of the 2006 and 2008 observations,
and nine for the 2010 observation). All exposures taken at a given dither position 
are included in the corresponding average image except for a few cases where 
the frames were of extremely poor quality for one of several reasons 
(e.g., clouds or laser collision with neighboring telescopes).  
As done in our previous efforts, we create three independent subset images 
of equivalent quality in order to determine astrometric and photometric 
uncertainties for the speckle and AO central 10$\arcsec$ images. Likewise, subset 
images are created for each of the individual dither positions in the mosaics.

\subsubsection{Star Lists}
\label{sec:starlists}
Stars are identified and their relative positions and brightnesses are 
extracted from all images using the PSF fitting algorithm
{\em StarFinder} \citep{stf}, which is optimized for AO observations of crowded
stellar fields to identify and characterize stars in the field of view.
A model PSF for each image is iteratively constructed based on a set of 
bright stars in the field that have been pre-selected by the user. The model
PSF is then cross-correlated with the image in order to identify sources in the
field. The stars that are input for PSF construction are IRS 16C, 16NW, and 16NE 
for the speckle images, and IRS 16C, 16NW, 16SW, 16NE, 29, 33E, S1-23, S2-16, and 
S3-22 for the central 10$\arcsec$ AO images.  The set of PSF stars used for each 
image in the mosaic, on the other hand, depends on the position of that
image within the wide mosaic field of view. These stars include the
aforementioned sources for the central 10$\arcsec$ AO data set, as well as 
the following stars: IRS 1NE, 1SE, 2, 7, 9, 10EE, 10E3, 12N, 14SW, 14NE, 28, 34W, 
S5-183, S5-69, S8-3, S8-8, S9-3, S9-9, S10-2, S10-3, S11-4, S11-6, S9-5, S12-2,
S13-61. To identify sources, we use a {\em StarFinder} correlation threshold
of 0.8 in the average image and 0.6 in each of the three subset images.  
The initial star list for each epoch contains only those sources that are 
detected in the average image and in all three subset images.
The inaccuracies in the PSF model for the adaptive optics images
occasionally lead to spurious source detections near bright stars. We
therefore use the procedure described in Appendix A of \citet{yelda10}
to remove these false sources ($\sim$20\% of the sources identified). 
Altogether, we identify 162 and 1915 stars on average in the 
speckle ($\langle K_{lim} \rangle$ = 15.4) and AO data sets 
($\langle K_{lim} \rangle$ = 18.8), respectively. 

\begin{figure}[!t]
\begin{center}
\includegraphics[scale=0.65]{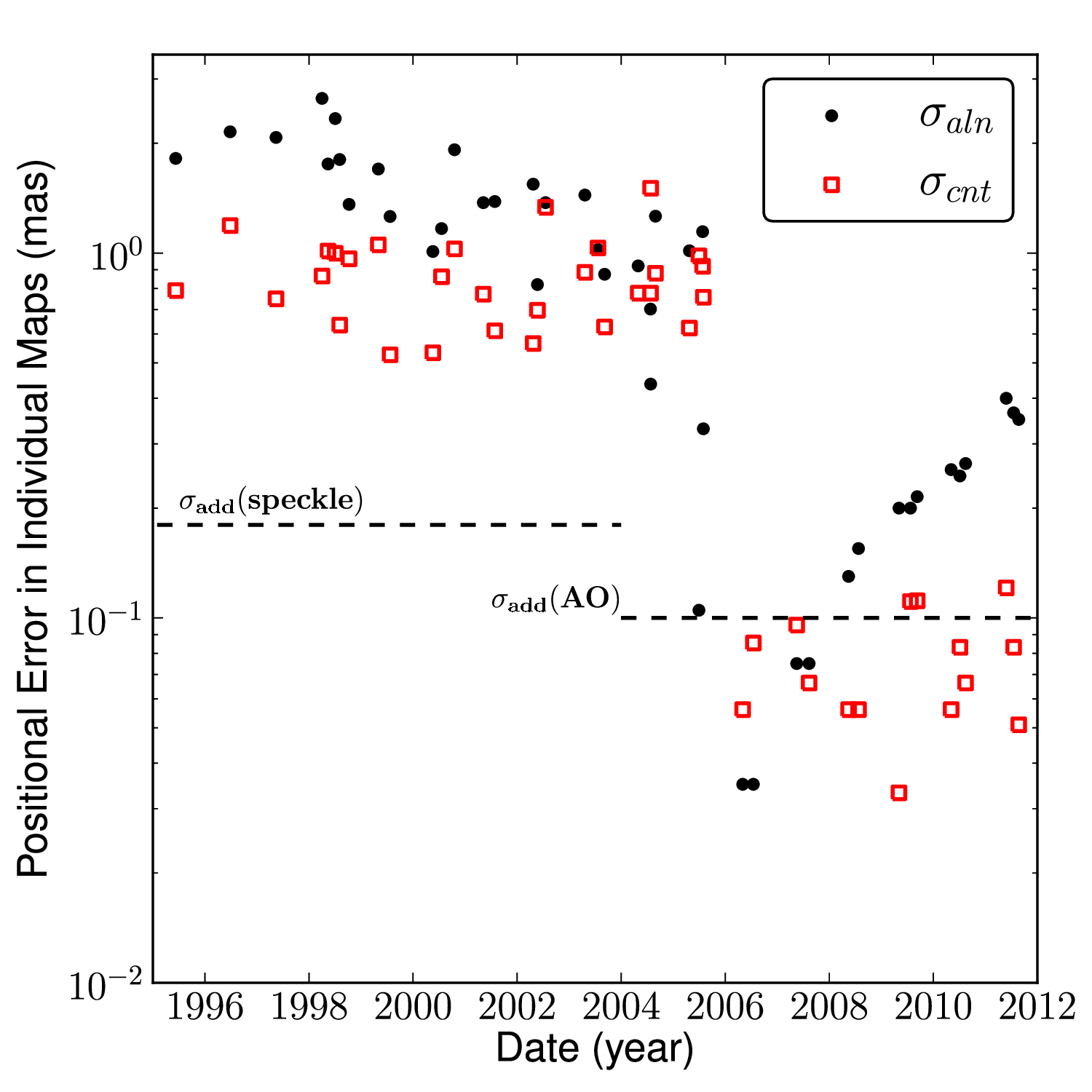}
\figcaption{Relative astrometric uncertainties, including alignment 
({\em filled black points}) and centroiding ({\em unfilled red 
squares}) uncertainties as a function of epoch for speckle data
from 1995-2005 and central 10$\arcsec$ AO data from 2004-2011.  
The median uncertainty of the 
young stars is reported for each epoch. Alignment errors are minimized near the 
reference epoch, 2006 June, and increase with time away from this epoch (see \S
\ref{sec:align}). 
All epochs with $\sigma_{aln} >$ 0.5 mas are from speckle imaging, where the 
higher uncertainties are a result of very few reference stars as compared to AO data.  
The additive errors for speckle and AO are shown as dashed lines.
}
\label{fig:posErr}
\end{center}
\end{figure}

There are two sources of statistical uncertainty associated with each positional 
measurement in the narrow-field images.  First is the centroiding uncertainty
($\sigma_{cnt}$), which is taken as the error on the mean of the positions for each
star in the three subset images.  Second, there is a term that appears to arise
from inaccuracies in the estimates of the PSF wings of neighboring sources
\citep{fritz10}. As
described in Appendix \ref{app:additive}, we follow a procedure similar to
\citet{clarkson11}, and estimate this additive error term
($\sigma_{add}$) to be 0.18 mas and 0.10 mas for the speckle and central 
10$\arcsec$ observations, respectively. Figure \ref{fig:posErr} shows the
centroiding and additive errors for each of our speckle and central 10$\arcsec$
data sets.
In addition, three of the adaptive optics data sets were taken at either 
different positions or position angles than the rest of the AO observations 
and therefore are impacted by residual distortion left over after the distortion
correction is applied, as described in
\citet{yelda10}. We account for the effects of residual distortion in these
images by performing a local distortion correction (Appendix \ref{app:localDist}),
which adds 0.5-1.4 mas errors to these epochs. The centroiding uncertainties in the
speckle data are typically a factor of $\sim$5 larger than the additive error
and therefore dominate the error budget. For the AO data, these two error terms
are comparable ($\sigma_{cnt}\sim$0.1 mas).

% Generated by python code:
% syelda_spie2012.alnErrVsEpoch()

The speckle observations were taken in stationary mode, and so the field rotated
over the course of the night. This led to a field of view with varying numbers
of frames contributing to each pixel in the final image. As a result, stars near the
edges of the FOV had relatively poor astrometric measurements.
To account for this effect, we require that each 
source be at a location in the average map that was covered by at least 80\%
of the frames contributing to that map\footnote{We note that this step was 
done after the cross-epoch coordinate transformation discussed in 
\S\ref{sec:align}.}. This prevented edge effect problems for
these data sets, which had much less uniform coverage than the AO data sets.
In total, 459 combined detections from 45 stars were removed, which is 
equivalent to 47\% of all young star measurements over all speckle epochs.

Final star lists for the wide field mosaics require additional steps and a
different treatment of the uncertainties. Star lists are created for each tile 
in the mosaics as done with the central 10$\arcsec$ AO data. The full mosaic
star list is then constructed by sequentially stitching together the lists 
from each tile following a procedure similar to that in \citet{anderson10}.
We begin by first transforming the stars' positions from the central tile to their 
positions in the Sgr A*-radio rest frame, in which $\sim$1200 stars down to a
K-band limiting magnitude of $K_{lim} \sim$ 16 were measured over the central 
22$\arcsec$ $\times$ 22$\arcsec$ in \citet{yelda10} and are updated here (see next 
section). We note that our wide field mosaics include fainter ($K_{lim} \sim$ 18) and 
more distant (FOV $\sim$ 30$\arcsec$ $\times$ 30$\arcsec$) stars than what was measured 
in \citet{yelda10}. Once the central tile is transformed, a new reference list of 
positions is created in the following way.  For stars that are matched,
their positions and their associated errors are 
updated. The new positions are taken as the weighted average of the positions in 
the existing reference list and the transformed star list
\footnote{Distortion errors include the statistical error
($\sim$0.05 pix) in the optical distortion model and the residual distortion 
term ($\sim$0.1 pix), both of which are described in \citet{yelda10}.}.  
The new positional errors are taken as the average of the errors.
For the stars that do not have positions in the Sgr A*-radio
frame (i.e., those fainter than $K \sim$ 16 or outside the 22$\arcsec$ $\times$ 22$\arcsec$ 
FOV), we include their transformed positions and their original errors (centroiding
and distortion errors) in the new reference list.
This new list then serves as the reference list for the stitching of the next tile in 
the sequence. This procedure is repeated until all tiles are aligned. 
After the central 
field from the 9-point dither observations is first aligned, the tiles from the 
4-point dither (if they were taken) are aligned (in the order: SW, NE, SE, NW).
This is followed by the alignment of the remaining tiles from the 9-point 
dither observations (in the order: E, W, N, S, NE, SE, NW, SW).  After completing
the full alignment, we refine this intermediate star list
by once again transforming each tile's star list to it a final time. In this 
instance, the averaging is done once all tiles are transformed and the
intermediate reference list of positions is not included in the averaging.
Each of the alignments performed in these steps involves a 2nd order 
polynomial transformation, which consists of 12 coefficients.

\subsubsection{Cross-Epoch Coordinate Transformations}
\label{sec:align}
In order to measure relative positions and proper motions, stellar positions 
from each epoch must be transformed to a common reference coordinate system. 
This procedure is complicated by the fact that stars available for performing 
the transformation have detectable proper motions. 
Previous Galactic center astrometric reference frames were constructed by
minimizing the net displacement of reference stars between star lists, a
procedure which 
implicitly assumes that these stars have no net motion over the field
\citep[the ``cluster'' reference frame; e.g.,][]{eckart97,ghez98,ghez08,
gillessen09}\footnote{We note that \citet{gillessen09} define a reference
frame using a combination of the cluster and maser reference frames.}.
However, net motion is known to exist in the GC, including an overall rotation
of the late-type star cluster in the plane of the Galaxy 
\citep{trippe08,schodel09,yelda10}, as well as coherent motion in a clockwise, 
young stellar disk in the central parsec 
\citep{levin03,genzel03,paumard06,lu09,bartko09}.  Neglecting to account for
this motion results in degeneracies between the transformation parameters and
the measured stellar velocities.
It is therefore important to understand the motion of these stars if they are to
be used in the construction of a stable reference frame.  

The absolute positions and proper motions presented in \citet{yelda10} of $>$1200 
Galactic center stars offers an opportunity to construct a stable astrometric 
reference frame for this work. Astrometric measurements of these stars
were determined relative to Sgr A* in a reference frame constructed by tying
infrared astrometry of seven SiO masers to their precise radio measurements 
\citep[the ``maser'' reference frame;][]{reid07,yelda10}. Here we update the 
positions and velocities of these ``secondary'' standards using a slightly 
modified version of the analysis described in \citet{yelda10}. Specifically, we 
now use mosaicked star lists as opposed to mosaicked images. The nine tiles are 
stitched together in the following order: C, E, W, N, S, NE, SE, NW, SW. The
final positions and their uncertainties are computed 
using a similar procedure as described in \S\ref{sec:starlists}.
The Sgr A*-rest reference frame was otherwise created in the 
same way as in \citet{yelda10}. The updated positions and proper motions relative 
to Sgr A* for 1210 stars are presented in Appendix \ref{app:new_refFrame}.

The alignment of the stars' positions across all epochs is a multi-step process.
The star lists from the deep central AO and speckle images are transformed to
the coordinate system defined by the 2006 June AO image using a second-order 
polynomial transformation. This epoch was chosen as the reference epoch, $t_{ref}$ 
because it is one of the deepest of our data sets 
($K_{lim}$=18.5 versus $K_{lim}$=15.7 for our earlier fiducial epoch of 
2004 July).  In the alignment of each epoch, $t_e$, we first propagate the
positions of the secondary astrometric standards from $t_{ref}$
to the expected positions in $t_e$ using their known absolute proper motions.  
We then find the best-fit transformation from the measured positions in $t_e$ 
of the astrometric standard stars
to their expected positions. This use of velocity information allows us to
use all the astrometric standards, regardless of spectral type, and removes 
the degeneracy between frame transformations and the stellar velocities.
Uncertainties from this transformation ($\sigma_{aln}$) are characterized using a 
half-sample bootstrap. These alignment errors are a 
function of time from the reference epoch and of the number of reference stars used 
in the transformation. As seen in Figure \ref{fig:posErr}, $\sigma_{aln}$ is 
minimized near the reference epoch and is larger for the speckle epochs
($\sigma_{aln} >$ 0.5 mas), which have on average $\sim$6$\times$ fewer reference stars
than are available in AO epochs.

Given the high stellar density environment of the Galactic center, it is important 
to consider the effects of source confusion \citep{ghez08,gillessen09,fritz10}.  
Stellar positions can be affected by unknown, underlying sources
that have not previously been detected, or they may be affected by
known sources that, when passing sufficiently close to a star, get detected
as only one source instead of two.  While it is not possible to account for the
former case, we can determine when a star's positional measurement is biased
by another known source. Using preliminary acceleration fits (see \S 
\ref{sec:pm_acc}), the distance between every pair of stars in the narrow-field
data is computed. For epochs
in which the predicted positions of two stars come within 60 mas of one
another (roughly the FWHM of our images), but only one star is actually detected, 
we exclude that detection as it is likely confused by the undetected source. 
Ten young stars (of the total 116)
in this work were affected by confusion between 1 and 11 times, although IRS16CC
was confused in 26 epochs by a $K \sim$ 13 mag star that has come within $\sim$30
mas since 2004. A total of 79 positional measurements were removed 
due to confusion, leaving 1756 positions for the narrow-field sources combined.

The mosaic star lists are aligned in a similar way as described above, but separately 
from the deep central and speckle data.  The reference epoch chosen for the alignment
of these three star lists was the 2008 observation, as this was the mid-point of
these data sets. Young stars that are outside the central 
10$\arcsec$ field of view and that are identified in all three mosaics are included
in the orbital analysis.  In other words, the astrometry obtained from the narrow
field data sets takes precedence over the mosaic astrometry.  The final analysis
includes astrometry for 69 young stars from the central AO + speckle data sets and
47 young stars from the wide field mosaics, bringing the total number of young
stars in this work to 116 and a total of 1897 positional measurements.

\subsubsection{Proper Motion and Acceleration Measurements}
\label{sec:pm_acc}
All the $x$ and $y$ positions are independently fit as a function of time 
with kinematic models.  For the central 10$\arcsec$ field, each star is fit
with two models: (1) proper motion only and (2) proper
motion and acceleration. Stars detected beyond the central 
10$\arcsec$ field (i.e., those in the wide mosaic fields) have just three 
positional measurements and are therefore only fit for velocities.
The reference time, $t_{0}$, for the position,
velocity, and acceleration measurements of each star is chosen as the mean
time of all epochs, weighted by the star's positional uncertainties.
The velocity fits take on the form
\begin{equation}
x(t) = x_{0} + v_{x,0}(t - t_{0}) 
\end{equation}
\begin{equation}
y(t) = y_{0} + v_{y,0}(t - t_{0}), 
\end{equation}
and the acceleration fits are of the form
\begin{equation}
x(t) = x_{0} + v_{x,0}(t - t_{0}) + \frac{1}{2}a_{x,0}(t - t_{0})^2
\end{equation}
\begin{equation}
y(t) = y_{0} + v_{y,0}(t - t_{0}) + \frac{1}{2}a_{y,0}(t - t_{0})^2.
\end{equation}

\begin{figure}[h]
\epsscale{0.75}
\plotone{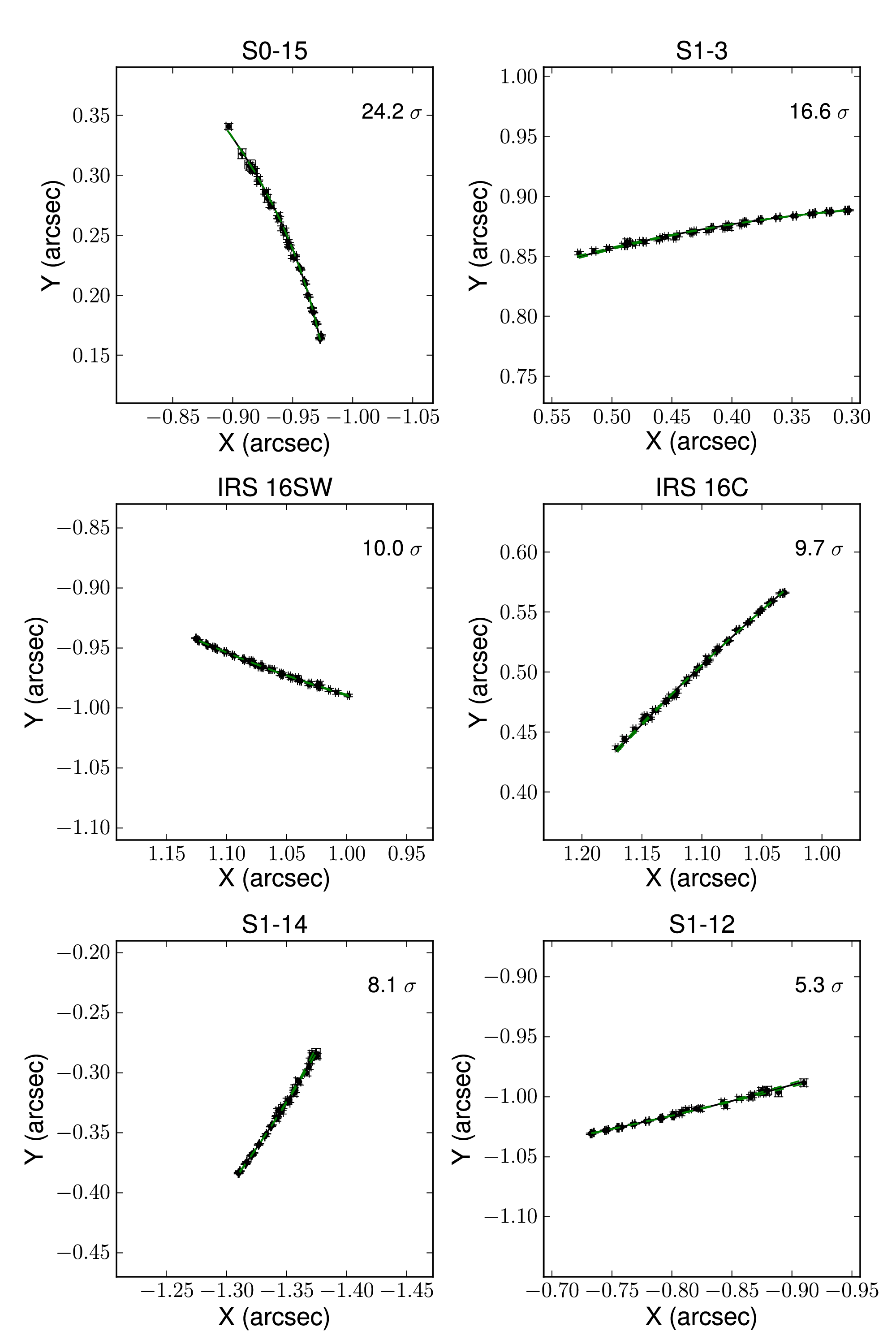}
\figcaption{Positional measurements for the six stars beyond 
a projected radius of 0.8$\arcsec$ with reliable acceleration detections.
Positions are measured relative to Sgr A*, with $X$ and $Y$ increasing to the 
East and North, respectively. Positional uncertainties do not include errors in the
transformation to absolute coordinates (i.e., plate scale, location of Sgr A*,
or position angle).  The best-fit acceleration model is shown for each source 
({\em black solid}) along with 1$\sigma$ error bars ({\em green dashed}).
The significance of each star's acceleration in the radial direction is shown 
in the upper right corner of each panel. The physical area is the same in each
panel (0$\farcs$28 $\times$ 0$\farcs$28).
}
\label{fig:accFits}
\end{figure}

\begin{figure}[h]
\epsscale{0.75}
\plotone{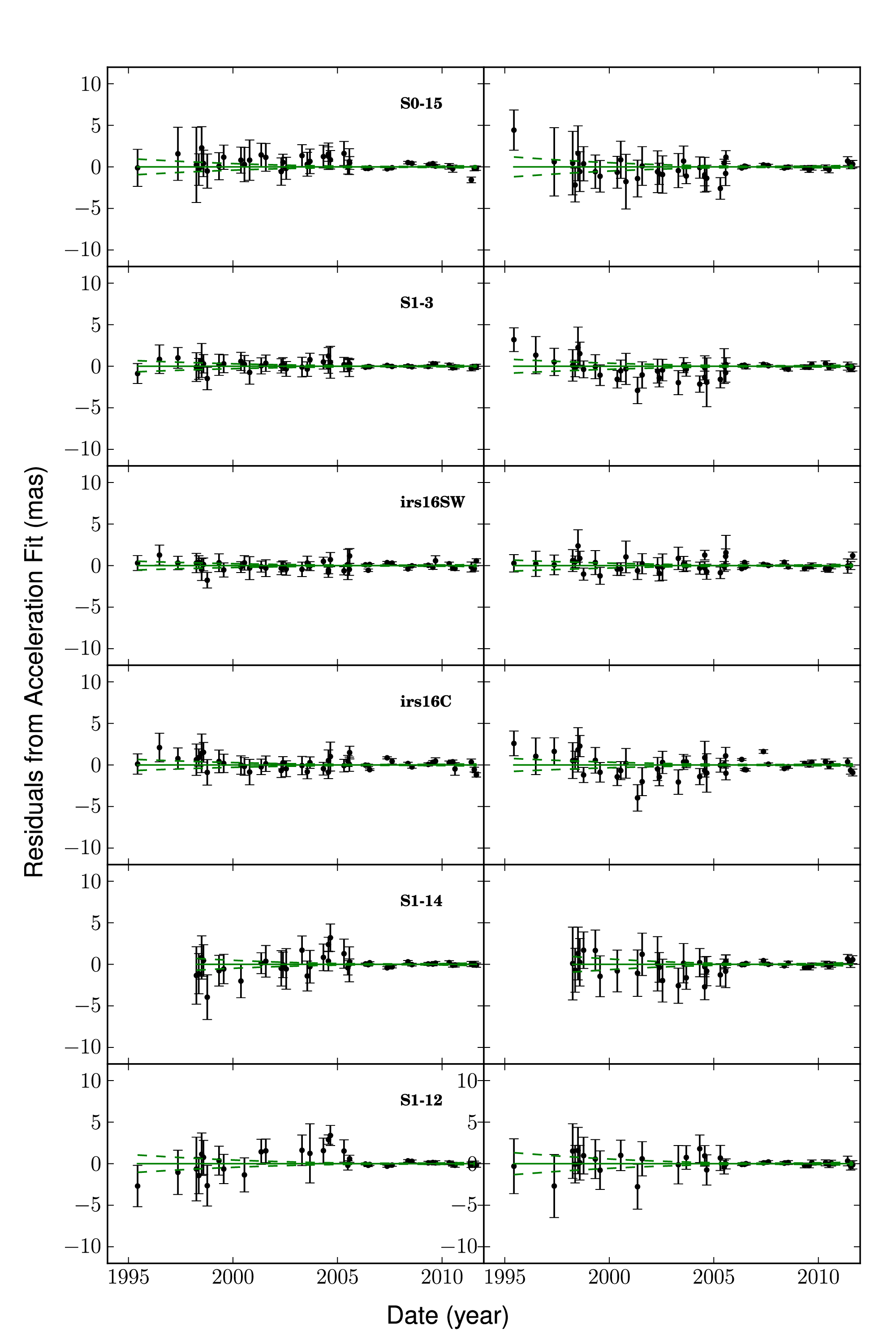}
\figcaption{Residuals in X ({\em left}) and Y ({\em right}) after subtracting the 
best-fit acceleration curves shown in 
Figure \ref{fig:accFits} for each of the six accelerating sources. 
}
\label{fig:accResid}
\end{figure}

Whether a star has measurable accelerated motion depends on several factors, 
including its distance from the supermassive black hole, the time baseline over 
which it is detected, and the precision with which its positions are measured.
To avoid including stars with non-physical accelerations, we require 
(1) the radial acceleration estimates to be significant at the $>$5$\sigma$ level
($N$ = 7 stars) and (2) the tangential acceleration estimates to be insignificant at the 
$<$5$\sigma$ level, which eliminates one star\footnote{This star, S1-8, shows a
significant tangential acceleration (~6$\sigma$) in our analysis, which we do not
believe is real. We are in the process of implementing a new data reduction 
technique known as 
speckle holography \citep{schodel13} on our speckle observations (instead of 
simple shift-and-add), and our preliminary analysis shows no significant
non-physical accelerations.}.
With these criteria, we measure physical accelerations 
for the following six stars (beyond a projected radius 
of 0$\farcs$8): S0-15, S1-3, IRS 16C, S1-12, S1-14, IRS 16SW. The acceleration 
fits and residuals from the fits for these six stars are shown in Figures 
\ref{fig:accFits} and \ref{fig:accResid}, respectively. This increases the 
number of acceleration measurements beyond 1$\arcsec$ over our previous work in 
\citet{lu09} by a factor of six, or equivalently, an additional five stars,
three of which are reported by \citet{gillessen09}. Furthermore, 
\citet{gillessen09} reports only one other physical acceleration from S1-2, 
which does not pass our criteria due to the large number of epochs in which it is
affected by source confusion (N = 15 confused epochs). The most distant star from 
the SMBH for which an acceleration measurement is made is IRS 16SW, located at 
$R$ = 1$\farcs$5 ($\sim$0.06 pc), which is well outside the inner edge of the 
stellar disk. For all other sources, the proper motion fit is used.
We present the positions, proper motions, and accelerations
for our sample in Table \ref{tab:yng_pm_table}.

% Generated by python code:
% syelda_yngstars.plot_accel_fit()

% Generated by python code:
% syelda_yngstars.plotAccelResidual()

\begin{figure}
\epsscale{1.0}
\plotone{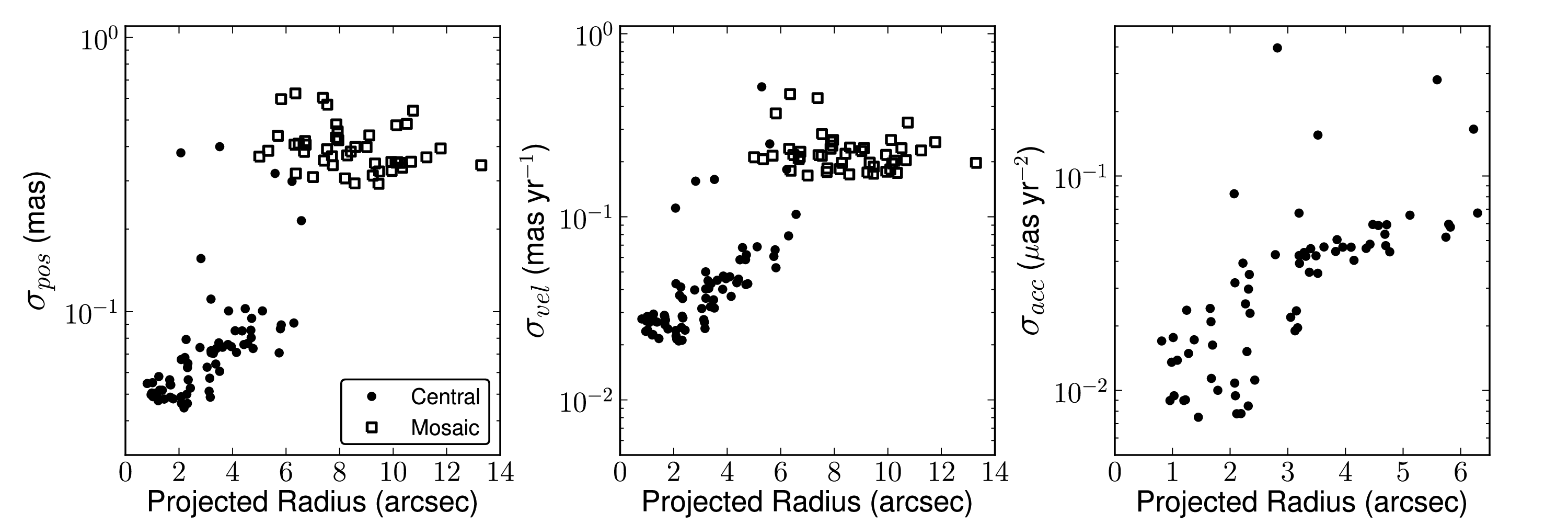}
\figcaption{Observed position ({\em left}), proper motion ({\em middle}), and acceleration
({\em right}) uncertainties as a function of projected radius. The average 
uncertainty along the $X$ and $Y$ coordinates are plotted.  The filled circles
mark the stars in our narrow-field data set and unfilled squares
indicate the stars in the wide-field mosaic data, which have projected radii 
$R$ $>$ 5$\arcsec$. 
}
\label{fig:astrometryR2d}
\end{figure}

\begin{figure}
\epsscale{0.7}
\plotone{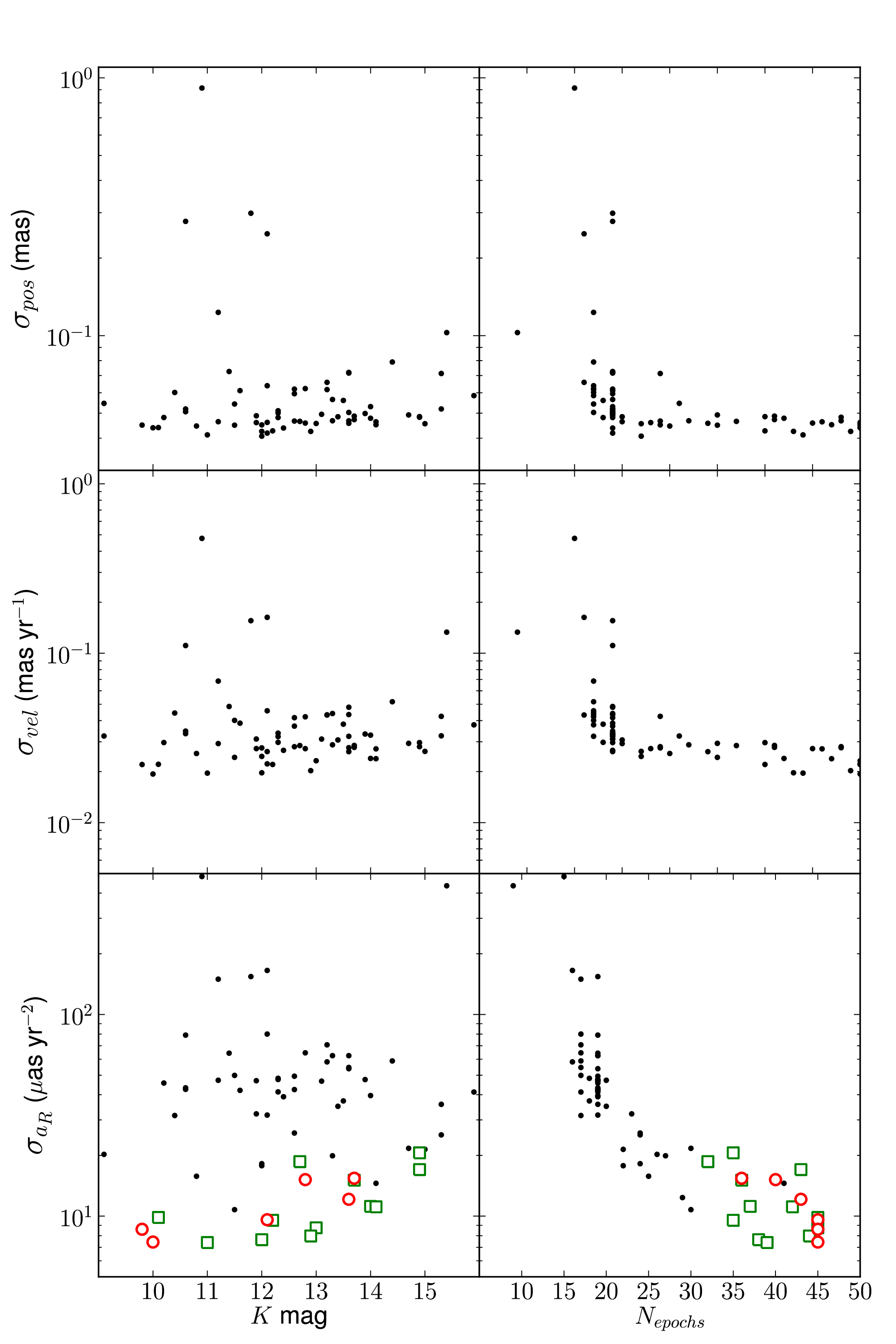}
\figcaption{Position ({\em top}), proper motion ({\em middle}), and radial acceleration 
({\em bottom}) uncertainties as a function of K magnitude ({\em left}) 
and number of epochs ({\em right}) for our sample of young stars beyond a projected 
radius of 0$\farcs$8 and in the central 10$\arcsec$ AO data set.  Note that the
acceleration uncertainties are shown in $\mu$as yr$^{-2}$. The astrometric
uncertainties are estimated from either the proper motion or 
acceleration fit to each star's individual positions over time.  
Stars with acceleration detections and acceleration constraints are shown as 
open red circles and open green squares, respectively.
The figures show that our astrometric uncertainties have only slight dependence on
stellar magnitudes and a strong dependence on the number of epochs a star
was detected in.
}
\label{fig:pvaErr}
\end{figure}

The position, proper motion, and acceleration uncertainties from the fitting procedure 
as a function of projected radius are shown in Figure \ref{fig:astrometryR2d}.
The smallest uncertainties are measured for stars with $R <$ 2$\farcs$5, which
are detected in both speckle and AO data sets and therefore have the longest 
time baseline. The observed increase in errors with radius is a result of 
alignment uncertainties and the number of epochs.
For the central 10$\arcsec$ sources, the median errors in positions 
and proper motions are 0.05 mas and 0.03 mas yr$^{-1}$, respectively.
The position and proper motion measurements of stars at large radii and detected
in only the wide mosaics have typical uncertainties of 0.4 mas and 
0.23 mas yr$^{-1}$, respectively. These relatively high uncertainties
are a result of having only three measurements and a four-year baseline.  
We also show the astrometric uncertainties as a function of $K$ magnitude and
number of epochs for stars in the central 10$\arcsec$ data set in Figure 
\ref{fig:pvaErr}.  The figure shows that the uncertainties have little 
dependence on magnitude but strongly correlate with the number of epochs a star
is detected in. Acceleration uncertainties for the six stars 
with reliable acceleration measurements and the 12 stars with 3$\sigma$
acceleration upper limits (see below) are highlighed in the bottom panel of 
Figure \ref{fig:pvaErr}.
The average acceleration uncertainty among 18 these stars is 10 $\mu$as yr$^{-2}$ 
($\sim$0.4 km s$^{-1}$ yr$^{-1}$), which is a factor of six improvement over
our earlier efforts in \citet{lu09}.  These measurements match and
sometimes exceed the highest astrometric precision that has been reported to
date \citep{gillessen09}. For completeness, we show the radial
velocity uncertainties for all 116 young stars in the sample and indicate
the source of the measurement that we use in our analysis 
(i.e., Keck/OSIRIS or VLT/SINFONI).   

% Generated by python code:
% sythesis_sim.posErr_vs_radius()

% Generated by python code:
% syelda_spie2012.pos_vel_acc_err()

% Generated by python code:
% sythesis.plot_velocity() for the RV error plot
\begin{figure}
\epsscale{1.0}
\plottwo{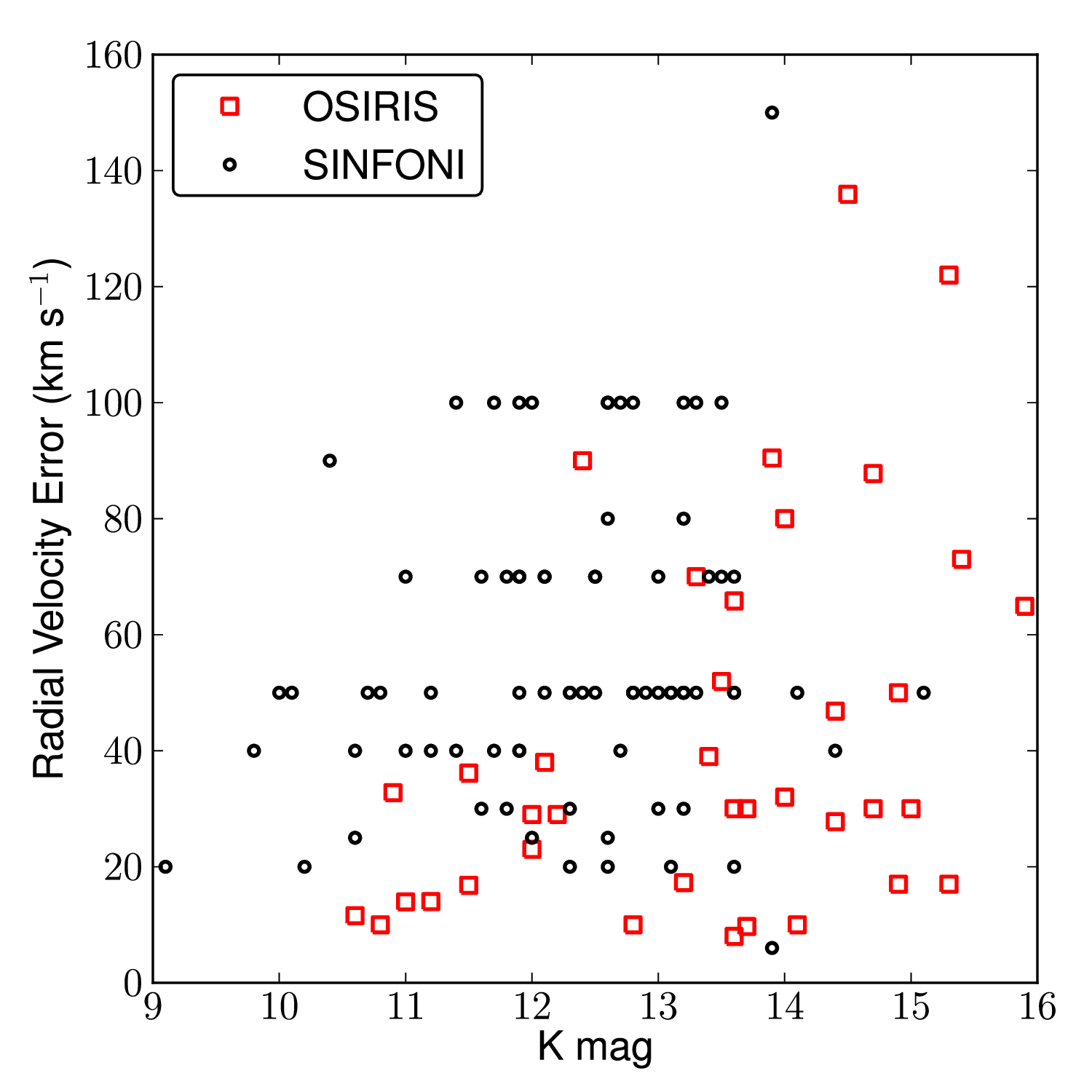}{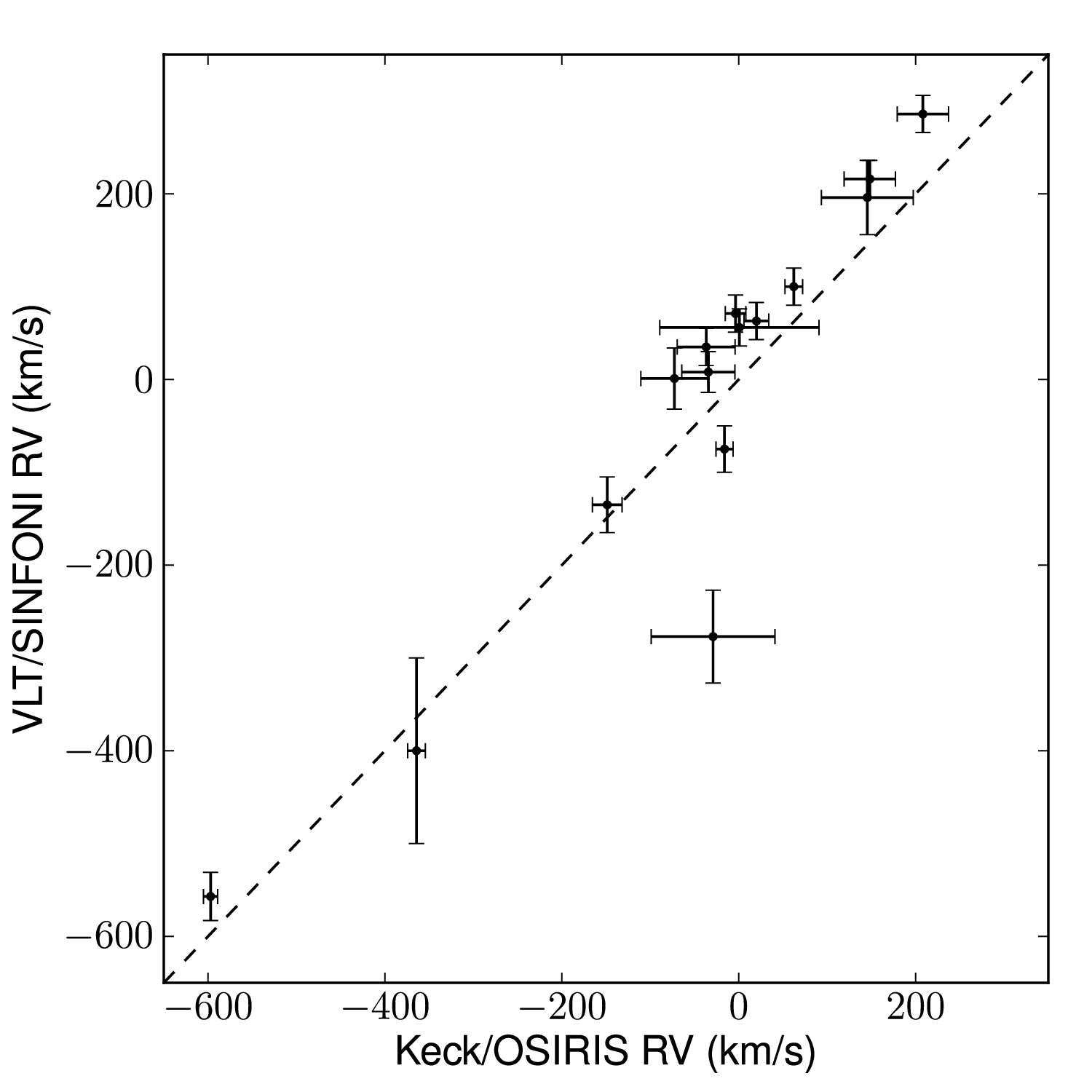}
\figcaption{{\em Left:} Line-of-sight velocity uncertainties plotted against K magnitude for 
the Keck/OSIRIS ({\em red squares}) and VLT/SINFONI ({\em black circles}) measurements.
{\em Right:} Comparison of radial velocity measurements for the 15 common stars in 
the two data sets.
}
\label{fig:rvKmag}
\end{figure}

\subsection{Radial Velocities}
Each OSIRIS radial velocity estimate is made by fitting a Gaussian model to the 
Br$\gamma$ line profile and comparing the wavelength of the best-fit peak to the 
rest wavelength of $\lambda_0$ = 2.1661 $\mu$m.
The velocities are then transformed to the local standard of rest (LSR) reference
frame by correcting for the Earth's rotation and motion around the sun, and for
the Sun's peculiar motion. RV uncertainties ($\sigma_{RV}$) from OSIRIS are estimated 
as the rms of the line profile fits from three independent subsets of the original
data and range from $\sim$10-90 km s$^{-1}$ 
($\langle \sigma_{RV} \rangle$ = 42 km s$^{-1}$). For the star S1-24 ($K$ = 11.5), 
however, only a single frame was obtained, which had a signal-to-noise ratio of $SNR$ = 74.
Given this relatively high $SNR$, we choose to include this RV measurement and we estimate
its uncertainty using an empirically-derived relation between $SNR$ and $\sigma_{RV}$:
\begin{equation}
\sigma_{RV} = 175.4 \times SNR^{-0.367}.
\end{equation}
Further details on the RV extraction process are reported in \citet{ghez08}.
The RV values and their uncertainties are reported in Table \ref{tab:yng_pm_table}.  

If multiple RV measurements for a star exist, the Keck measurements take precedence,
followed by the VLT measurements reported in \citet{bartko09} and \citet{paumard06}. 
This results in a sample of RV measurements, of which 38 were made with OSIRIS,
78 were taken from either \citet{bartko09} or \citet{paumard06} 
($\langle \sigma_{RV} \rangle$ = 57 km s$^{-1}$). The 15 stars 
that are in common between the Do et al. OSIRIS observations and the VLT 
observations are shown in the right panel of Figure \ref{fig:rvKmag}, and the
measurements between the two telescopes are
consistent within their uncertainties except for IRS 13E1.
Using the VLT radial velocities for these 15 stars does not change the results 
presented here. The \citet{do13} observations were designed to
sample the eastern portion of the clockwise disk in order to maximize the 
number of young star identifications.  With observed $K$ magnitudes ranging from
$\sim$10-16, the Do et al. samples include O stars at the bright end and B1V stars at 
the faint end.  
In contrast, \citet{bartko09} include mainly WR and O-type stars
($K <$ 14), all 90 of which are included in our sample. The uncertainties in the 
RV measurements from both OSIRIS and SINFONI are shown in Figure \ref{fig:rvKmag}.
We note that there was no attempt to extract radial velocities of the bright
WR stars from the OSIRIS data.

\clearpage
\begin{deluxetable}{lrccrrrrrrrrrr}
\rotate
\tabletypesize{\scriptsize}
\tablewidth{0pt}
\tablecaption{Kinematic Data of Galactic Center Young Stars\label{tab:yng_pm_table}}
\tablehead{
  \colhead{Name} &
  \colhead{K} &
  \colhead{N} &
  \colhead{Epoch \tablenotemark{a}} &
  \colhead{$\Delta RA$} &
  \colhead{$\sigma_{RA}$} &
  \colhead{$\Delta DEC$} &
  \colhead{$\sigma_{DEC}$} &
  \colhead{$v_{ra}$} &
  \colhead{$v_{dec}$} &
  \colhead{$v_z$} &
  \colhead{$v_z$ \tablenotemark{b}} &
  \colhead{$a_R$ \tablenotemark{c}} & 
  \colhead{Disk Prob.} \\ 
  \colhead{} &
  \colhead{(mag)} &
  \colhead{epochs} &
  \colhead{(year)} &
  \colhead{(arcsec)} &
  \colhead{(mas)} &
  \colhead{(arcsec)} &
  \colhead{(mas)} &
  \colhead{(mas/yr)} &
  \colhead{(mas/yr)} &
  \colhead{(km/s)} &
  \colhead{Ref} &
  \colhead{(mas/yr$^2$)} &
  \colhead{(1-L)} 
}
\startdata
\multicolumn{14}{c}{Stars with Acceleration Detections} \\
     S1-3 & 12.1 & 45 & 2007.7 &   0.357 & 0.047 &   0.888 & 0.053 &  -13.70 $\pm$   0.02 &    1.73 $\pm$   0.03 &   -72 $\pm$  38 &   1 & -0.160 $\pm$ 0.010 & 0.415 \\ 
    S0-15 & 13.6 & 43 & 2007.6 &  -0.962 & 0.046 &   0.214 & 0.055 &   -3.47 $\pm$   0.02 &  -10.82 $\pm$   0.03 &  -597 $\pm$   8 &   1 & -0.293 $\pm$ 0.012 & 0.950 \\ 
   irs16C &  9.8 & 45 & 2007.5 &   1.068 & 0.047 &   0.539 & 0.051 &   -8.54 $\pm$   0.02 &    7.72 $\pm$   0.02 &   158 $\pm$  40 &   2 & -0.083 $\pm$ 0.009 & 0.447 \\ 
    S1-12 & 13.7 & 36 & 2007.8 &  -0.777 & 0.047 &  -1.016 & 0.056 &   10.84 $\pm$   0.02 &   -2.42 $\pm$   0.03 &   -34 $\pm$  30 &   1 & -0.082 $\pm$ 0.015 & 0.487 \\ 
    S1-14 & 12.8 & 40 & 2007.8 &  -1.332 & 0.047 &  -0.349 & 0.056 &    4.96 $\pm$   0.02 &   -7.39 $\pm$   0.03 &  -364 $\pm$  10 &   1 & -0.123 $\pm$ 0.015 & 0.497 \\ 
  irs16SW & 10.0 & 45 & 2007.3 &   1.091 & 0.046 &  -0.952 & 0.050 &    7.37 $\pm$   0.02 &    2.97 $\pm$   0.02 &   470 $\pm$  50 &   2 & -0.074 $\pm$ 0.007 & 0.082 \\ 
\multicolumn{14}{c}{Linearly-Moving Stars with Acceleration Constraints} \\
    S0-14 & 13.7 & 36 & 2007.9 &  -0.764 & 0.047 &  -0.277 & 0.063 &    2.16 $\pm$   0.02 &   -0.93 $\pm$   0.03 &   -16 $\pm$   9 &   1 & $>$ -0.120 & 0.011 \\ 
     S1-1 & 13.0 & 45 & 2007.7 &   1.027 & 0.046 &   0.037 & 0.052 &    5.62 $\pm$   0.02 &    1.62 $\pm$   0.03 &   536 $\pm$  30 &   2 & $>$ -0.025 & 0.000 \\ 
  irs16NW & 10.1 & 45 & 2007.6 &   0.064 & 0.044 &   1.223 & 0.051 &    5.79 $\pm$   0.02 &    1.21 $\pm$   0.03 &   -15 $\pm$  50 &   2 & $>$ -0.045 & 0.000 \\ 
    S1-33 & 14.9 & 35 & 2008.0 &  -1.246 & 0.051 &  -0.007 & 0.065 &   -0.34 $\pm$   0.02 &    5.55 $\pm$   0.03 &     3 $\pm$  17 &   1 & $>$ -0.110 & 0.000 \\ 
    S1-18 & 14.9 & 43 & 2007.8 &  -0.773 & 0.050 &   1.508 & 0.058 &   -7.63 $\pm$   0.02 &    1.13 $\pm$   0.03 &  -249 $\pm$  50 &   1 & $>$ -0.050 & 0.022 \\ 
    S1-22 & 12.7 & 32 & 2007.8 &  -1.588 & 0.049 &  -0.509 & 0.060 &    7.73 $\pm$   0.02 &   -2.60 $\pm$   0.03 &  -235 $\pm$ 100 &   2 & $>$ -0.122 & 0.201 \\ 
     S2-4 & 12.2 & 35 & 2007.6 &   1.498 & 0.045 &  -1.459 & 0.048 &    7.94 $\pm$   0.02 &    3.07 $\pm$   0.03 &   208 $\pm$  29 &   1 & $>$ -0.023 & 0.205 \\ 
     S2-7 & 14.1 & 42 & 2007.7 &   0.943 & 0.045 &   1.853 & 0.053 &   -6.61 $\pm$   0.02 &    1.83 $\pm$   0.03 &   -94 $\pm$  50 &   3 & $>$ -0.053 & 0.228 \\ 
     S2-6 & 12.0 & 38 & 2007.6 &   1.641 & 0.044 &  -1.332 & 0.048 &    7.85 $\pm$   0.02 &    2.33 $\pm$   0.02 &   148 $\pm$  29 &   1 & $>$ -0.020 & 0.068 \\ 
irs16SW-E & 11.0 & 39 & 2007.6 &   1.880 & 0.042 &  -1.120 & 0.047 &    5.87 $\pm$   0.02 &    3.89 $\pm$   0.02 &   366 $\pm$  70 & 2,3 & $>$ -0.028 & 0.047 \\ 
    S2-22 & 12.9 & 44 & 2007.7 &   2.304 & 0.044 &  -0.214 & 0.048 &   -1.63 $\pm$   0.02 &    6.42 $\pm$   0.02 &    49 $\pm$  50 &   2 & $>$ -0.040 & 0.000 \\ 
    S2-58 & 14.0 & 37 & 2007.9 &   2.146 & 0.053 &  -1.134 & 0.052 &   -0.73 $\pm$   0.02 &    6.62 $\pm$   0.03 &    63 $\pm$  32 &   1 & $>$ -0.040 & 0.000 \\ 
\multicolumn{14}{c}{Linearly-Moving Stars} \\
     S1-2 & 14.7 & 30 & 2007.8 &   0.046 & 0.045 &  -1.011 & 0.054 &   13.10 $\pm$   0.02 &   -0.12 $\pm$   0.03 &    34 $\pm$  30 &   1 &            & 0.286\\ 
      S1-8 & 14.1 & 41 & 2007.7 &  -0.606 & 0.042 &  -0.898 & 0.050 &    9.05 $\pm$   0.02 &   -5.24 $\pm$   0.03 &  -171 $\pm$  10 &   1 &            & 0.500\\ 
     S1-21 & 13.3 & 27 & 2007.9 &  -1.650 & 0.042 &   0.109 & 0.051 &    4.02 $\pm$   0.02 &   -4.79 $\pm$   0.03 &   -29 $\pm$  70 &   1 &            & 0.132\\ 
     S1-19 & 13.6 & 29 & 2007.7 &   0.411 & 0.042 &  -1.623 & 0.050 &    8.41 $\pm$   0.02 &   -3.11 $\pm$   0.03 &  -164 $\pm$  30 &   1 &            & 0.205\\ 
     S1-24 & 11.5 & 30 & 2007.7 &   0.728 & 0.042 &  -1.631 & 0.048 &    2.64 $\pm$   0.02 &   -6.16 $\pm$   0.03 &   116 $\pm$  36 &   1 &            & 0.000\\ 
   irs16CC & 10.6 & 19 & 2001.3 &   1.997 & 0.263 &   0.545 & 0.292 &   -1.72 $\pm$   0.11 &    6.79 $\pm$   0.11 &   241 $\pm$  25 & 2,3 &            & 0.391\\ 
    irs29N & 10.4 & 17 & 2009.0 &  -1.560 & 0.051 &   1.381 & 0.069 &    4.75 $\pm$   0.04 &   -5.05 $\pm$   0.05 &  -190 $\pm$  90 & 2,3 &            & 0.034\\ 
    irs33N & 11.2 & 20 & 2007.8 &  -0.037 & 0.041 &  -2.222 & 0.052 &    3.62 $\pm$   0.02 &   -5.54 $\pm$   0.03 &    20 $\pm$  14 &   1 &            & 0.000\\ 
     S2-50 & 15.3 & 24 & 2008.1 &   1.696 & 0.069 &  -1.503 & 0.074 &    2.25 $\pm$   0.04 &    2.22 $\pm$   0.04 &   -56 $\pm$ 122 &   1 &            & 0.275\\ 
     S2-17 & 10.8 & 25 & 2007.9 &   1.323 & 0.041 &  -1.871 & 0.048 &    9.24 $\pm$   0.02 &    0.15 $\pm$   0.03 &    62 $\pm$  10 &   1 &            & 0.296\\ 
     S2-16 & 11.9 & 23 & 2008.0 &  -1.052 & 0.041 &   2.066 & 0.051 &   -9.16 $\pm$   0.02 &   -0.61 $\pm$   0.03 &  -100 $\pm$  70 & 2,3 &            & 0.730\\ 
     S2-21 & 13.4 & 20 & 2007.8 &  -1.641 & 0.042 &  -1.658 & 0.055 &    9.35 $\pm$   0.03 &   -3.33 $\pm$   0.04 &  -109 $\pm$  39 &   1 &            & 0.476\\ 
     S2-19 & 12.6 & 24 & 2007.8 &   0.398 & 0.043 &   2.311 & 0.050 &   -8.28 $\pm$   0.02 &    1.32 $\pm$   0.03 &    41 $\pm$  20 & 2,3 &            & 0.346\\ 
     S2-74 & 13.1 & 19 & 2007.8 &   0.134 & 0.045 &   2.781 & 0.054 &   -8.84 $\pm$   0.03 &    1.55 $\pm$   0.03 &    36 $\pm$  20 & 2,3 &            & 0.276\\ 
     S2-76 & 15.4 &  9 & 2010.2 &  -0.225 & 0.090 &   2.811 & 0.116 &    2.81 $\pm$   0.12 &    1.86 $\pm$   0.15 &   -28 $\pm$  73 &   1 &            & 0.000\\ 
   irs16NE &  9.1 & 26 & 2008.0 &   2.888 & 0.052 &   0.981 & 0.058 &    3.05 $\pm$   0.03 &   -8.90 $\pm$   0.04 &   -10 $\pm$  20 & 2,3 &            & 0.000\\ 
      S3-2 & 12.0 & 24 & 2007.9 &   3.076 & 0.041 &   0.555 & 0.049 &    4.31 $\pm$   0.02 &    1.30 $\pm$   0.03 &  -446 $\pm$  23 &   1 &            & 0.000\\ 
      S3-3 & 15.0 & 22 & 2008.1 &   3.082 & 0.040 &  -0.645 & 0.051 &    3.76 $\pm$   0.02 &    4.57 $\pm$   0.03 &    43 $\pm$  30 &   1 &            & 0.000\\ 
      S3-5 & 12.0 & 22 & 2008.0 &   2.953 & 0.038 &  -1.153 & 0.044 &    2.87 $\pm$   0.02 &    4.94 $\pm$   0.03 &   327 $\pm$ 100 & 2,3 &            & 0.430\\ 
     S3-96 & 14.4 & 17 & 2008.9 &  -3.133 & 0.068 &  -0.627 & 0.090 &   -0.07 $\pm$   0.04 &    5.64 $\pm$   0.06 &    40 $\pm$  40 & 2,3 &            & 0.000\\ 
     S3-19 & 11.9 & 19 & 2007.9 &  -1.566 & 0.043 &  -2.786 & 0.055 &    7.90 $\pm$   0.03 &   -1.30 $\pm$   0.04 &  -114 $\pm$  50 & 2,3 &            & 0.317\\ 
    irs33E & 10.2 & 19 & 2008.0 &   0.691 & 0.041 &  -3.127 & 0.055 &    6.85 $\pm$   0.02 &   -1.06 $\pm$   0.04 &   170 $\pm$  20 & 2,3 &            & 0.026\\ 
     S3-25 & 13.9 & 19 & 2007.6 &   1.424 & 0.046 &   2.959 & 0.054 &   -7.08 $\pm$   0.03 &    0.77 $\pm$   0.04 &   -84 $\pm$   6 &   2 &            & 0.195\\ 
     S3-26 & 12.3 & 19 & 2007.8 &  -2.588 & 0.045 &  -2.069 & 0.056 &    5.80 $\pm$   0.03 &    2.12 $\pm$   0.04 &    63 $\pm$  30 & 2,3 &            & 0.085\\ 
     S3-30 & 12.4 & 19 & 2008.0 &   1.661 & 0.039 &  -2.937 & 0.049 &   -0.72 $\pm$   0.02 &    4.32 $\pm$   0.03 &     0 $\pm$  90 &   1 &            & 0.018\\ 
   irs13E1 & 10.6 & 19 & 2007.8 &  -2.971 & 0.046 &  -1.647 & 0.056 &   -3.87 $\pm$   0.03 &   -1.98 $\pm$   0.04 &    -3 $\pm$  11 &   1 &            & 0.000\\ 
    S3-190 & 14.0 & 19 & 2008.2 &  -3.186 & 0.048 &   1.423 & 0.058 &   -3.27 $\pm$   0.03 &   -2.29 $\pm$   0.04 &  -244 $\pm$  80 &   1 &            & 0.215\\ 
     S3-10 & 12.1 & 19 & 2008.0 &   3.340 & 0.039 &  -1.113 & 0.045 &   -0.14 $\pm$   0.02 &    5.50 $\pm$   0.03 &   305 $\pm$  70 &   2 &            & 0.427\\ 
   irs13E4 & 11.8 & 19 & 2008.1 &  -3.231 & 0.303 &  -1.403 & 0.294 &   -5.77 $\pm$   0.17 &    1.70 $\pm$   0.14 &    56 $\pm$  70 & 2,3 &            & 0.000\\ 
   irs13E2 & 10.6 & 19 & 2007.7 &  -3.190 & 0.047 &  -1.726 & 0.057 &   -6.78 $\pm$   0.03 &    1.43 $\pm$   0.04 &    40 $\pm$  40 & 2,3 &            & 0.000\\ 
    S3-314 & 15.3 & 19 & 2007.9 &   3.829 & 0.045 &  -0.090 & 0.058 &    3.08 $\pm$   0.03 &    4.16 $\pm$   0.04 &    11 $\pm$  17 &   1 &            & 0.000\\ 
    S3-331 & 13.6 & 19 & 2008.0 &  -1.238 & 0.060 &   3.650 & 0.084 &    5.70 $\pm$   0.04 &    4.64 $\pm$   0.05 &  -167 $\pm$  20 & 2,3 &            & 0.000\\ 
    S3-374 & 12.3 & 19 & 2007.7 &  -2.757 & 0.046 &  -2.835 & 0.056 &   -0.49 $\pm$   0.03 &   -3.78 $\pm$   0.04 &    20 $\pm$  20 &   2 &            & 0.000\\ 
     S4-36 & 12.6 & 19 & 2007.8 &  -3.685 & 0.056 &   1.794 & 0.063 &   -5.37 $\pm$   0.03 &   -3.92 $\pm$   0.04 &  -154 $\pm$  25 & 2,3 &            & 0.345\\ 
     S4-71 & 12.3 & 18 & 2008.0 &   0.769 & 0.040 &  -4.076 & 0.056 &    0.12 $\pm$   0.02 &   -4.28 $\pm$   0.04 &    60 $\pm$  50 &   2 &            & 0.000\\ 
    irs34W & 11.6 & 19 & 2007.9 &  -4.066 & 0.059 &   1.570 & 0.063 &   -2.66 $\pm$   0.04 &   -4.89 $\pm$   0.04 &  -290 $\pm$  30 & 2,3 &            & 0.441\\ 
    S4-169 & 13.5 & 18 & 2007.8 &   4.417 & 0.046 &   0.274 & 0.066 &   -2.28 $\pm$   0.03 &    4.42 $\pm$   0.05 &   145 $\pm$  51 &   1 &            & 0.555\\ 
     irs3E & 11.4 & 19 & 2007.8 &  -2.338 & 0.064 &   3.816 & 0.081 &    4.61 $\pm$   0.04 &    1.19 $\pm$   0.05 &   107 $\pm$ 100 &   3 &            & 0.000\\ 
    irs7SE & 13.3 & 19 & 2007.3 &   2.976 & 0.051 &   3.469 & 0.062 &    5.84 $\pm$   0.04 &    0.22 $\pm$   0.05 &  -150 $\pm$ 100 & 2,3 &            & 0.000\\ 
    S4-258 & 12.6 & 19 & 2007.5 &  -4.392 & 0.062 &  -1.630 & 0.062 &   -4.70 $\pm$   0.04 &    2.45 $\pm$   0.05 &   330 $\pm$  80 & 2,3 &            & 0.000\\ 
    S4-262 & 15.9 & 17 & 2008.1 &   4.280 & 0.048 &  -1.939 & 0.069 &   -1.25 $\pm$   0.03 &   -5.09 $\pm$   0.05 &    43 $\pm$  64 &   1 &            & 0.000\\ 
   irs34NW & 13.2 & 16 & 2007.7 &  -3.766 & 0.062 &   2.839 & 0.070 &   -5.94 $\pm$   0.04 &   -3.33 $\pm$   0.05 &  -150 $\pm$  30 & 2,3 &            & 0.519\\ 
    S4-287 & 13.6 & 17 & 2007.8 &   0.125 & 0.041 &  -4.767 & 0.059 &    2.97 $\pm$   0.02 &    1.57 $\pm$   0.04 &   -51 $\pm$  65 &   1 &            & 0.058\\ 
    S4-364 & 11.7 &  3 & 2007.8 &   2.224 & 0.380 &   4.481 & 0.355 &    5.77 $\pm$   0.20 &   -2.30 $\pm$   0.22 &  -134 $\pm$  40 & 2,3 &            & 0.000\\ 
     S5-34 & 13.6 & 19 & 2007.6 &  -4.329 & 0.073 &  -2.731 & 0.070 &   -3.61 $\pm$   0.04 &   -1.68 $\pm$   0.05 &   -40 $\pm$  70 &   2 &            & 0.000\\ 
     irs1W & 10.9 & 15 & 2008.3 &   5.255 & 0.455 &   0.620 & 1.372 &   -1.35 $\pm$   0.21 &    9.66 $\pm$   0.74 &   -36 $\pm$  32 &   1 &            & 0.000\\ 
    S5-235 & 13.2 &  3 & 2007.8 &   2.781 & 0.391 &   4.553 & 0.381 &   -1.14 $\pm$   0.20 &   -3.77 $\pm$   0.22 &  -115 $\pm$  50 &   2 &            & 0.000\\ 
    S5-237 & 13.2 &  3 & 2007.8 &   5.500 & 0.403 &   1.002 & 0.361 &   -1.33 $\pm$   0.23 &    6.44 $\pm$   0.18 &    35 $\pm$  17 &   1 &            & 0.000\\ 
    S5-236 & 13.1 &  3 & 2008.4 &  -5.547 & 0.442 &  -1.282 & 0.433 &    4.43 $\pm$   0.22 &    1.95 $\pm$   0.21 &   155 $\pm$  50 &   2 &            & 0.000\\ 
    S5-183 & 11.5 & 17 & 2007.4 &   4.604 & 0.048 &  -3.431 & 0.061 &   -4.29 $\pm$   0.03 &   -1.90 $\pm$   0.05 &  -148 $\pm$  16 &   1 &            & 0.000\\ 
    S5-187 & 13.2 & 17 & 2007.5 &  -1.712 & 0.053 &  -5.532 & 0.071 &   -0.97 $\pm$   0.03 &   -3.76 $\pm$   0.05 &    10 $\pm$  50 &   2 &            & 0.000\\ 
    S5-231 & 12.0 &  3 & 2008.4 &   5.813 & 0.848 &   0.097 & 0.343 &    0.02 $\pm$   0.56 &    6.04 $\pm$   0.17 &    24 $\pm$  25 &   2 &            & 0.000\\ 
    S5-191 & 12.8 & 17 & 2007.9 &   3.184 & 0.051 &  -4.872 & 0.073 &   -1.35 $\pm$   0.03 &   -3.54 $\pm$   0.05 &   140 $\pm$  50 &   2 &            & 0.000\\ 
     S6-89 & 12.1 & 16 & 2009.2 &   5.445 & 0.225 &   3.013 & 0.272 &    3.05 $\pm$   0.16 &   -6.01 $\pm$   0.16 &  -135 $\pm$  70 &   2 &            & 0.000\\ 
     irs9W & 12.1 & 17 & 2007.4 &   2.882 & 0.051 &  -5.593 & 0.077 &    5.62 $\pm$   0.03 &    3.58 $\pm$   0.06 &   140 $\pm$  50 & 2,3 &            & 0.422\\ 
     S6-90 & 12.3 &  3 & 2007.9 &  -3.954 & 0.408 &   4.924 & 0.405 &   -0.50 $\pm$   0.24 &   -2.93 $\pm$   0.23 &  -350 $\pm$  50 & 2,3 &            & 0.005\\ 
     S6-96 & 12.8 &  3 & 2006.9 &  -6.045 & 0.563 &  -1.940 & 0.688 &   -1.35 $\pm$   0.37 &    8.38 $\pm$   0.57 &   -35 $\pm$  50 &   2 &            & 0.000\\ 
     S6-81 & 11.0 &  3 & 2008.1 &   6.360 & 0.348 &   0.267 & 0.290 &   -2.16 $\pm$   0.20 &    5.34 $\pm$   0.15 &   -14 $\pm$  13 &   1 &            & 0.000\\ 
     S6-95 & 13.2 &  3 & 2008.1 &  -2.420 & 0.411 &   6.004 & 0.410 &    4.02 $\pm$   0.22 &    0.77 $\pm$   0.22 &  -305 $\pm$ 100 & 2,3 &            & 0.000\\ 
     S6-63 & 11.2 & 17 & 2007.9 &   1.852 & 0.074 &  -6.306 & 0.172 &    6.04 $\pm$   0.05 &    1.80 $\pm$   0.09 &   110 $\pm$  50 &   2 &            & 0.611\\ 
     S6-93 & 12.8 &  3 & 2008.1 &   4.448 & 0.396 &   4.973 & 0.369 &    4.69 $\pm$   0.21 &   -0.85 $\pm$   0.20 &   -80 $\pm$ 100 & 2,3 &            & 0.000\\ 
    S6-100 & 13.9 &  3 & 2008.0 &   1.562 & 0.418 &   6.524 & 0.419 &   -4.87 $\pm$   0.22 &    2.87 $\pm$   0.20 &  -300 $\pm$ 150 &   3 &            & 0.016\\ 
     S6-82 & 13.5 &  3 & 2007.5 &   6.715 & 0.401 &  -0.470 & 0.412 &    2.07 $\pm$   0.19 &    5.68 $\pm$   0.26 &    86 $\pm$ 100 & 2,3 &            & 0.184\\ 
     S7-30 & 13.9 &  3 & 2008.4 &   6.469 & 0.332 &  -2.682 & 0.287 &   -2.63 $\pm$   0.19 &   -3.32 $\pm$   0.15 &   -87 $\pm$  90 &   1 &            & 0.000\\ 
    S7-161 & 13.6 &  3 & 2009.5 &  -7.376 & 0.605 &   0.061 & 0.599 &   -2.16 $\pm$   0.49 &   -2.55 $\pm$   0.40 &  -120 $\pm$  50 &   2 &            & 0.333\\ 
     S7-16 & 12.5 &  3 & 2008.1 &   1.621 & 0.358 &  -7.236 & 0.354 &    2.49 $\pm$   0.21 &    3.30 $\pm$   0.23 &   160 $\pm$  50 &   2 &            & 0.264\\ 
     S7-19 & 13.2 &  3 & 2007.8 &  -3.794 & 0.411 &   6.507 & 0.372 &    4.22 $\pm$   0.24 &    3.52 $\pm$   0.19 &   -65 $\pm$  50 &   2 &            & 0.000\\ 
    S7-180 & 13.4 &  3 & 2008.9 &  -7.360 & 0.558 &  -1.637 & 0.580 &   -3.92 $\pm$   0.28 &    0.48 $\pm$   0.29 &   120 $\pm$  70 &   2 &            & 0.000\\ 
     S7-10 & 11.4 &  3 & 2008.1 &  -1.105 & 0.399 &   7.635 & 0.338 &   -5.14 $\pm$   0.19 &   -1.87 $\pm$   0.16 &   -92 $\pm$  40 & 2,3 &            & 0.474\\ 
     S7-36 & 14.4 &  3 & 2008.5 &   6.363 & 0.376 &  -4.415 & 0.309 &    2.63 $\pm$   0.20 &    2.33 $\pm$   0.16 &    26 $\pm$  46 &   1 &            & 0.180\\ 
    S7-216 & 10.7 &  3 & 2008.4 &  -7.731 & 0.449 &   1.424 & 0.413 &    1.82 $\pm$   0.24 &    6.59 $\pm$   0.23 &    60 $\pm$  50 &   2 &            & 0.000\\ 
     S7-20 & 13.3 &  3 & 2008.4 &  -3.700 & 0.474 &   6.955 & 0.490 &    4.11 $\pm$   0.26 &    2.89 $\pm$   0.26 &   -45 $\pm$  50 &   2 &            & 0.000\\ 
    S7-228 & 11.8 &  3 & 2008.3 &  -7.741 & 0.477 &   1.708 & 0.432 &    2.19 $\pm$   0.26 &    3.67 $\pm$   0.23 &   150 $\pm$  30 &   2 &            & 0.000\\ 
    S7-236 & 12.5 &  3 & 2007.7 &  -7.093 & 0.451 &   3.598 & 0.393 &   -3.65 $\pm$   0.28 &   -2.59 $\pm$   0.25 &  -170 $\pm$  70 &   2 &            & 0.681\\ 
     S8-15 & 13.0 &  3 & 2008.2 &  -1.603 & 0.343 &   8.043 & 0.270 &   -3.30 $\pm$   0.20 &   -2.64 $\pm$   0.16 &  -130 $\pm$  50 &   2 &            & 0.387\\ 
      S8-7 & 11.9 &  3 & 2008.3 &  -3.688 & 0.397 &  -7.415 & 0.345 &    4.53 $\pm$   0.20 &    0.54 $\pm$   0.19 &    30 $\pm$ 100 &   2 &            & 0.516\\ 
    S8-181 & 11.6 &  3 & 2007.9 &  -7.620 & 0.393 &  -3.580 & 0.374 &   -2.15 $\pm$   0.24 &   -1.90 $\pm$   0.20 &    70 $\pm$  70 & 2,3 &            & 0.005\\ 
      S8-4 & 11.0 &  3 & 2008.3 &  -0.021 & 0.324 &   8.560 & 0.263 &   -0.92 $\pm$   0.18 &    3.81 $\pm$   0.16 &  -138 $\pm$  40 & 2,3 &            & 0.000\\ 
    S8-196 & 12.4 &  3 & 2008.4 &  -8.087 & 0.426 &  -2.896 & 0.371 &    0.16 $\pm$   0.26 &   -0.14 $\pm$   0.21 &   190 $\pm$  50 &   2 &            & 0.000\\ 
    S9-143 & 12.6 &  3 & 2008.1 &  -8.365 & 0.416 &  -3.347 & 0.380 &   -0.33 $\pm$   0.25 &   -1.29 $\pm$   0.21 &    40 $\pm$ 100 &   2 &            & 0.112\\ 
     S9-20 & 13.2 &  3 & 2007.8 &   4.304 & 0.490 &  -8.031 & 0.388 &    2.55 $\pm$   0.27 &    1.48 $\pm$   0.21 &   180 $\pm$  80 & 2,3 &            & 0.227\\ 
     S9-23 & 13.6 &  3 & 2008.2 &  -1.277 & 0.342 &   9.151 & 0.285 &   -3.89 $\pm$   0.18 &   -3.43 $\pm$   0.17 &  -185 $\pm$  50 &   2 &            & 0.156\\ 
     S9-13 & 13.1 &  3 & 2008.2 &  -3.019 & 0.360 &   8.821 & 0.334 &    2.00 $\pm$   0.19 &    3.68 $\pm$   0.20 &  -160 $\pm$  50 &   2 &            & 0.000\\ 
      S9-1 & 12.6 &  3 & 2008.4 &   9.450 & 0.321 &   0.281 & 0.265 &   -1.79 $\pm$   0.18 &   -2.59 $\pm$   0.16 &  -230 $\pm$ 100 & 2,3 &            & 0.000\\ 
    S9-114 & 10.8 &  3 & 2008.3 &  -6.509 & 0.341 &  -6.886 & 0.308 &    2.08 $\pm$   0.20 &    3.22 $\pm$   0.18 &   160 $\pm$  50 &   2 &            & 0.000\\ 
    S9-283 & 12.5 &  3 & 2008.1 &  -9.605 & 0.368 &  -2.539 & 0.335 &    0.80 $\pm$   0.23 &    1.00 $\pm$   0.21 &    30 $\pm$  70 & 2,3 &            & 0.000\\ 
      S9-9 & 11.7 &  3 & 2008.4 &   5.650 & 0.329 &  -8.182 & 0.323 &   -0.91 $\pm$   0.18 &   -1.18 $\pm$   0.17 &   130 $\pm$ 100 & 2,3 &            & 0.000\\ 
    S10-50 & 14.7 &  3 & 2008.2 &   9.586 & 0.386 &  -3.160 & 0.314 &   -0.48 $\pm$   0.21 &   -3.88 $\pm$   0.16 &    96 $\pm$  87 &   1 &            & 0.000\\ 
   S10-136 & 13.0 &  3 & 2007.9 &  -8.624 & 0.488 &  -5.289 & 0.469 &   -1.76 $\pm$   0.27 &    6.04 $\pm$   0.26 &   -70 $\pm$  70 & 2,3 &            & 0.000\\ 
     S10-5 & 11.9 &  3 & 2008.3 &  -1.574 & 0.347 &  10.039 & 0.355 &   -1.67 $\pm$   0.18 &   -1.18 $\pm$   0.21 &  -180 $\pm$  70 & 2,3 &            & 0.047\\ 
     S10-4 & 11.2 &  3 & 2008.1 &   0.078 & 0.351 &  10.254 & 0.347 &   -2.09 $\pm$   0.19 &    1.56 $\pm$   0.21 &  -250 $\pm$  40 & 2,3 &            & 0.000\\ 
    S10-32 & 14.4 &  3 & 2008.3 &  10.200 & 0.354 &  -1.694 & 0.317 &    3.41 $\pm$   0.18 &    3.68 $\pm$   0.16 &   161 $\pm$  27 &   1 &            & 0.366\\ 
    S10-34 & 14.5 &  3 & 2008.4 &   8.877 & 0.483 &  -5.626 & 0.485 &    1.08 $\pm$   0.23 &    3.76 $\pm$   0.24 &  -107 $\pm$ 135 &   1 &            & 0.059\\ 
     S10-7 & 12.7 &  3 & 2008.0 &   9.709 & 0.332 &   4.428 & 0.372 &   -0.34 $\pm$   0.21 &   -4.24 $\pm$   0.20 &  -150 $\pm$  40 & 2,3 &            & 0.000\\ 
    S10-48 & 15.1 &  3 & 2007.2 &  -0.533 & 0.486 &  10.732 & 0.596 &    2.01 $\pm$   0.22 &    1.56 $\pm$   0.44 &  -205 $\pm$  50 &   3 &            & 0.000\\ 
    S11-21 & 13.5 &  3 & 2008.0 &   2.566 & 0.378 &  10.947 & 0.353 &   -2.08 $\pm$   0.22 &   -1.78 $\pm$   0.24 &  -160 $\pm$  70 &   2 &            & 0.075\\ 
     S11-5 & 11.9 &  3 & 2007.9 &   1.370 & 0.367 &  11.693 & 0.421 &   -0.26 $\pm$   0.21 &    2.64 $\pm$   0.30 &   -65 $\pm$  40 & 2,3 &            & 0.000\\ 
     S13-3 & 11.9 &  3 & 2008.3 &  11.895 & 0.347 &   5.932 & 0.337 &    1.31 $\pm$   0.20 &    2.28 $\pm$   0.19 &  -190 $\pm$  40 &   3 &            & 0.000\\ 
 \enddata
\tablecomments{All uncertainties are 1$\sigma$ relative
errors and do not include errors in the plate scale, 
location of Sgr A*, or position angle.}
\tablenotetext{a}{Epoch taken as the mean of the imaging observations, 
weighted by positional uncertainties for each star.} 
\tablenotetext{b}{Radial velocity data obtained from 
observations in (1) \citet{do09} and \citet{do13},
(2) \citet{bartko09}, and (3) \citet{paumard06}. Note that some 
RV measurements reported in \citet{bartko09} were first reported 
in \citet{paumard06}.} 
\tablenotetext{c}{Accelerations were fit for stars falling 
within the central 10$\arcsec$ field of view only. 
For stars with acceleration limits, the 
positions and velocities are from the linear fits and 
the acceleration limits are from the acceleration fits. 
}\end{deluxetable}

\clearpage

\subsection{Orbital Analysis}
\label{sec:orbitAnalysis}
With six kinematic variables measured ($x_0$, $y_0$, $z$, $v_x$, $v_y$, $v_z$, $a_R$),
the standard Keplerian orbital elements (inclination $i$, angle to the ascending node 
$\Omega$, time of periapase passage $T_0$, longitude of periapse $\omega$, period $P$, 
and eccentricity $e$) can be estimated if the central potential is known
(see \citet{lu09} Appendix B for conversion equations).
The description of the central potential used in this analysis is based on
a spherically symmetric mass, $M_{tot}$, located at a distance $R_0$, and composed 
of the mass of the central SMBH, $M_{BH}$, and an extended mass component from the
nuclear star cluster, $M_{ext}$. In this work, we rederive the $M_{BH}$ and $R_0$ 
from the orbit of the 16-year period central-arcsecond star, S0-2, using
(1) the astrometry from the aligned star lists reported here to ensure that the 
position of Sgr A* is identified in the same reference frame as our stellar 
kinematic measurements, and (2) all radial velocities used by \citet{ghez08}
and newly acquired data from OSIRIS since that work\footnote{Since the analysis
presented here was carried out, the black hole properties were rederived after 
including the most recent imaging observation from Keck in 2012 May as well as all
currently published radial velocity data \citep{meyer12}. The 
newly-derived black hole mass and distance are consistent with the values we 
use to within 1$\sigma$.}; this results
in a $M_{BH}$ estimate of 4.6 $\pm$ 0.7 $\times$10$^6$ \msun and an
$R_0$ estimate of 8.23 $\pm$ 0.67 kpc. 
We base $M_{ext}$ on the work of \citet{schodel09} and take it to be
$M_{ext}(r$ = 1 pc) $\sim$ 1 $\pm$ 0.4 $\times$ 10$^6$ \msun, where the error 
is the difference in extended mass estimates from their isotropic and 
anisotropic velocity models.  If $M_{ext}$ is modeled as
\begin{equation}
M_{ext} = 4\pi\int \rho(r)r^2dr,
\end{equation}
where the mass density, $\rho(r)$, is a power-law of the form
\begin{equation}
\rho(r) = \rho_0\bigg(\frac{r}{5 pc}\bigg)^{-1},
\end{equation}
then $\rho_0$ = 3.2 $\pm$ 1.3 $\times$ 10$^4$ \msun pc$^{-3}$. 
For the radial range of our data (assuming $z$ = 0), this leads to
$M_{ext} <$ 5 $\times$ 10$^5$ \msun, which is an order of magnitude
smaller than the mass of the SMBH and should therefore have a minimal effect on
the orbital estimates, but we include it for completeness.

\begin{figure}
\epsscale{0.7}
\plotone{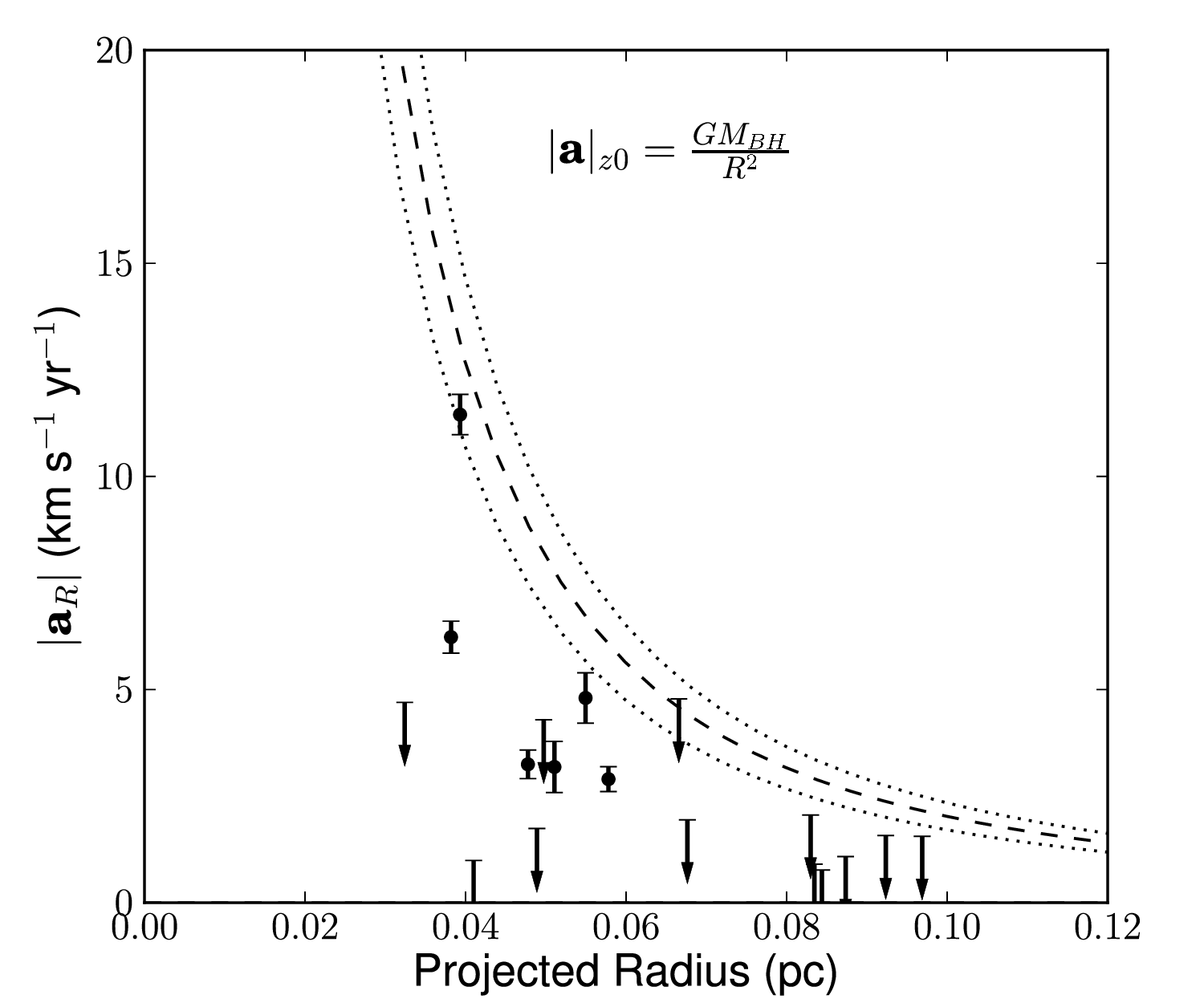}
\figcaption{Accelerations along the radial coordinate as a function
of the stars' projected radius, assuming $R_0$ = 8.23 kpc.  
The theoretical maximum acceleration 
($|a|_{z0}$) for the nominal black hole mass of 4.6$\times$10$^6$ \msun is 
shown as the dashed curve, with the 1$\sigma$ upper and lower boundaries shown
as dotted curves. We detect six significant accelerations 
out to R=1$\farcs$5 (0.06 pc), shown with 1$\sigma$ error bars.
These sources have known line-of-sight distances and therefore have the
best determined orbital solutions. Stars with 3$\sigma$ acceleration upper 
limits below the theoretical maximum acceleration are shown as downward pointing 
arrows and have strong constraints on their line-of-sight distances.
}
\label{fig:acc_r2d}
\end{figure}

Our orbital analysis breaks down into the following three categories based on
the information content contained in the acceleration measurement:
1) stars with significant acceleration detections (N = 6),
2) stars with acceleration upper limits 
below or equivalent to the nominal theoretical maximum acceleration and 
for which a lower limit to the line-of-sight distance can be estimated 
(i.e., inferred from the lack of acceleration; N = 12),
and 3) all other stars (N = 98). 
Stars with acceleration detections and with useful upper limits are shown in
Figure \ref{fig:acc_r2d}. For stars that show significant deviations from linear
motion in the plane of the sky, the measured $a_R$ is converted to a line
of sight distance through the following relationship:
\begin{equation}
\label{eq:arad}
a_R = \frac{-GM_{tot}(r)R}{r^3} \simeq \frac{-GM_{tot}(R)R}{(R^2 + z^2)^{3/2}}.
\end{equation}
While we do not know the
line-of-sight distance (and therefore the full 3D distance) {\em a priori},
we use the star's projected radius ($R$) as the star's 3D radius ($r$)
to determine $M_{ext}$, which is a lower limit on the true extended mass. 
This allows Equation \ref{eq:arad} to be rearranged to obtain $z$,
\begin{equation}
\label{eq:zz}
|z| = \bigg[\bigg(\frac{GM_{tot}(R)R}{a_R}\bigg)^{2/3} - R^2\bigg]^{1/2},
\end{equation}
We note that there is a sign ambiguity 
in the line-of-sight distance, which results in degenerate orbital solutions.

% Generated by python code:
% syelda_yngstars.accelLimit()

With two possible $z$ solutions in hand, two sets of orbital elements are 
found. The probability density functions (PDFs) for each are constructed by carrying 
out a Monte-Carlo simulation in which 10$^5$ artificial data sets are created. In each 
data set, we sample the six kinematic measurements from Gaussian distributions, which 
have a mean and 1$\sigma$ width corresponding to the variables's measured values and 
uncertainties. The gravitational potential parameters, $M_{BH}$, $R_0$, $x_0$, and $y_0$,
are sampled from a 4-dimensional PDF based on the orbit of S0-2, and $\rho_0$ is sampled
from a Gaussian distribution centered on the value quoted above.
We note that in a given trial, all stars' orbits are determined using the same
gravitational potential. Figure \ref{fig:orbs} shows the $e$, $i$, and $\Omega$ 
PDFs as a function of $z$ for the six accelerating sources.
The PDFs are constrained to small regions of parameter space 
for positive and negative $z$. Each of the degenerate sets of solutions have 1$\sigma$
widths in $i$ and $\Omega$ of less than 7$\deg$ and in eccentricity of less than 
0.14 for each of these stars. 

% Generated by python code:
% syelda_yngStars.orb_pdfs_accelerators()
\begin{figure}
\begin{center}
\includegraphics[scale=0.5]{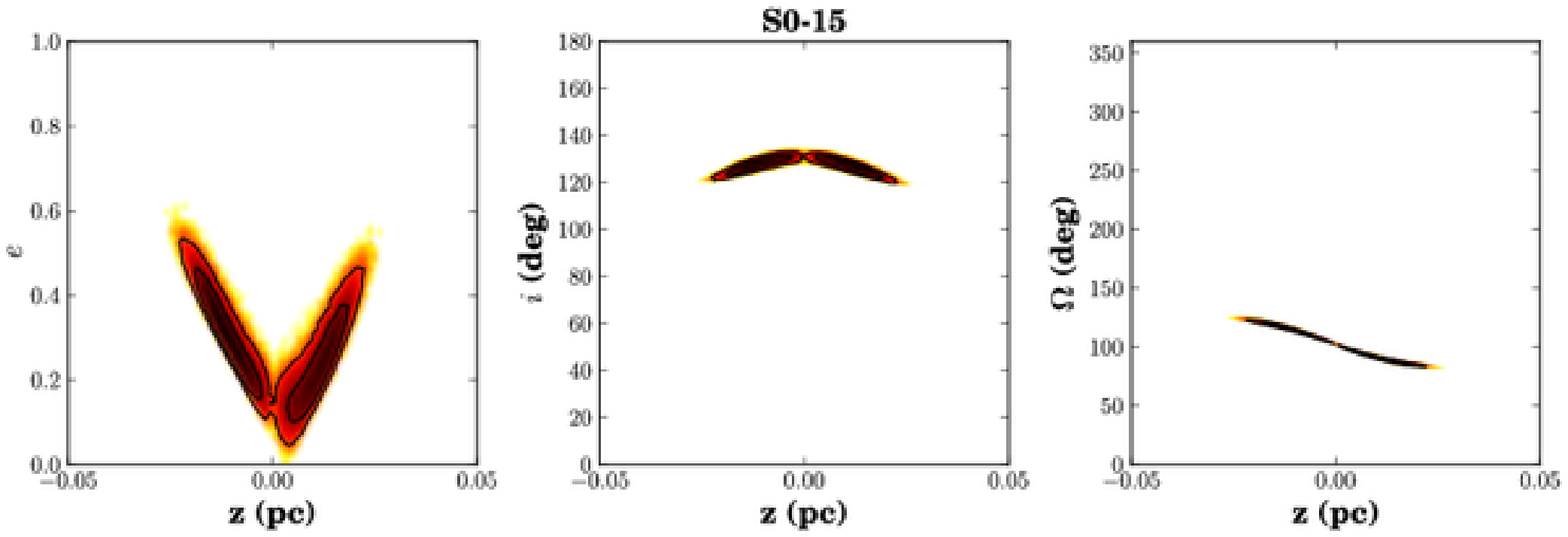}
\includegraphics[scale=0.5]{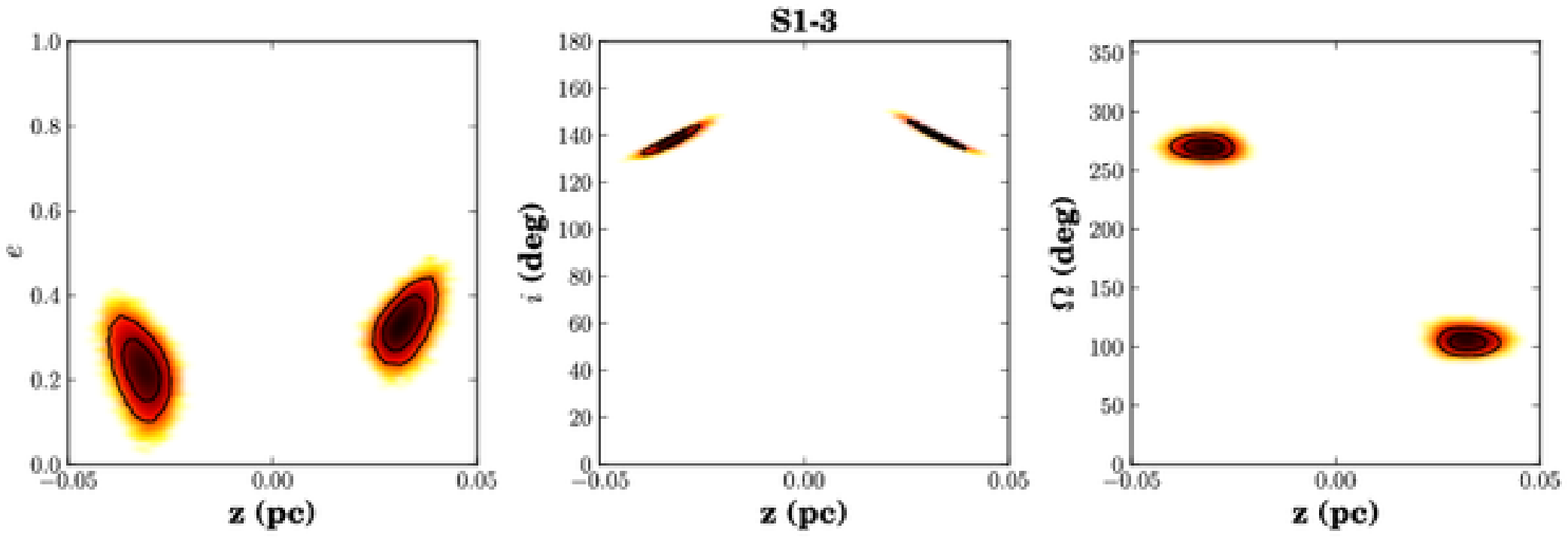}
\includegraphics[scale=0.5]{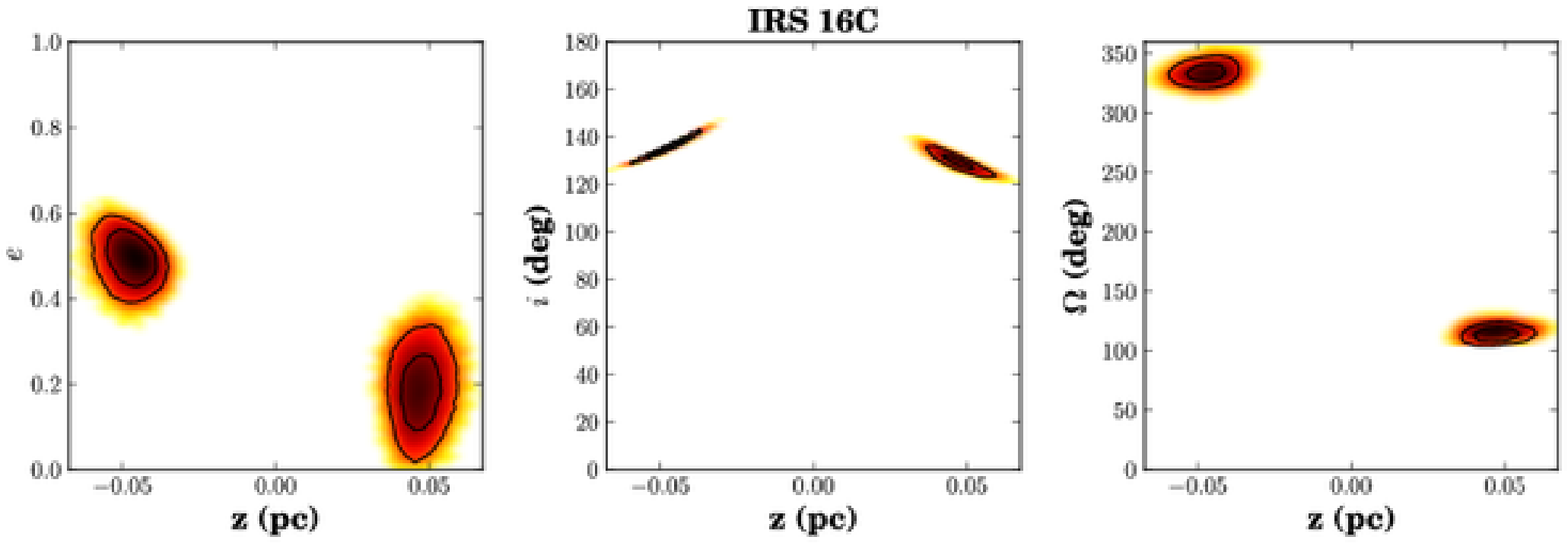}
\end{center}
\figcaption{The probability distribution functions for eccentricity ({\em left}),
inclination ({\em middle}), and angle to the ascending node ({\em right}) 
as a function of the line-of-sight distance for the six stars with significant 
accelerations in the plane of the sky. The absolute value of the line-of-sight 
distance, $|z|$, is precisely determined for each of these stars from their measured 
accelerations. The sign ambiguity of $z$ results in the degenerate set of solutions.
The stars S0-15 and S1-14 have solutions consistent with $z$=0.
The 1$\sigma$ and 2$\sigma$ contours of the PDFs are overplotted as solid lines.
}
\end{figure}

\begin{figure}[htb]
\ContinuedFloat
\begin{center}
\includegraphics[scale=0.5]{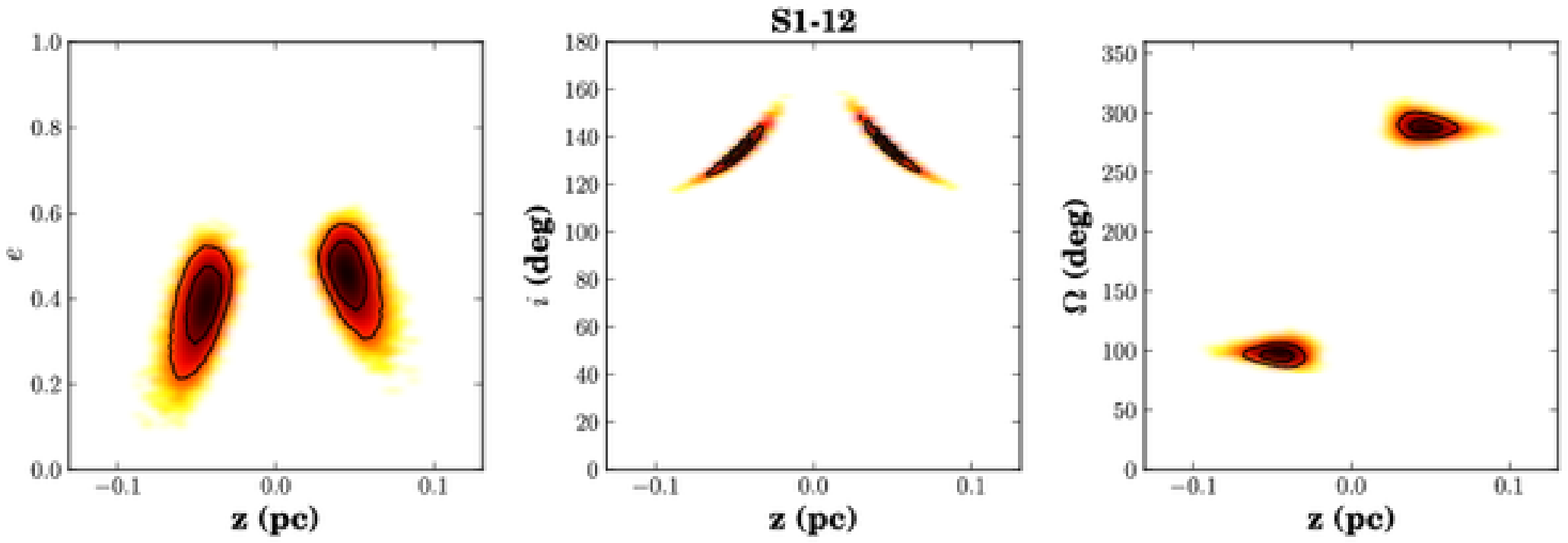}
\includegraphics[scale=0.5]{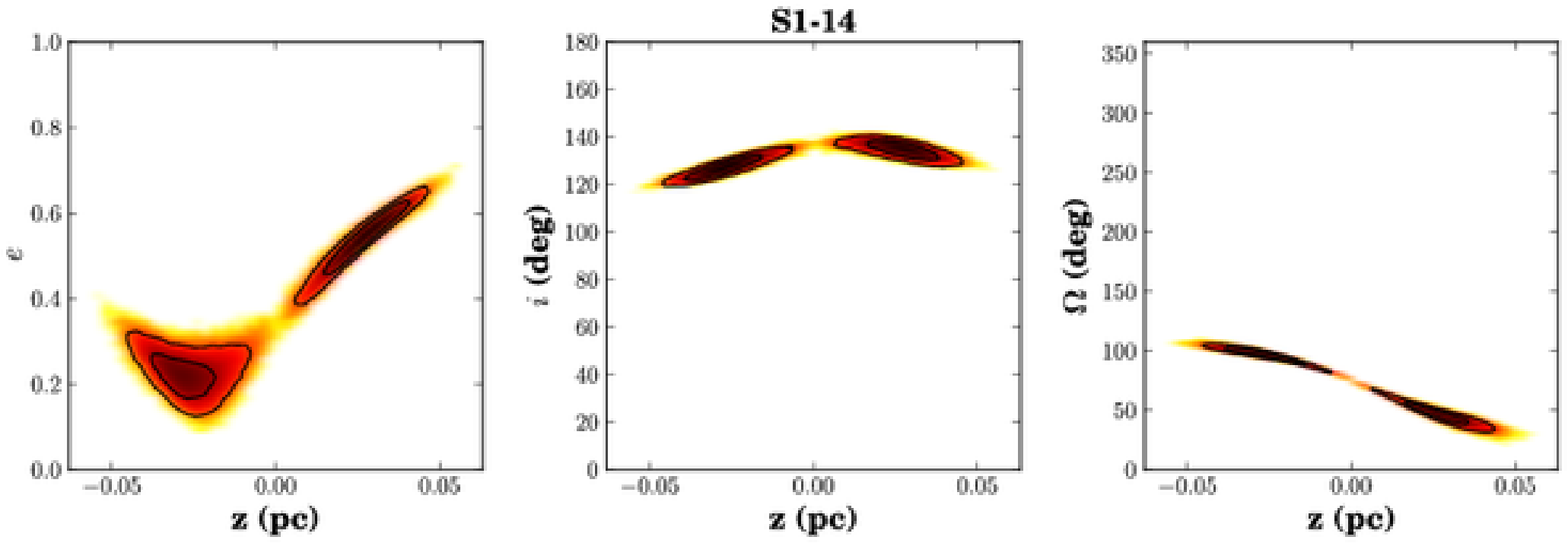}
\includegraphics[scale=0.5]{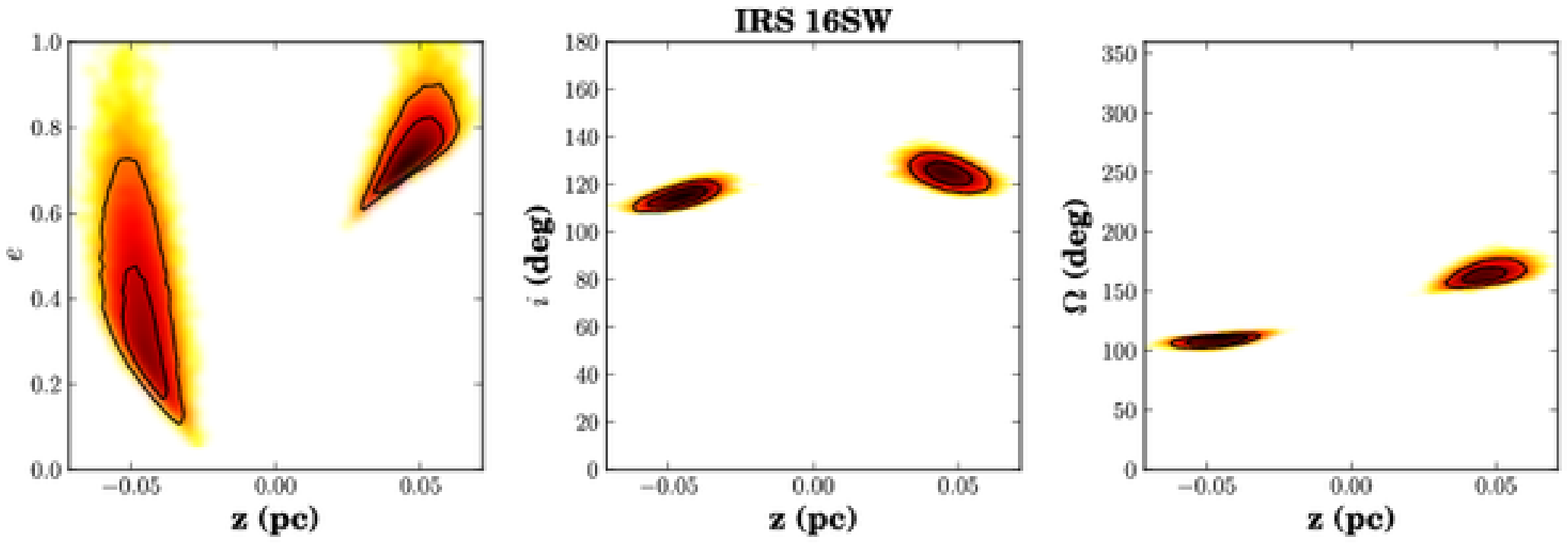}
\end{center}
\figcaption{(Continued)
}
\label{fig:orbs}
\end{figure}
\clearpage

Accelerations that are consistent with zero can also provide 
constraints on the line-of-sight distance. The maximum acceleration a star 
can have is $|a|_{z0} = GM_{tot}(R)/R^2$, which is equivalent to the acceleration the 
star would have if its line-of-sight distance were $z$ = 0. 
A star with a 3$\sigma$ acceleration upper limit, $|a|_{3\sigma}$, that is less 
than $|a|_{z0}$ must therefore have a 3-dimensional position that is larger than
its observed projected position (i.e., $|z| > |z(a_{3\sigma})|$). 
Thus, the non-detection of 
an acceleration translates to a lower limit on the line-of-sight distance. 
Furthermore, the minimum acceleration allowed, $|a|_{bound}$, is set by the
assumption that the star is bound.  For stars with 3$\sigma$ upper 
limits below $|a|_{z0}$, 
we carry out the simulations similarly to those above, with the exception that
we sample from a uniform acceleration distribution between $a_{bound}$ and 
$a_{3\sigma}$. For all other stars, including those outside the central 10$\arcsec$ 
field (i.e., stars from the mosaic fields), we sample from a uniform acceleration 
distribution between $a_{bound}$ and $a_{z0}$.  We include the $e$, $i$, and 
$\Omega$ PDFs for the 12 stars with 3$\sigma$ upper limits in an electronic
Appendix.

\section{Results}
\label{sec:results}
Compared to our earlier efforts in \citet{lu09}, we have increased 
1) the radial extent of our observations from 3$\arcsec$ to 13.3$\arcsec$,
2) the number of young stars in our sample from 32 to 116, 
3) the number of reliable acceleration detections from one to six, and
4) the number of useful acceleration upper limits (which constrain the orbital
parameters) from seven to 12.
Taken together, these improvements provide tighter constraints on the orbits
of the young stars as well as any kinematic structures present. We construct
various ensemble distributions from the real data in \S\ref{sec:2dPDF}.
We also incorporate simulations of mock data sets, which are run through our 
orbital analysis and combined into the same distribution functions. The results
are compared to the real data in order to model the true underlying distributions and
to explore any biases introduced by measurement uncertainties
and assumptions in our analyses (\S\ref{sec:compare2sims}).

\subsection{Observed Global Kinematic Structures}
\label{sec:2dPDF}
\begin{figure}[!h]
\epsscale{0.7}
\plotone{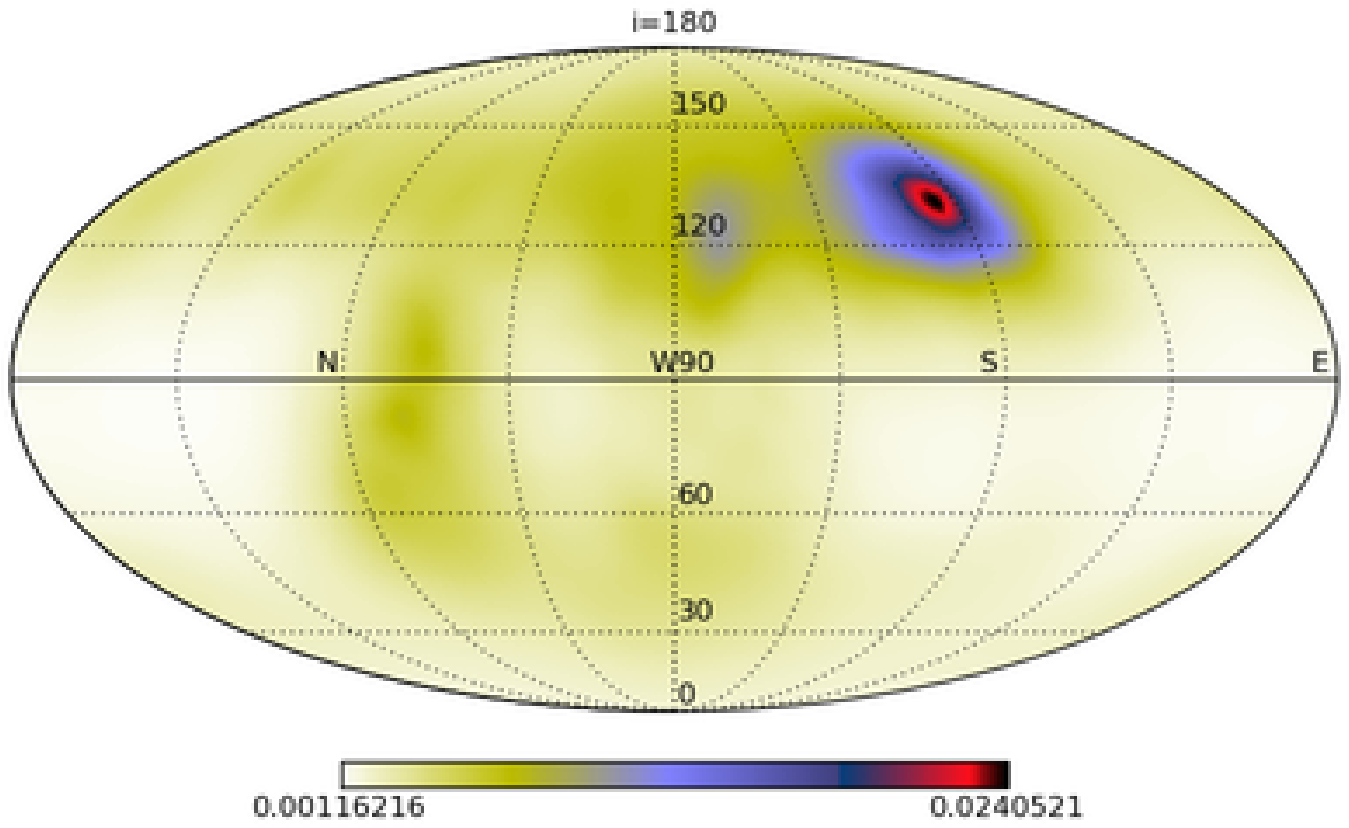}
\figcaption{Density of normal vectors to the orbital planes (in stars deg$^{-2}$) of 
all 116 stars in the sample. The direction of the normal vector is described by the 
inclination ($i$; horizontal lines spaced 30$\deg$ apart) and the angle to the 
ascending node ($\Omega$; longitudinal lines spaced 45$\deg$ apart, with the line marked
E representing $\Omega$ = 0$\deg$).  An overdensity of 
0.024 stars deg$^{-2}$ occurs at ($i$, $\Omega$) = (130$\deg$, 96$\deg$).
}
\label{fig:diskDensity}
\end{figure}

Kinematic structures are identified by constructing a
density map of the normal vectors to the stars' orbital planes.
The direction of the normal vector is described by inclination,
$i$, and the position angle of the ascending node, $\Omega$.
The density at each location (in stars deg$^{-2}$) in the PDF($i$, $\Omega$) map is 
computed for the six nearest normal vectors within a given trial in the MC 
simulation \citep{lu09}. These values are then averaged over all 10$^5$ MC trials 
to produce an average density map. Figure \ref{fig:diskDensity} shows the density 
map for all 116 stars in our observed sample using the HEALpix framework 
\citep[$N_{pixels}$ = 49152 equal-area pixels;][]{gorski05}.
A clear peak of 0.024 stars deg$^{-2}$ is found at ($i$, $\Omega$) = 
(130$\deg$, 96$\deg$) and the half-width at half maximum (HWHM) from the peak 
density is 15$\deg$.
This peak corresponds to the clockwise disk reported in many earlier publications,
including the original work by \citet{levin03}.
The direction of the disk plane differs by 15$\deg$ ($\sim$3$\sigma$) from that in 
\citet{lu09} due to our use of an improved model for the optical distortion in our
images \citep{yelda10} and is within 1$\sigma$ agreement with \citet{paumard06} and 
\citet{bartko09}. 

\begin{figure}
\epsscale{1.0}
\plotone{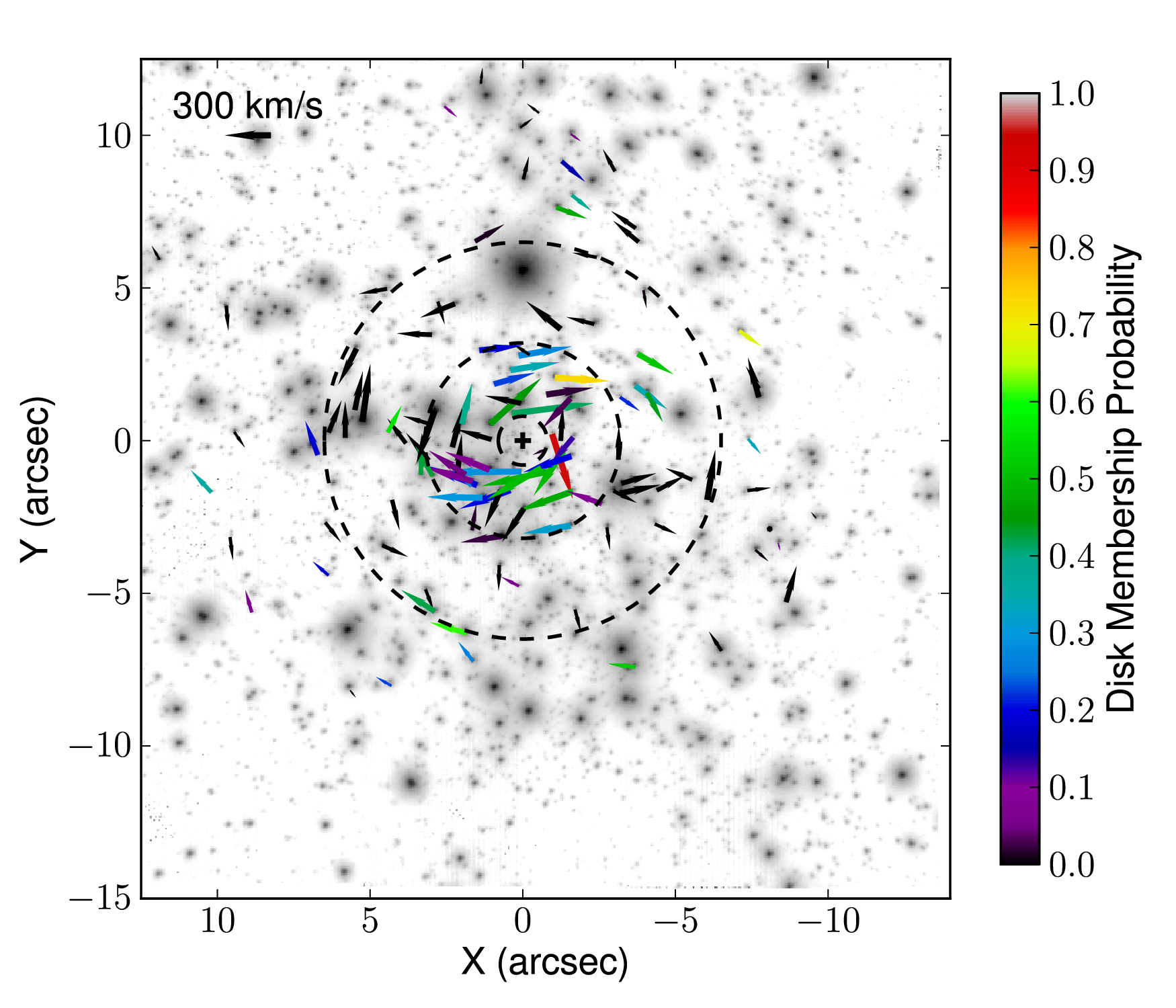}
\figcaption{Velocity vectors of all 116 stars in the sample. Sgr A* is marked as a cross 
in the center. The arrows are color-coded according to their disk membership
probability. The dashed circles mark the three radial bins discussed in 
\S \ref{sec:signif} and are located at $R$ = 0$\farcs$8, 3$\farcs$2, and 6$\farcs$5.
}
\label{fig:velVectors}
\end{figure}

While the existence of the clockwise disk has been well-established prior to
this work, it is important to estimate disk membership probabilities for each star 
in order to properly characterize the disk properties. The probabilities are calculated
following \citet{lu09} and are $1 - L_{non-disk}$, where $L_{non-disk}$ is the 
likelihood that the star is {\em not} part of the disk, and is computed as
\begin{equation}
L_{non-disk} = 1 - \frac{\int_{disk} PDF(i, \Omega)dSA}{\int_{peak} PDF(i, \Omega)dSA}
\end{equation}
\begin{equation}
\int_{disk}dSA = \int_{peak}dSA,
\end{equation}
where $SA$ is the solid angle measured as the contour at which the density drops to
50\% of its peak value ($\sim$0.2 sr or FWHM=15.2$\deg$).
The disk membership probabilities ($1 - L_{non-disk}$) are
given in Table \ref{tab:yng_pm_table} and the stars' proper motion vectors 
are color-coded according to these probabilities in Figure \ref{fig:velVectors}.
We note that five of the six accelerating sources are among the most likely disk
members ($1 - L_{non-disk} >$ 0.4; see Figure \ref{fig:probNodes} in Appendix \ref{app:nodes}).
In \citet{lu09}, non-disk candidates were identified at the 3$\sigma$
significance level ($L_{non-disk} >$ 0.9973), which would result in 58 stars that
are not disk stars, and the remaining stars would be considered as
disk members ($N_{candidates}$ = 58) and include all six stars with significant
acceleration detections.  
Using this metric, the fraction of stars in our sample that are candidate disk 
members is 50\%, which is consistent with earlier work.  The true disk fraction, 
however, is likely to be smaller than this and is explored below using mock data sets.

% Generated by python code:
% plot_disk_healpix.go()

% Generated by python code:
% sythesis.plot_velocity()

\begin{figure}[htb]
\epsscale{1.0}
\plotone{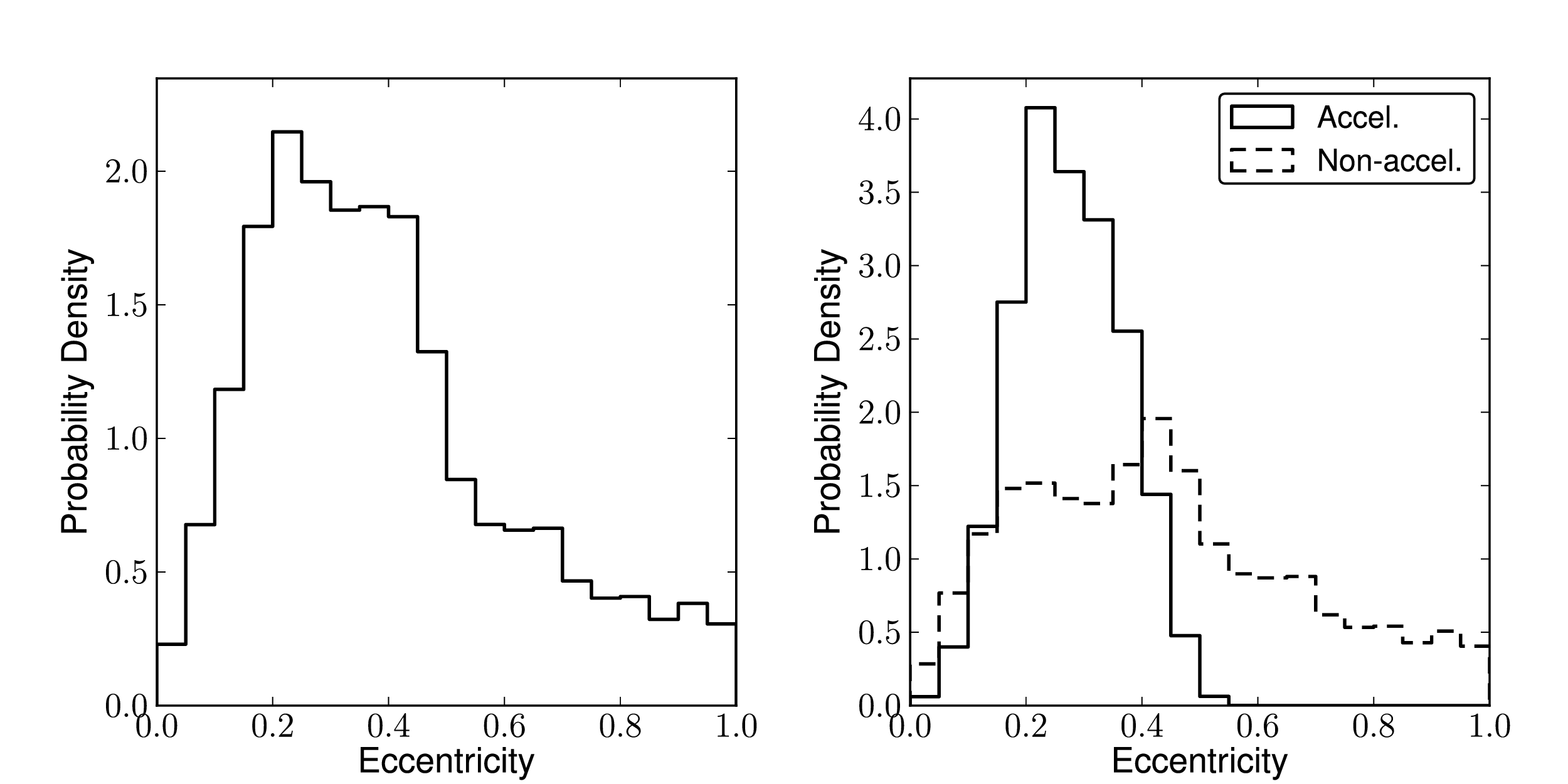}
\figcaption{{\em Left:} Eccentricity distribution of the clockwise disk.
All orbital solutions falling within 15.2$\deg$ of the disk are included, thereby 
weighting the distributions by disk membership probability. 
{\em Right:} Eccentricity distributions shown separately for likely disk 
members with acceleration detections ({\em solid}) and without ({\em dashed}). 
}
\label{fig:eccPDF}
\end{figure}

The stars in the clockwise disk are found to have non-circular orbits. 
In Figure \ref{fig:eccPDF}, we plot the solutions of the accelerating 
stars separately from those of the non-accelerating stars. Only orbital solutions that 
fall within 15.2$\deg$ of the disk solution are included for each star, thus weighting 
the distribution by disk membership probabilities. Both eccentricity distributions are 
clearly offset from $e$ = 0, with an average for the accelerating stars of
$\langle e \rangle$ = 0.27 $\pm$ 0.09 and the non-accelerating stars 
$\langle e \rangle$ = 0.43 $\pm$ 0.24.
The uncertainty on the eccentricity reported here is the standard deviation
of the distribution ($\sigma_{e,measured}$).  Below we explore the impact and possible
bias of measurement uncertainty (since the eccentricity is a positive definite quantity) 
and of the uniform acceleration prior on what is observed.

% Generated by python code:
% sythesis.plotOrbAccelerators()

The scale height of the disk ($h/r$) can be estimated using the velocity 
dispersion perpendicular to the disk plane ($\sigma_{\vec{n}}$) and the average 
magnitude of the 3-dimensional velocity ($\langle v \rangle$) as:
\begin{equation}
\frac{h}{r} = \frac{\sigma_{\vec{n}}}{\langle v \rangle},
\end{equation}
where the velocity dispersion is corrected for the bias due to measurement 
uncertainties and is weighted by disk membership probability.
This quantity can be related to a dispersion angle, $\Delta\theta$, about the 
disk plane: $h/r \sim \sqrt{1/2}\Delta\theta$ \citep{beloborodov06}. 
We find $\sigma_{\vec{n}}$ = 33 $\pm$ 4 km s$^{-1}$ and 
a scale height of $h/r$ = 0.10 $\pm$ 0.01, which gives a 
dispersion angle of $\Delta\theta$ = 8.0$\deg$ $\pm$ 1.0$\deg$, consistent
with earlier estimates \citep{paumard06,lu09,bartko09}. The data were also
separated into two radial bins (at $R$ = 3.3$\arcsec$) that have equal total 
weights (i.e., total disk membership probabilities) and the scale height
within each radial bin was computed. We find for the inner ($N$ = 42 stars) 
and outer ($N$ = 74 stars) bins $h / r$ = 0.10 $\pm$ 0.02 and 
$h / r$ = 0.07 $\pm$ 0.01, respectively.  Thus, the dispersion angle of the disk 
does not get larger with radius.

\subsection{Comparison with Simulations}
\label{sec:compare2sims}
\subsubsection{Mock Data Sets}
\label{sec:sims}
To explore the impacts of measurement error and the acceleration prior 
assumptions used in our orbital analysis, we construct mock data sets 
with known underlying kinematic properties.
Both stars with a common orbital plane (i.e., in a stellar disk) and
stars on randomly-oriented orbits are modeled, allowing us to quantify our 
ability to reconstruct orbital elements from mock data and subsequently 
identify kinematic structures and their members.   

In each set of simulations performed, we create mock kinematic data 
($x$, $y$, $v_x$, $v_y$, $v_z$, $a_x$, $a_y$), add errors to each of these
variables, and run our MC orbital simulations in the same way that the observed 
data are treated. These mock data are generated by
assuming a true orbit (with elements $P$, $e$, $i$, $\Omega$, $\omega$, and $T_0$)
around a point mass of 4.6 $\times$ 10$^6$ \msun. We choose to use a point mass
since including the extended mass in the analysis of the real data did not make
a difference in the final results. For all stars in all simulations, 
$T_0$ is uniformly sampled from 1995 (the beginning of our observations) to 
1995 + $P$, and $\omega$ from 0$\deg$ to 360$\deg$. For stars on orbits 
that are randomly oriented, we sample from uniform distributions in $cos(i)$
from 0 to 1 and in $\Omega$ from 0$\deg$ to 360$\deg$.
We assume the surface density profile found by \citet{do13} for the young
stars beyond $R$ = 1$\arcsec$, $\Sigma(R) \propto R^{-1.14}$, which when combined with 
the black hole mass gives the period of the orbit.
The eccentricities are sampled from a thermal distribution ($n(e) \sim e$). 
When simulating disk stars, the semi-major axes are randomly sampled such that the 
resulting surface density profile in the disk plane follows the observed profile, 
$\Sigma(R) \propto R^{-1.9}$
\citep{paumard06,lu09,bartko09}. The orientation of the disk plane is set to 
that of the observed disk, ($i$, $\Omega$) = (130$\deg$, 96$\deg$). The 
distribution of orbital eccentricities for the disk stars depends on the simulation. 
From these simulated 
orbits, we select the 3D positions, velocities, and accelerations at a 
particular ``observation'' time, which we take as 2004.2, the mean time of our 
actual observations. Mock accelerations are only determined for stars within 
5$\arcsec$ of the black hole, consistent with our treatment of the real data. 
We consider only those simulated stars whose projected positions are within the 
field of view covered by the Keck and VLT spectroscopic observations.

The noise added to the mock data is based on the
observed measurement uncertainties as a function of distance from the black hole,
as astrometric uncertainties tend to increase with radius (see
Figure \ref{fig:astrometryR2d}).
We determine the minimum and maximum uncertainties in position, velocity, and 
acceleration of the known young stars in our sample in 1$\arcsec$ radial intervals.
In each trial, the uncertainties assigned to a simulated star are randomly selected from a
uniform distribution between these boundaries for the appropriate radial interval
(dependent on the simulated star's projected radius). We then run 10$^4$ MC trials 
in which we sample from the mock data and the assigned uncertainties for each 
simulated star. This results in a 6-dimensional PDF representing
the probability distributions for the six orbital elements.
For simplicity, we only use acceleration information if the star's simulated
acceleration is significant (5$\sigma$), given its assigned uncertainty. 
For the remaining stars, a uniform acceleration prior is used, imposing the same
boundaries of the minimum acceleration allowed given a bound orbit and the maximum
acceleration given the star's projected radius.

\begin{deluxetable}{lccccc}
\tabletypesize{\scriptsize}
\tablewidth{0pt}
\tablecaption{Mock Data Sets}
\tablehead{
  \colhead{ID} & 
  \colhead{$N_{cases}$} & 
  \colhead{$N_{stars}$} & 
  \colhead{$R (\arcsec)$} & 
  \colhead{$f_{disk}$} & 
  \colhead{$e_{disk}$}}
\startdata
  1   & 1000 &  116 & 0.8 $\;$-$\;$ 14.0  &        0.00 &   -          \\
  1a  & 1000 &   40 & 0.8 $\;$-$\;$  3.2  &        0.00 &   -          \\
  1b  & 1000 &   40 & 3.2 $\;$-$\;$  6.5  &        0.00 &   -          \\
  1c  & 1000 &   40 & 6.5 $\;$-$\;$ 14.0  &        0.00 &   -          \\
  2   &   13 &  100 & 0.8 $\;$-$\;$ 14.0  &        1.00 & 0.00 - 0.5  \\
  3   &  110 &  120 & 0.8 $\;$-$\;$ 14.0  & 0.05 - 0.55 & 0.32         \\
\enddata 
\label{tab:mockData}

\end{deluxetable}

Table \ref{tab:mockData} summarizes the mock data sets created and we describe
their details here: 
\begin{enumerate}
\item {\em Significance of Kinematic Structures (\S \ref{sec:signif}):}
The statistical significance of a density peak in the PDF($i$, $\Omega$) map is 
quantified through a comparison to the density expected from a population of stars with 
randomly-oriented orbits. To this end, we create 1000 separate data sets, each of which 
includes 116 stars on randomly-oriented orbits (ID 1 in Table \ref{tab:mockData}), and 
run our orbital analysis. Likewise, 1000 data sets are generated for an isotropic 
population of 40 stars within each of three radial bins (0$\farcs$8 - 3$\farcs$2, 
3$\farcs$2 - 6$\farcs$5, and 6$\farcs$5 - 13$\farcs$3) for the purposes of
quantifying the significance of substructures as a function of radius.

\item {\em The Eccentricity Distribution of Disk Stars (\S \ref{sec:ecc}):} 
The eccentricity distribution of the stellar disk population is explored with mock 
data sets, each of which consists of a disk of 100 stars, each having the same 
eccentricity vector of magnitude $e_0$ and a direction that is randomly oriented 
within the disk plane, and with other orbital parameters as described above (ID 2).  
Thirteen data sets with the following eccentricities are modeled: $e_0$ =
[0.0, 0.05, 0.1, 0.15, 0.2, 0.25, 0.27, 0.3, 0.32, 0.35, 0.4, 0.45, 0.5].
These data sets are run through our orbital analysis and the resulting eccentricity
distributions are compared to that of the observed distributions in order to determine 
the eccentricity to be used for simulations of disk stars. The models with 
$e_0 \sim$ 0.3 give the most similar eccentricity distributions (in a 
least-squares sense; see \S \ref{sec:ecc}), and we therefore choose to use 
$e_0$ = 0.3 for modeling disk stars (see next item).

\item {\em Fraction of Stars in Disk (\S \ref{sec:diskfrac}):}
Orbits of both disk stars and stars with isotropically-distributed orbital planes 
are generated to estimate the true disk fraction (ID 3). Eleven disk fractions
are tested, from $f_{disk}$ = 5\% to $f_{disk}$ = 55\%, spaced every 5\% and for
a total of 116 stars.  For each disk fraction, 10 independent mock data sets are 
generated, resulting in 110 sets in total. The disk stars' orbital properties 
are $i$ = 130$\deg$, $\Omega$ = 96$\deg$, and $e$ = 0.32.
The eccentricity was chosen based on the orbit simulations on the
mock data above (ID 2)\footnote{An earlier analysis of the observations
revealed that a disk with $e$ = 0.32 produced the most similar eccentricity
distributions. The updated value of $e$ = 0.3 does not significantly affect
the results.}.

\end{enumerate}

\subsubsection{Significant Kinematic Features}
\label{sec:signif}
\begin{figure}[!h]
\epsscale{0.7}
\plotone{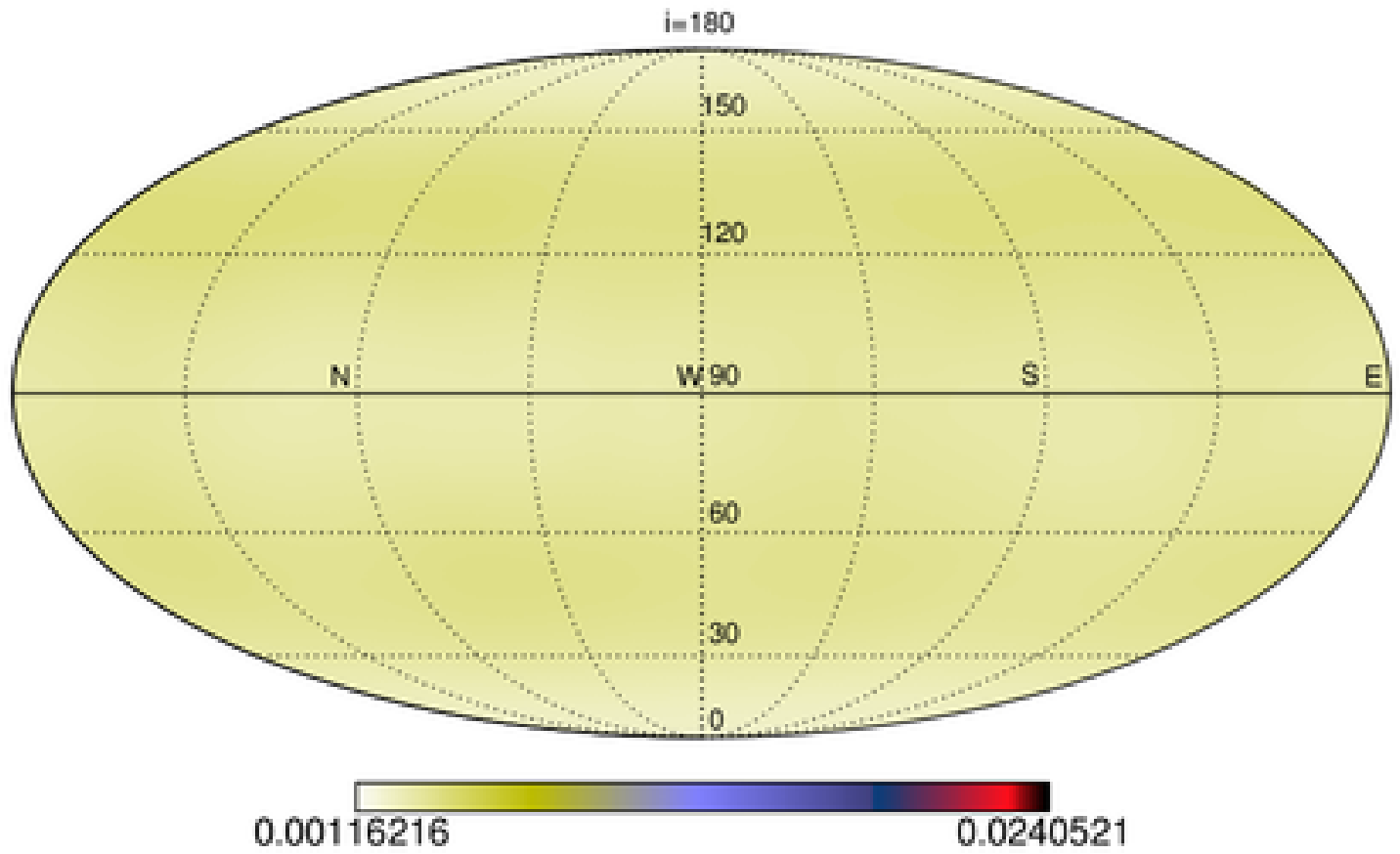}
\figcaption{Average PDF($i$, $\Omega$) map in units of stars deg$^{-2}$ for an isotropic 
population of 116 stars. The distribution of normal vectors is relatively uniform
over the sky, with a slight deficit of edge-on orbits ($i$ = 90$\deg$) due to the 
uniform acceleration prior.
}
\label{fig:ave_iso_heal_map}
\end{figure}

To quantify the significance of peaks in the observed PDF($i$, $\Omega$) map, the
peak density is compared to the density of normal vectors expected for an 
isotropic population.
Density maps were produced for all 1000 simulated data sets (ID 1 in
Table \ref{tab:mockData}). Figure \ref{fig:ave_iso_heal_map} 
shows the average PDF($i$, $\Omega$) map for the isotropic data. 
The normal vectors are nearly uniformly distributed over the sky with a 
slight deficit of edge-on orbits ($i$ = 90$\deg$) due to the uniform acceleration 
prior. This prior results in smaller line-of-sight distances on average than the simulated 
stars' true distances, and small $|z|$ will favor face-on orbits over edge-on orbits. 
Due to this slight dependence
on inclination, we split the maps up into inclination bins spaced every 20$\deg$
from 0$\deg$-180$\deg$ and determine the peak density within each bin.
The average, $\rho_{iso,i}$, and standard deviation, $\sigma_{iso,i}$, of the peak 
densities over all 1000 simulations are then calculated for each inclination bin, $i$. 
We then quantify the significance of any density enhancements, $\rho_{peak}$, in our data as
\begin{equation}
\label{eq:sig}
S = \frac{\rho_{peak} - \rho_{iso,i}}{\sigma_{iso,i}},
\end{equation}
where the inclination bin, $i$, is selected based on the location of the observed peak
density.

An isotropically distributed set of stars 
has an average peak density of $\rho_{iso,i}$ = 0.006 $\pm$ 0.001 stars deg$^{-2}$.
The peak density of the observed normal vectors for the entire sample 
is $\rho_{peak}$ = 0.024 stars deg$^{-2}$, and has a significance of $S$ = 20.7.

% Generated by python code:
% sythesis.run_isotropic_results()

\begin{figure}
\centering
\includegraphics[scale=0.6]{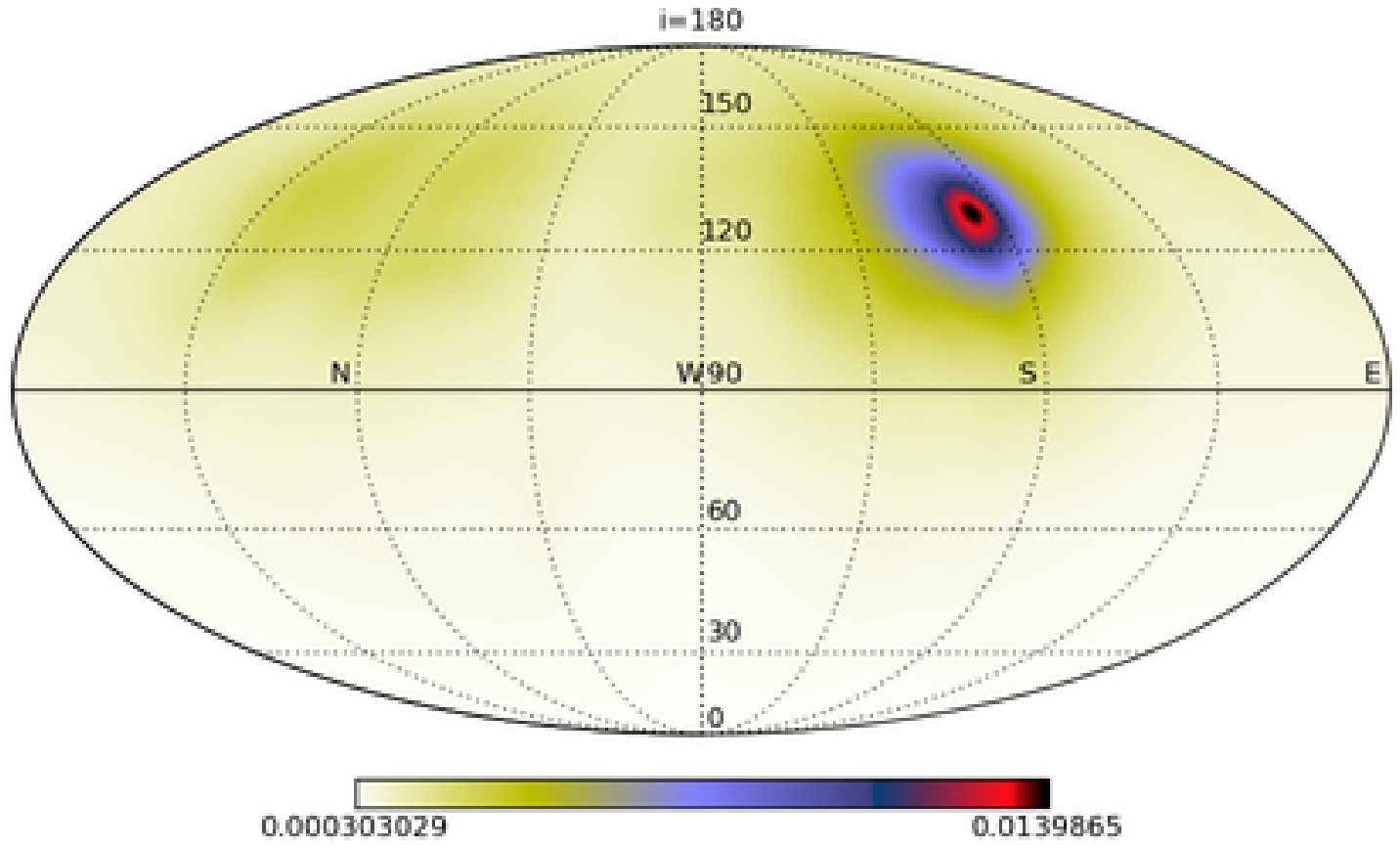}
\includegraphics[scale=0.6]{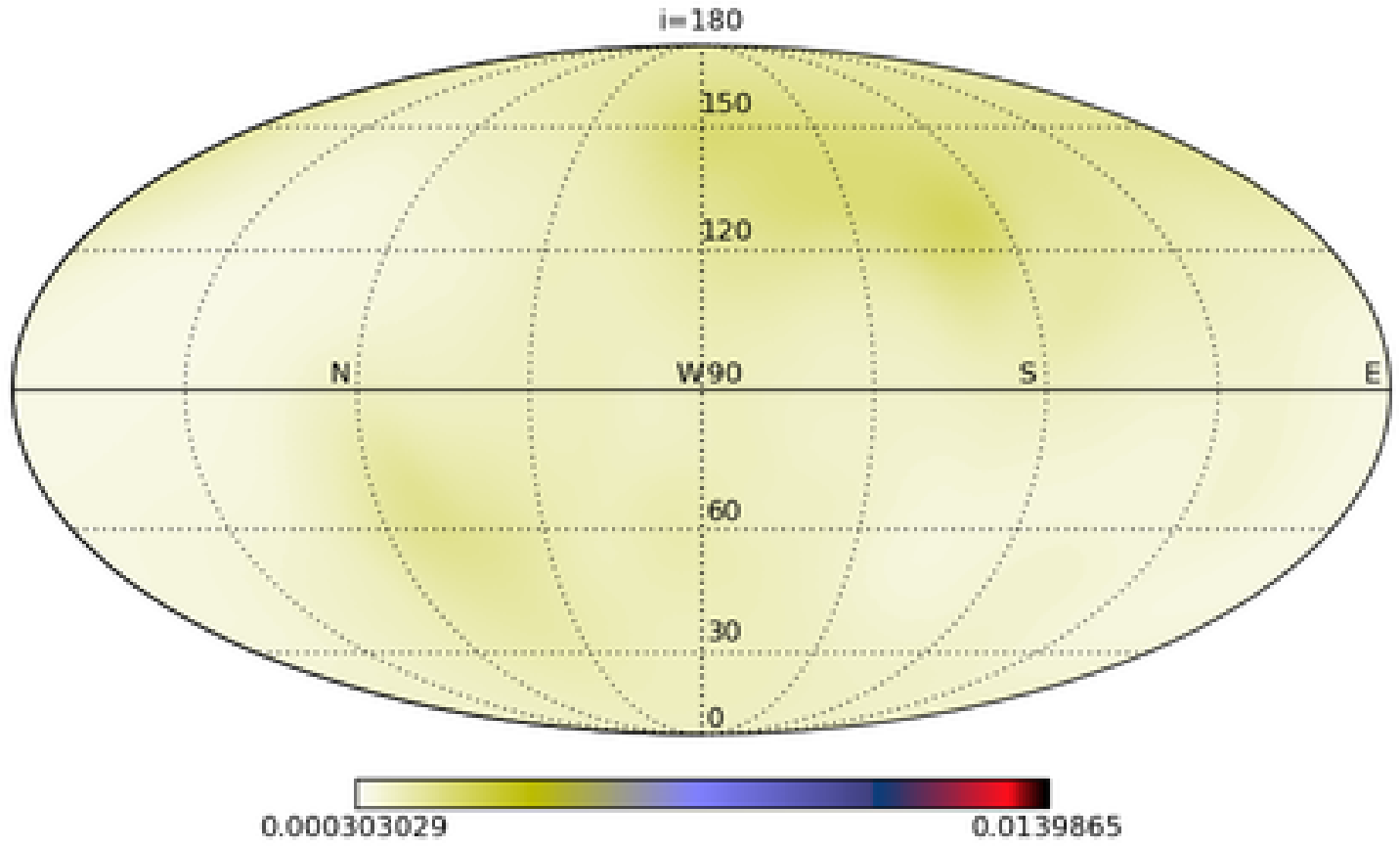}
\includegraphics[scale=0.6]{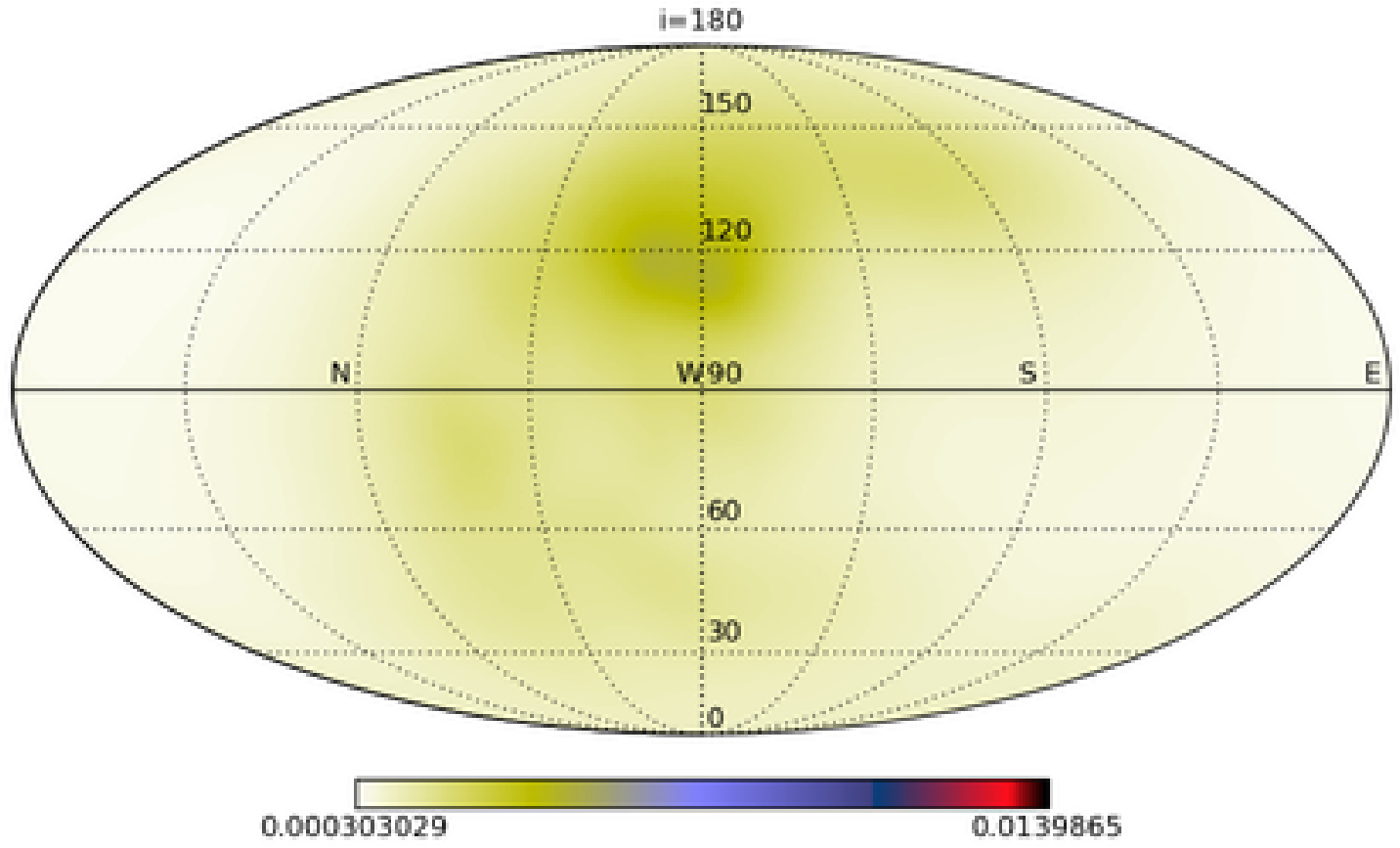}
\figcaption{Density of normal vectors for stars in the three separate radial
bins: 0$\farcs$8-3$\farcs$2 ({\em top}), 3$\farcs$2-6$\farcs$5 ({\em middle}),
and 6$\farcs$5-13$\farcs$3 ({\em bottom}). The clockwise disk feature at
($i$, $\Omega$) = (130$\deg$, 96$\deg$) is prominent in the inner 
radial bin and shows a decrease in density with radius.  The degenerate orbital 
solutions associated with the CW disk stars are seen as the slight density 
enhancement near ($i$, $\Omega$) $\sim$ (130$\deg$, 300$\deg$) in the top panel.
The middle radial interval shows hints of the CW disk and extended structure 
around this location.  In the outermost radial bin, a density
enhancement is seen at ($i$, $\Omega$) = (117$\deg$, 192$\deg$).
The same scaling is used in each plot to show the relative strength of the features.
Recall that the horizontal lines represent $i$ and are spaced 30$\deg$ apart
and the longitudinal lines represent $\Omega$ and are spaced 45$\deg$ apart,
with the line marked E representing $\Omega$ = 0$\deg$. 
}
\label{fig:diskRadial}
\end{figure}

The global structure of the disk can be described by studying its orientation as a 
function of radius. To this end, we group stars into three radial bins, selecting 
radial intervals such that roughly equal numbers of stars ($\sim$40) fall in each 
bin, similarly to the method of \citet{bartko09}. The radial intervals used are 
0$\farcs$8 - 3$\farcs$2, 3$\farcs$2 - 6$\farcs$5, and 
6$\farcs$5 - 13$\farcs$3\footnote{The edges of the three radial bins we use are slightly 
different than those used by \citet{bartko09} since the two studies contain 
different sample sizes and the radial bins were defined such that they each 
contained an equal number of stars.  In our work, we define the edges of the 
bins using $R$ = 0$\farcs$8, 3$\farcs$2, and 6$\farcs$5 and include $\sim$40 
stars per bin, whereas \citet{bartko09} use $R$ = 0$\farcs$8, 3$\farcs$5, 
and 7$\arcsec$ and had $\sim$30 stars per bin. This does not affect the overall 
conclusions.}. Figure \ref{fig:diskRadial} shows the resulting PDF($i$, $\Omega$) 
for each bin. The significance of a density enhancement found in either of these 
maps is determined as in Equation \ref{eq:sig}, but relative to
40 isotropically-distributed orbits within the radial bin of interest.

% Generated by python code:
% plot_disk_healpix.go()

The peak density of normal vectors in the inner radial bin ($N$ = 39 stars)
is 0.014 stars deg$^{-2}$ and is found at ($i$, $\Omega$) = (129$\deg$, 98$\deg$), 
consistent with the angles found when using the entire sample. The significance
of this feature is $S$ = 20.3 and is the only structure detected at these radii.
The middle radial bin ($N$ = 38) shows no significant peak ($S_{max}$ = 1.4),
which differs from what was found by \citet{bartko09}.
Finally, the outermost bin ($N$ = 39) shows an overdensity of 0.004 stars deg$^{-2}$ 
near ($i$, $\Omega$) = (117$\deg$, 192$\deg$), consistent with the feature seen at 
large radii by \citet{bartko09}, and has a significance of $S$ = 5.1. 
We caution, however, that this feature is a result of mainly three stars 
and that the outer radial bin is not sampled uniformly in azimuth.
While \citet{bartko09} report a significant CW feature in each of their
three radial bins at different angles, hence leading to the claim of a warp, we
do not detect any features at intermediate radii. 
Furthermore, the previously proposed counterclockwise disk is not detected
in any radial bin in this work. We therefore conclude that the population 
not on the clockwise disk 
(aside from the three stars with common orbital planes in the outer radial bin) 
is consistent with an isotropic distribution within the
measured uncertainties.

\subsubsection{Eccentricity Distribution of Candidate Disk Stars}
\label{sec:ecc}
\begin{figure}[!h]
\begin{center}
\centering
\includegraphics[scale=0.45]{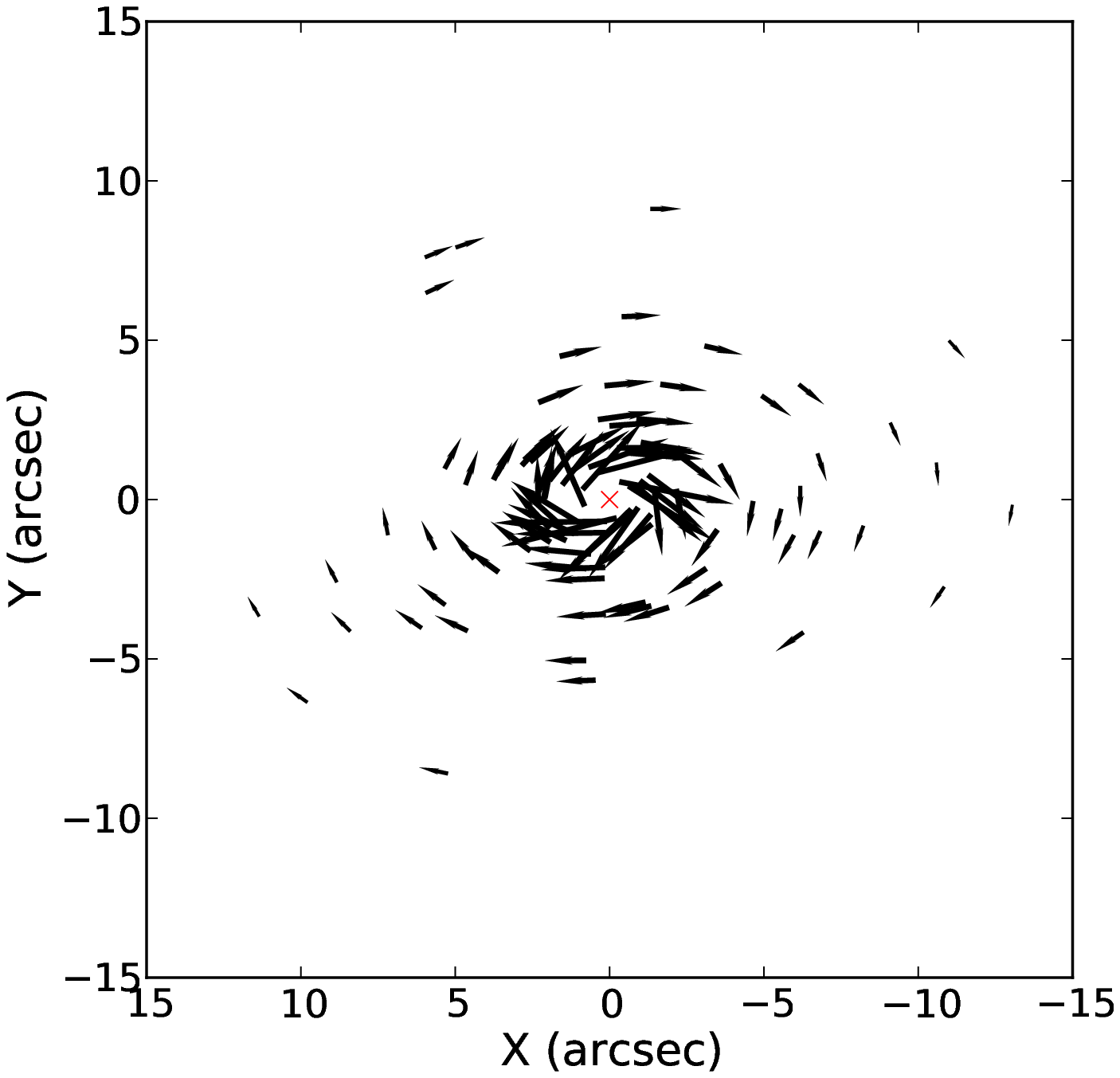}
\includegraphics[scale=0.45]{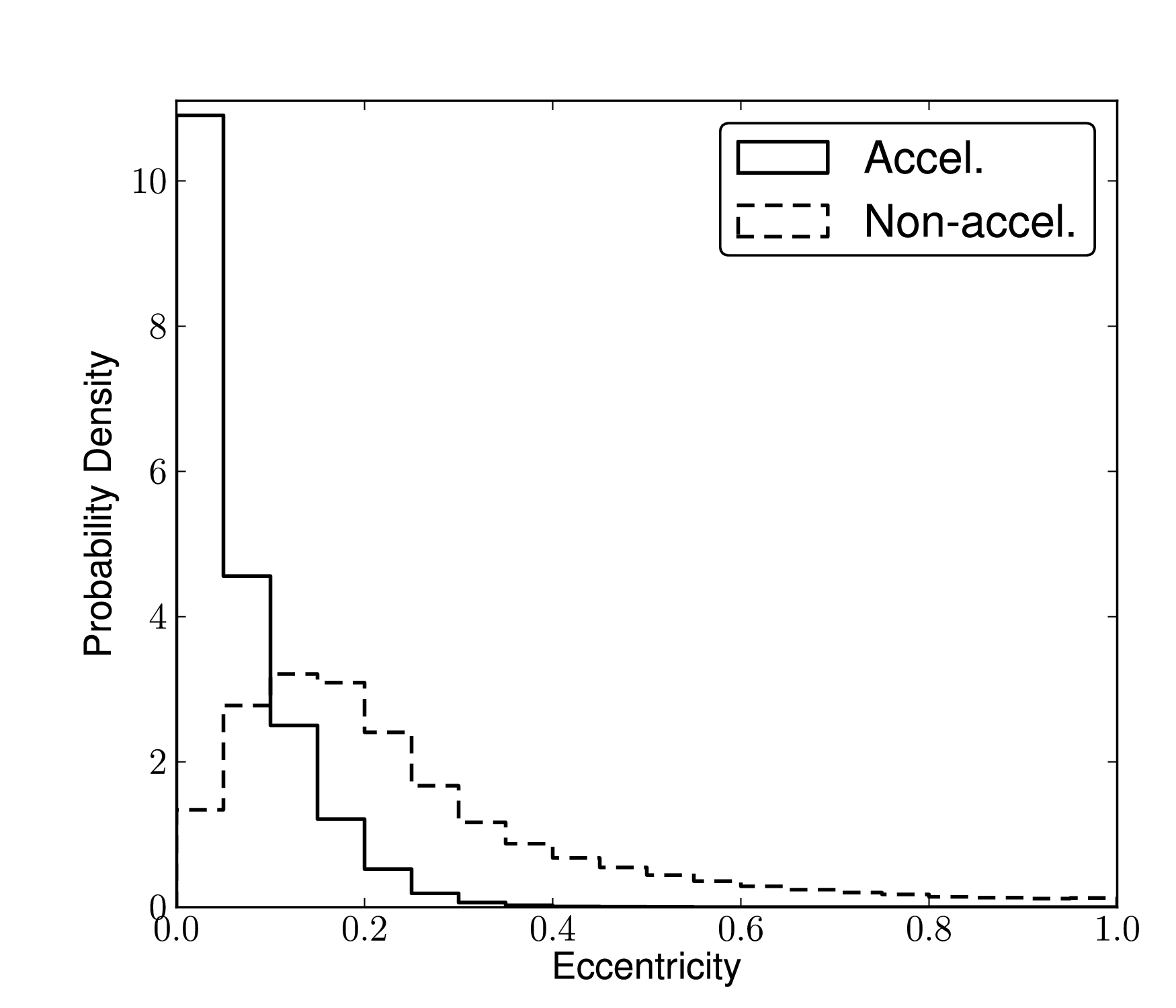}
\begin{minipage}[b]{1.00\linewidth}
\centering
\includegraphics[scale=0.45]{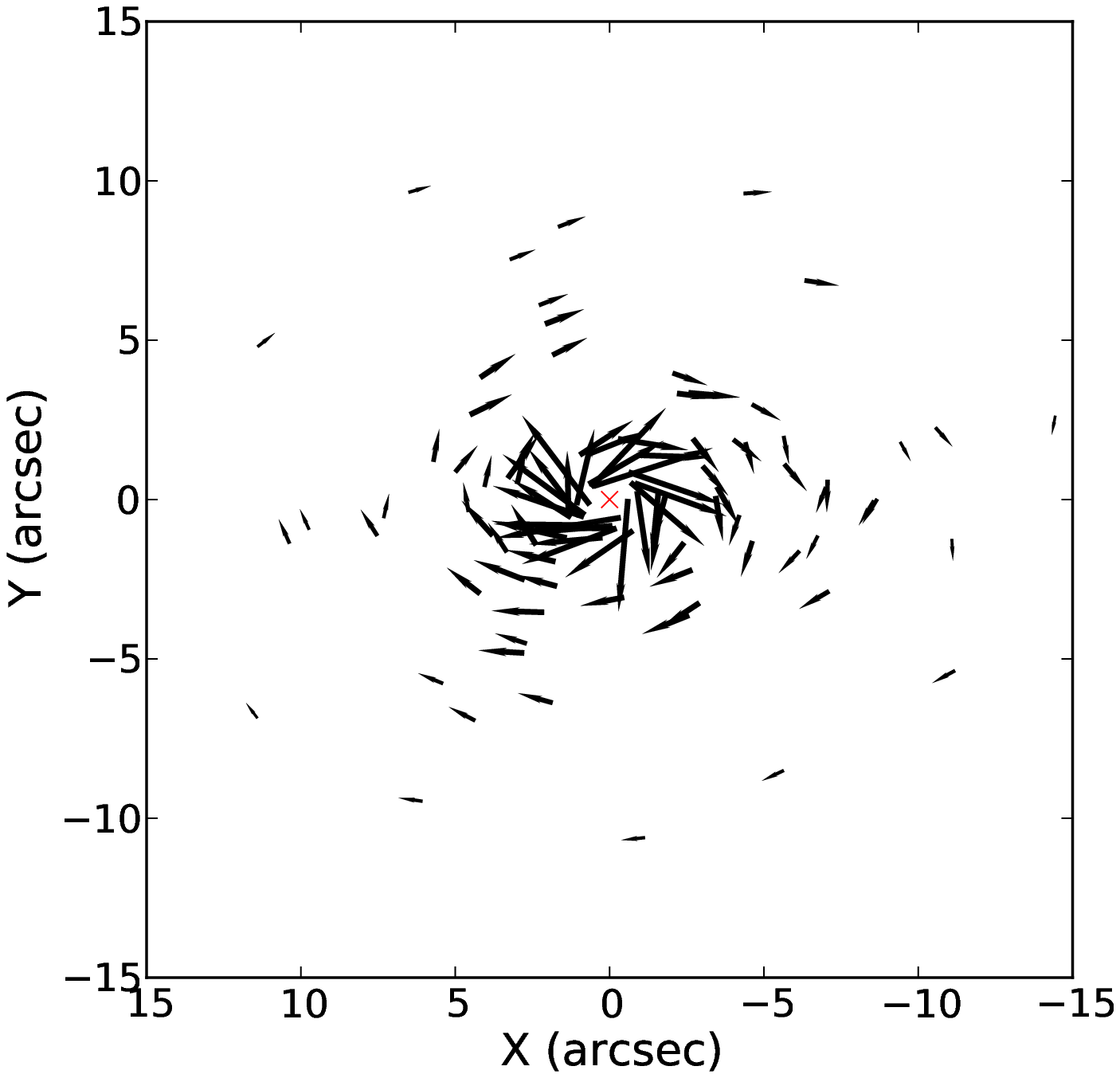}
\includegraphics[scale=0.45]{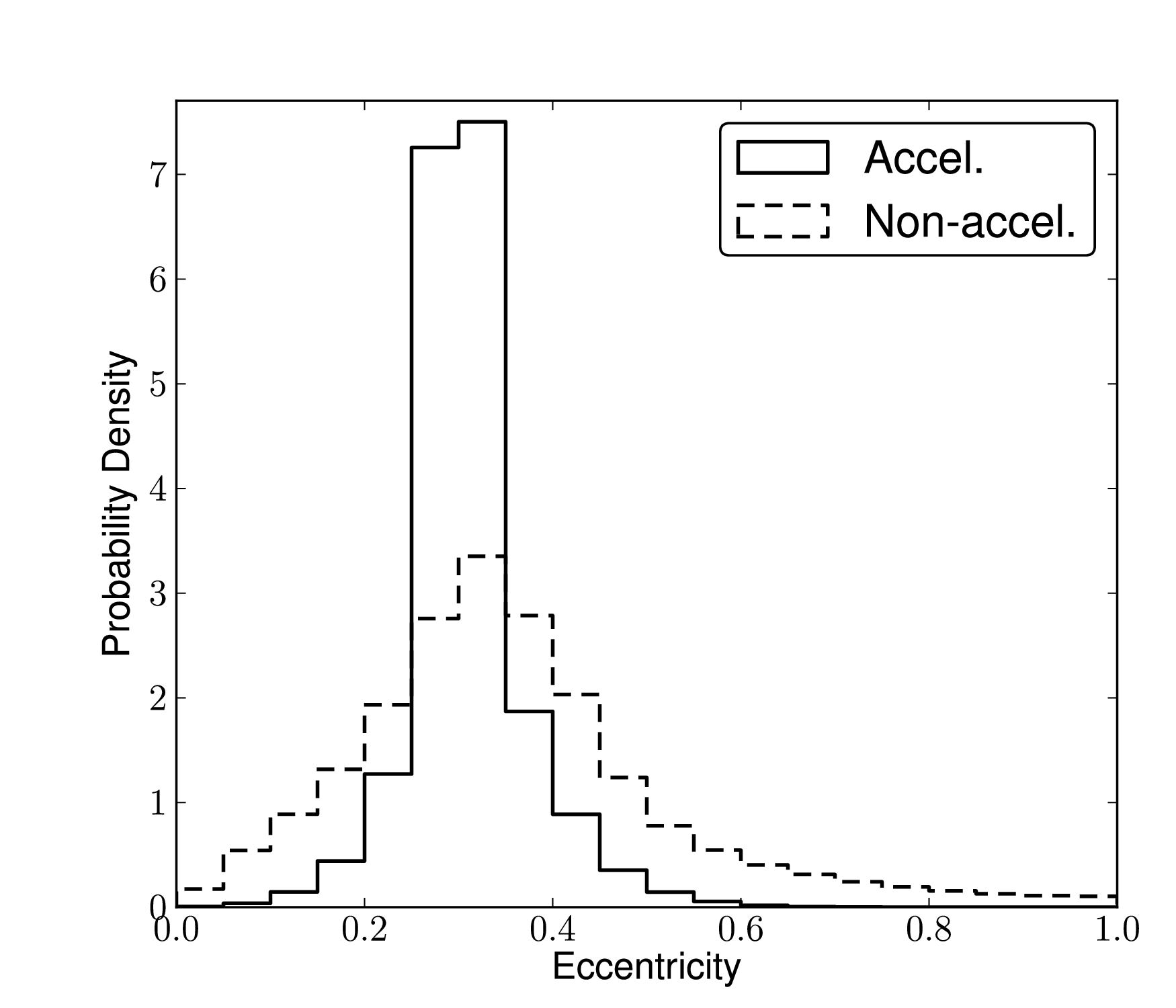}
\end{minipage}\\
\figcaption{Simulated circular ({\em top}) and eccentric ($e$=0.3; {\em bottom}) 
orbits each consisting of 100 stars within a disk with an orbital plane 
orientation similar to that of the observed disk.
For each simulation, the generated velocity vectors of the stars are shown on
the left, with the location of the black hole being marked as a red X at the center.
The eccentricity distributions of the accelerating ({\em solid}) and non-accelerating 
({\em dashed}) stars from each simulation are shown on the right. The orbits of the 
accelerating stars are more accurately determined, as expected.  Based on these
simulations, the observed eccentricity distribution in Figure \ref{fig:eccPDF} cannot 
be a result of measurement bias added to an intrinsically circular disk. 
}
\label{fig:mockEccBias}
\end{center}
\end{figure}

\begin{figure}[!h]
\epsscale{0.6}
\plotone{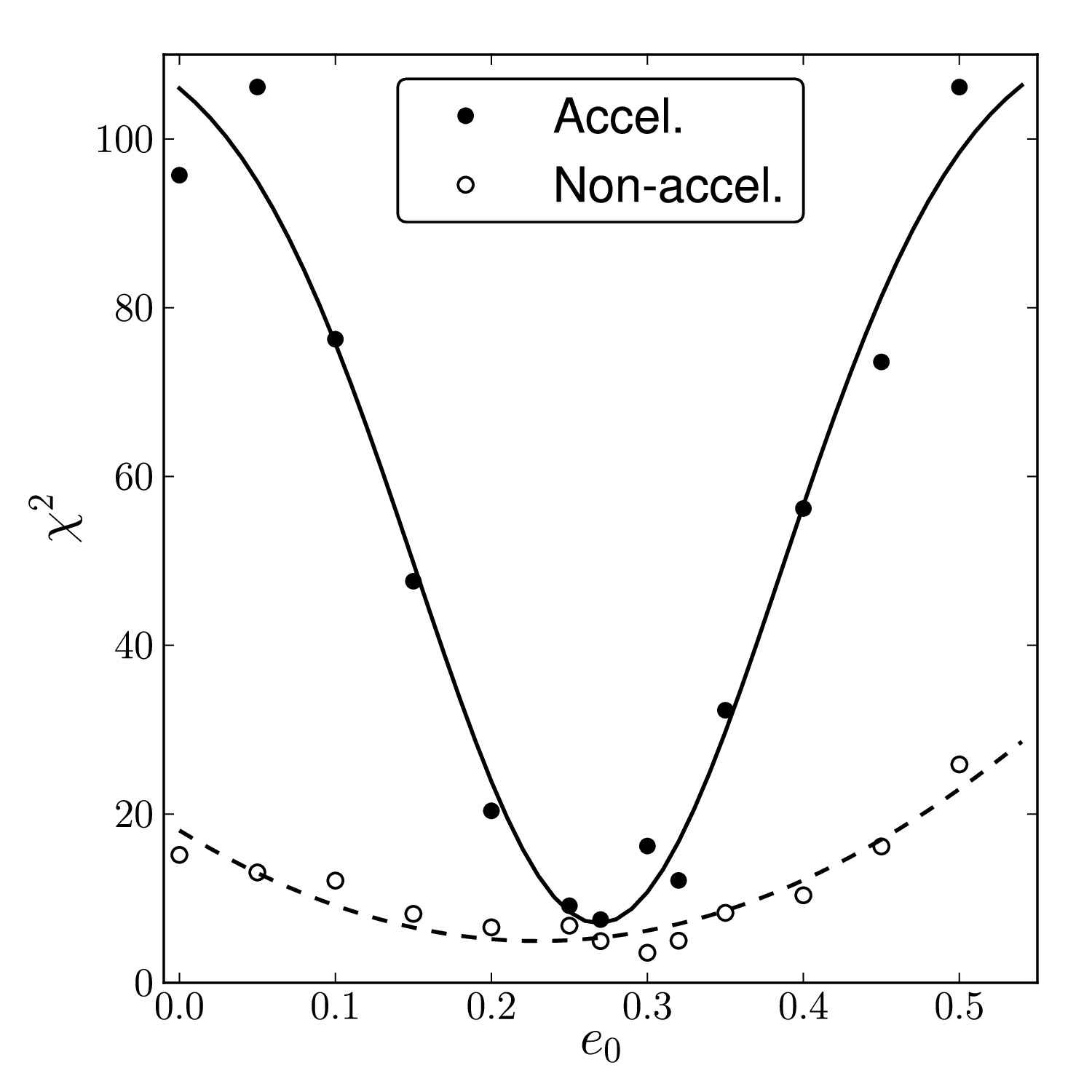}
\figcaption{$\chi^2$ as a function of initial eccentricity of simulated disk stars.
For each $e_0$ tested, $\chi^2$ was calculated by comparing the resulting 
normalized eccentricity distribution to that of the observed candidate disk stars
separately for the accelerating ({\em filled points}) and non-accelerating stars 
({\em unfilled points}) stars (we assume all errors are unity). Based on a Gaussian fit 
to these data, $\chi^2$ is minimized at $e_0$ = 0.27 for accelerating 
({\em solid curve}) and $e_0$ = 0.23 for non-accelerating ({\em dashed curve}) stars.
}
\label{fig:chi2ecc}
\end{figure}

The resulting eccentricity distributions from the orbital analysis on the simulated
disks with a range of input eccentricities (ID 2; see Figure \ref{fig:mockEccBias} for
two cases) were compared to the observed values.
The $\chi^2$ statistic was calculated separately for the eccentricity distributions
of accelerating and non-accelerating stars (Figure \ref{fig:chi2ecc}). The true 
eccentricity value at which $\chi^2$ is minimized based on a Gaussian fit to the data
is $e_0$ = 0.27 for the accelerating stars and $e_0$ = 0.23 for the non-accelerating 
stars. We therefore conclude that the young stars in the disk have a true eccentricity
of $e$ = 0.27.

Measurement uncertainties of the individual eccentricities can both bias the observed
average values and increase the width of the eccentricity distribution. For the simulated
case of $e_0$ = 0.3 with no intrinsic width, the observed average and rms values 
are 0.31 $\pm$ 0.06 and 0.42 $\pm$ 0.21 for the accelerating and non-accelerating stars,
respectively. This shows that there is a bias in the average eccentricity that is 
more substantial for the non-accelerating stars. Furthermore, if we treat the rms
values from the simulation as a bias term and subtract in quadrature from the observed
rms values, then it appears that most of the spread in the observed eccentricities
can be accounted for by measurement uncertainties. We obtain a formal estimate of
the intrinsic rms of 0.07 and 0.12 for the accelerating and non-accelerating stars,
respectively.

% Generated by python code:
% sythesis.ecc_bias_simulation(grid=True, plotsOnly=True)

% Generated by python code:
% sythesis.ecc_bias_simulation()

Based on these simulations, our final eccentricity estimate for the young stars is
$\langle e \rangle$ = 0.27 $\pm$ 0.07. This is the first time the
measurement bias has been quantified via simulations and explicitly accounted for 
in estimates of the eccentricities of stars on the clockwise disk. The 
eccentricities are slightly lower than previously determined
\citep{beloborodov06,lu09,bartko09}, which in part is because of the removal
of measurement bias in this work and the more precise eccentricity measurements 
of the accelerating stars. 

\clearpage
\subsubsection{True Disk Fraction}
\label{sec:diskfrac}
\begin{figure}[!h]
\epsscale{1.0}
\plottwo{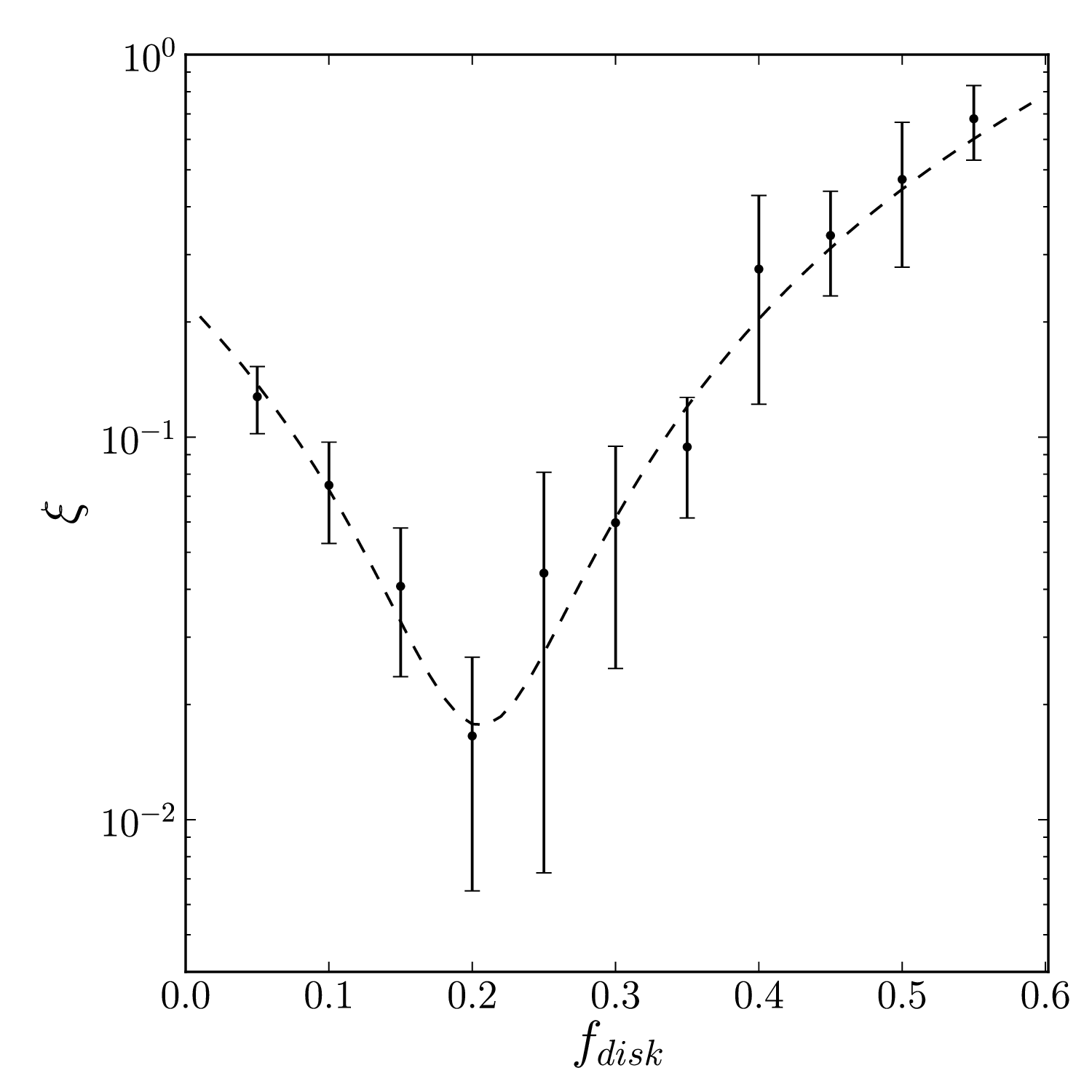}{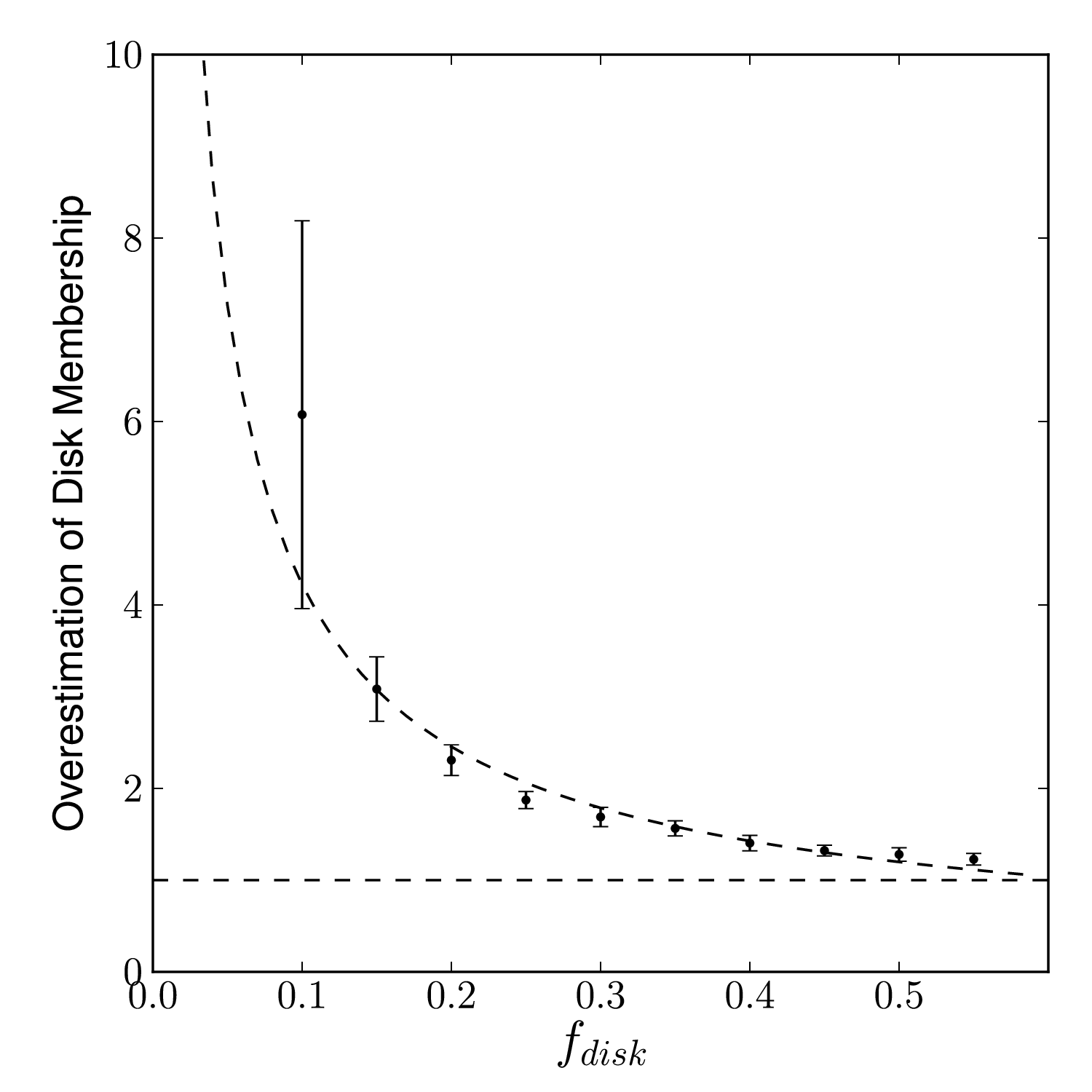}
\figcaption{
{\em Left}: Sum of the squared differences in density between the disk
fraction models and the observations (see \S\ref{sec:diskfrac}), averaged over the 
ten independent trials run for each disk fraction model. A 2nd order
polynomial was fit to the data ({\em dashed curve}) and gives a minimum $\xi$ value 
at $f_{disk}$ = 0.21, implying a true disk fraction in our sample of 21\%, or $\sim$24 
stars. {\em Right}: The level to which the true number of disk members is overestimated
in each disk-fraction model, again averaged over the ten trials run for each disk fraction
model. As a visual reference, a dashed line marks where the number of candidate disk 
stars equals the true number in each simulation. Note that the Y-axis is truncated for 
clarity. For $f_{disk}$ = 0.21, the number of disk candidates is overestimated by a 
factor of 2.4. 
}
\label{fig:diskfrac}
\end{figure}

\begin{figure}[!h]
\epsscale{0.7}
\plotone{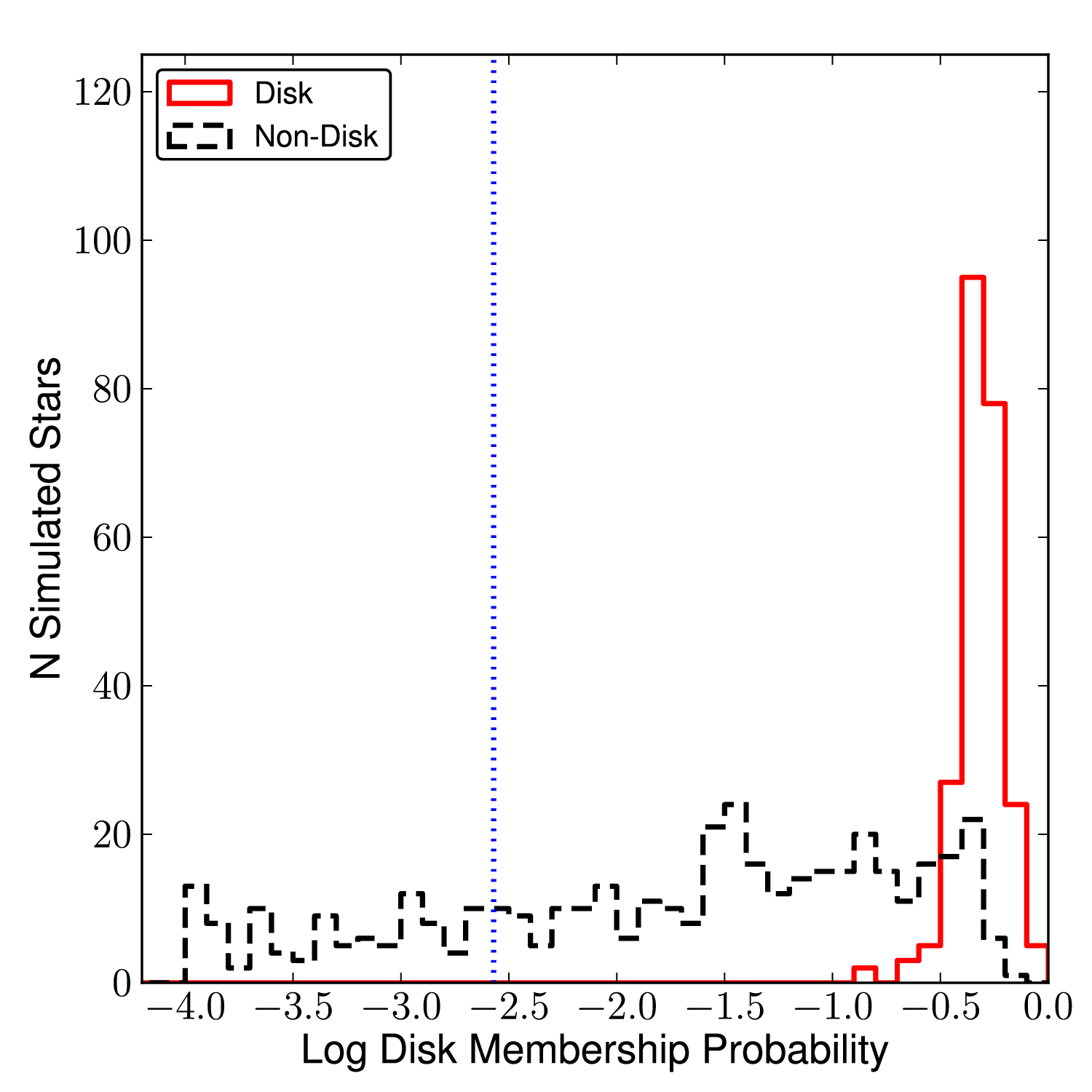}
\figcaption{Combined histogram of disk membership probabilities ($1 - L_{non-disk}$)
for the 10 simulations with $f_{disk}$ = 0.2. True disk members are shown as the
{\em red solid} histogram, while the isotropic population is shown in {\em black
dashed}. The vertical {\em blue dotted line} marks the criteria used for selecting
candidate disk members, namely $(1 - L_{non-disk}) >$ 0.0027.  While this conservative 
threshold identifies all true disk members as candidates, there is an abundance of 
contaminants, even at the highest probabilities. 
An additional 544 isotropic
stars were cut off to the left of the figure for clarity.
}
\label{fig:hist_frac20}
\end{figure}

Our ability to estimate the true fraction of disk members can be quantified using the 
orbital analysis of mock data involving a combination of disk and isotropic stars (ID 3 
in Table \ref{tab:mockData}).  For each disk fraction simulation, a density map
of ($i$, $\Omega$) is generated and compared to that of the observed data.
We compute the squared difference in density between the model and the observations at
each pixel, $j$, in the ($i$, $\Omega$) map that is within 30$\deg$ from the location
of the observed peak, and sum over all pixels ($N_{pix}$ = 3292).
We refer to this quantity as $\xi$:
\begin{equation}
\label{eq:diskfrac}
\xi = \sum^{N_{pix}}_j(\rho_{model,j} - \rho_{observed,j})^2.
\end{equation}
The left panel of Figure \ref{fig:diskfrac} shows these results, averaged over the
10 trials for each disk fraction, with a 2nd-degree polynomial fit to the data.
Based on this fit to the data, the disk fraction for which 
$\xi$ is minimized is $f_{disk}$ = 0.21 $\pm$ 0.02, where the uncertainty is taken
as the rms error on the minimum $\xi$ obtained from fits on each of the 10 trials. 
The method for identifying
disk stars, described in \S\ref{sec:2dPDF}, can also be tested using these disk fraction 
simulations. The right panel of Figure \ref{fig:diskfrac} shows the ratio of the estimated 
number of candidates from this method to the true number of 
disk members for each model, which reveals the degree of contamination from the non-members. 
After fitting a functional form to these data, we find that the number of disk members
is overestimated by a factor of $\sim$2.4 for $f_{disk}$ = 0.21. Thus, in
the observed data, the number of candidates ($N_{candidates}$ = 58) is an overestimate, and 
we take the true number of disk members to be 58 $/$ 2.4 $\sim$ 24. We note that 24 is
$\sim$21\% of our sample, which validates our finding of a disk fraction of 21\%.
For the 10 cases that were run using $f_{disk}$ = 0.20, the combined histogram of disk
membership probabilities is shown in Figure \ref{fig:hist_frac20}. With the conservative
cut used to select candidate disk members described in \S\ref{sec:2dPDF},
all true disk members are identified but we also see an abundance of contaminants from 
the isotropic population.  

From these simulations, we also find that that there is a small dependence of the disk 
membership probability on the position angle from the disk's line of nodes for stars 
with no acceleration information (see Appendix \ref{app:nodes}), whose orbital 
solutions are more sensitive to the prior for the line-of-sight distance 
(\S\ref{sec:orbitAnalysis}). If the disk membership
probability is to be used for identifying specific disk candidates in the observed 
sample, the stars' position angles must also be considered. The most likely 24 true 
disk members are listed in Table \ref{tab:nodes_app_table} of Appendix \ref{app:nodes}.

% Generated by python code:
% sythesis.disk_fraction_results() and disk_fraction_sumSquaredDiff()

% Generated by python code:
% sythesis.disk_fraction_results()

\section{Discussion}
\label{sec:disc}
We have performed a detailed kinematic analysis on the central parsec young 
star population using high precision astrometry over a longer time
baseline than in any other such study.  Combined with radial velocity 
measurements, we have confirmed the existence of the clockwise stellar 
disk and have shown there is no significant counterclockwise structure,
in agreement with \citet{lu09}. Through a series of orbital analyses on mock
data sets, we showed that 20\% of the stars in our sample are true members of the 
disk, a factor of more than two lower than previous estimates, which were based 
on disk candidacy alone \citep{lu09,bartko09}.
The intrinsic average eccentricity of disk members is $e$ = 0.27, 
and we find no significant detection of the disk beyond 3$\farcs$2.
Here we discuss the implications of these findings and explore the relationship
between the B stars and O/WR stars in our sample. 

\subsection{Disk Remnant}
Our orbital analysis of mock data sets reveals that the disk is made up of
20\% of the sample. Thus, assuming a single-disk origin, we are likely observing 
the remnants of what used to be a more densely populated disk.
For such a scenario, some dynamical mechanism(s) that can excite 
the orbits such that 80\% of the stars are no longer kinematically associated with 
the original disk must be invoked. It was shown that 2-body relaxation is not 
sufficient to explain the high inclinations relative to the clockwise disk 
\citep[e.g.,][]{cuadra08}. Vector resonant relaxation with the surrounding 
stellar cluster, on the other hand, can lead to a strongly warped disk 
\citep{kocsis11}. It is unclear, however, whether this mechanism can explain the 
observed properties of the stars both on and off the disk.
Massive perturbers, such as an IMBH can lead to strong scattering off the disk 
\citep{yu07}. One of the major challenges to this scenario, however,
is the lack of evidence for an IMBH at the Galactic center. One massive perturber
that is observed, however, is the circumnuclear disk (CND) located at 
$R \sim$ 1.5 pc \citep[CND;][]{christopher05}. \citet{subr09} first investigated 
the influence of the CND on a thin stellar disk and found that differential 
precession can lead to a configuration that is similar to what is 
observed. The effects of the CND will be most pronounced at the outermost portions 
of the stellar disk, erasing any observable disk-like structure at large radii while
leaving the innermost orbits untouched \citep{subr09,haas11a,haas11b}.  
This is qualitatively consistent with the observations reported here. 

\begin{figure}[!h]
\epsscale{0.6}
\plotone{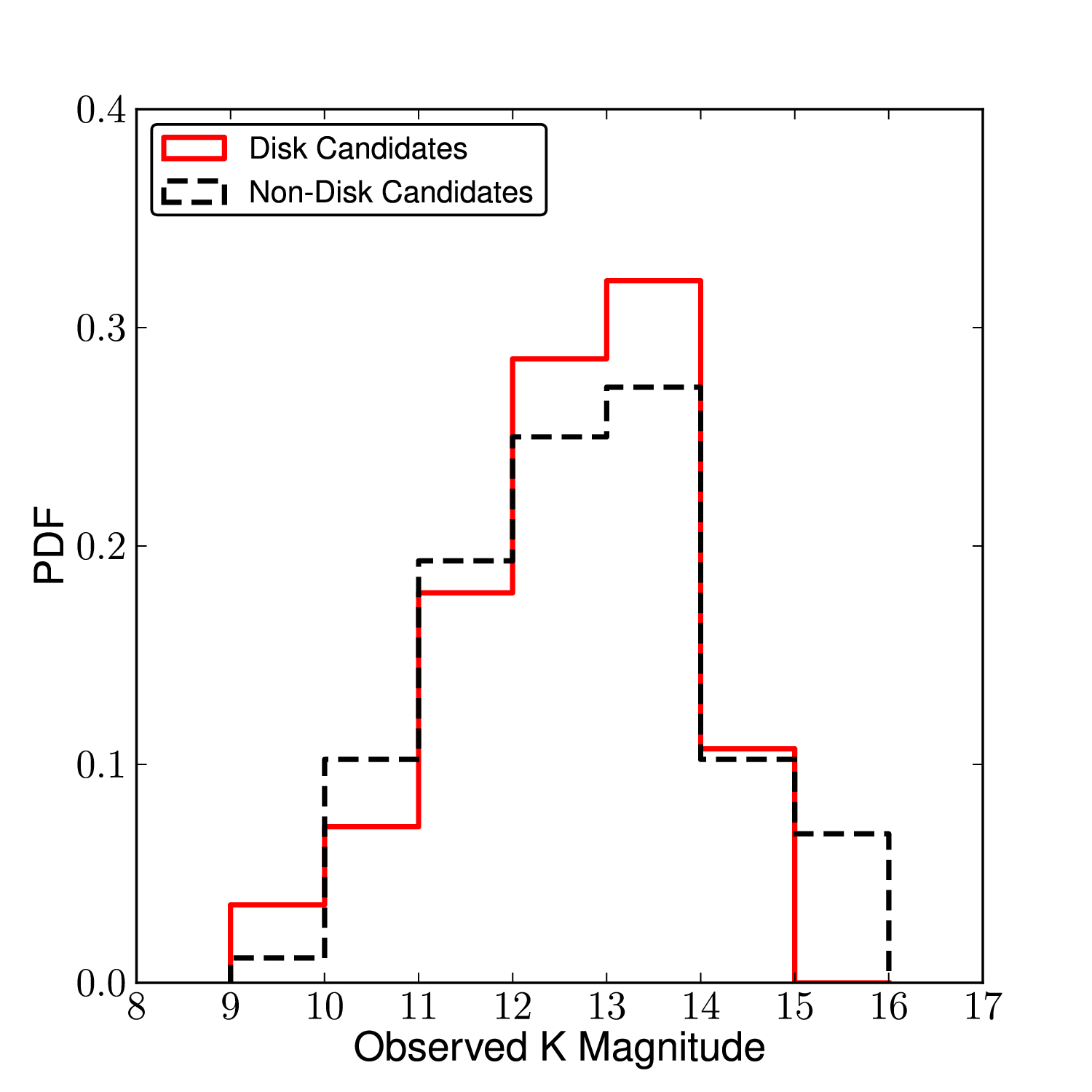}
\figcaption{Normalized distribution of K-band luminosities of the most likely
members (see Appendix
\ref{app:nodes}) of the disk ({\em red solid}) and the remaining stars 
({\em black dashed}).  A 2-sample KS test yields a probability of 87\% that
the most likely disk members have the same KLF as the remainder of the sample.
We caution that the K-band magnitudes in this figure are {\em observed} 
magnitudes and have not been corrected for completeness or extinction effects.
}
\label{fig:klf_topNodes}
\end{figure}

Figure \ref{fig:klf_topNodes} shows the K-band luminosity function (KLF) of the
sample, plotted separately for the most likely disk members ($N$ = 28) and non-members
($N$ = 88; see Appendix \ref{app:nodes}). A 2-sample KS test yields a
probability of 87\% that the distributions are the same, lending support to
a common origin scenario (although not necessarily a common {\em disk} origin). 
It is still unclear whether all of the stars formed in a single disk. Indeed, the 
existence of a second, less massive disk with counterclockwise orbiting stars has
remained controversial \citep{genzel03,paumard06,bartko09,lu09} and is not supported 
by the observations reported here. However, if two highly inclined disks of different
masses existed at one point in the GC, their mutual interaction 
would lead to the ultimate destruction of the lower mass disk within 5 Myr and
we would therefore not observe the structure today \citep{lockmann09a}. 
While this may explain the lack of counterclockwise structures in our analysis,
such a scenario would demand that two star formation events at or
near the GC occurred within 2 Myr of one another.

% Generated by python code:
% sythesis.klf_disk_on_off(top20pct=True)

\subsection{Eccentricity of Disk Stars}
\label{sec:discEcc}
The orbits of the disk members are found to be eccentric, with a
distribution that peaks near $e$ = 0.3. We show above that these results cannot be 
explained by a circular disk ($e$ = 0) whose eccentricity is biased upward due
to measurement error. Furthermore, the fact that five of the six stars with reliable
acceleration measurements are likely disk members and collectively have an
eccentricity of $\langle e \rangle$ = 0.27 $\pm$ 0.07 gives us confidence that the disk
is eccentric.  We also find that the distribution is unimodal, lacking the
high eccentricity bin ($e >$ 0.9) reported by \citet{bartko09}, which those
authors claim may have been a result of contamination by non-disk members.

The observed eccentricities can be used to constrain formation scenarios for the
disk. \citet{levin05} showed that the circular inspiral of a cluster anchored by
an IMBH can produce significant eccentricities for some of the cluster stars. 
Similarly, \citet{berukoff06} found that the stars' eccentricities 
will mirror the eccentricity of the cluster's IMBH
with a scatter of roughly $\pm$0.1-0.2. While these values are somewhat consistent 
with what we find for disk members in this work, we caution that 
the cluster-infall scenario suffers from many 
theoretical and observational challenges, as discussed in \S\ref{sec:intro}.

\begin{figure}[h!]
\epsscale{0.7}
\plotone{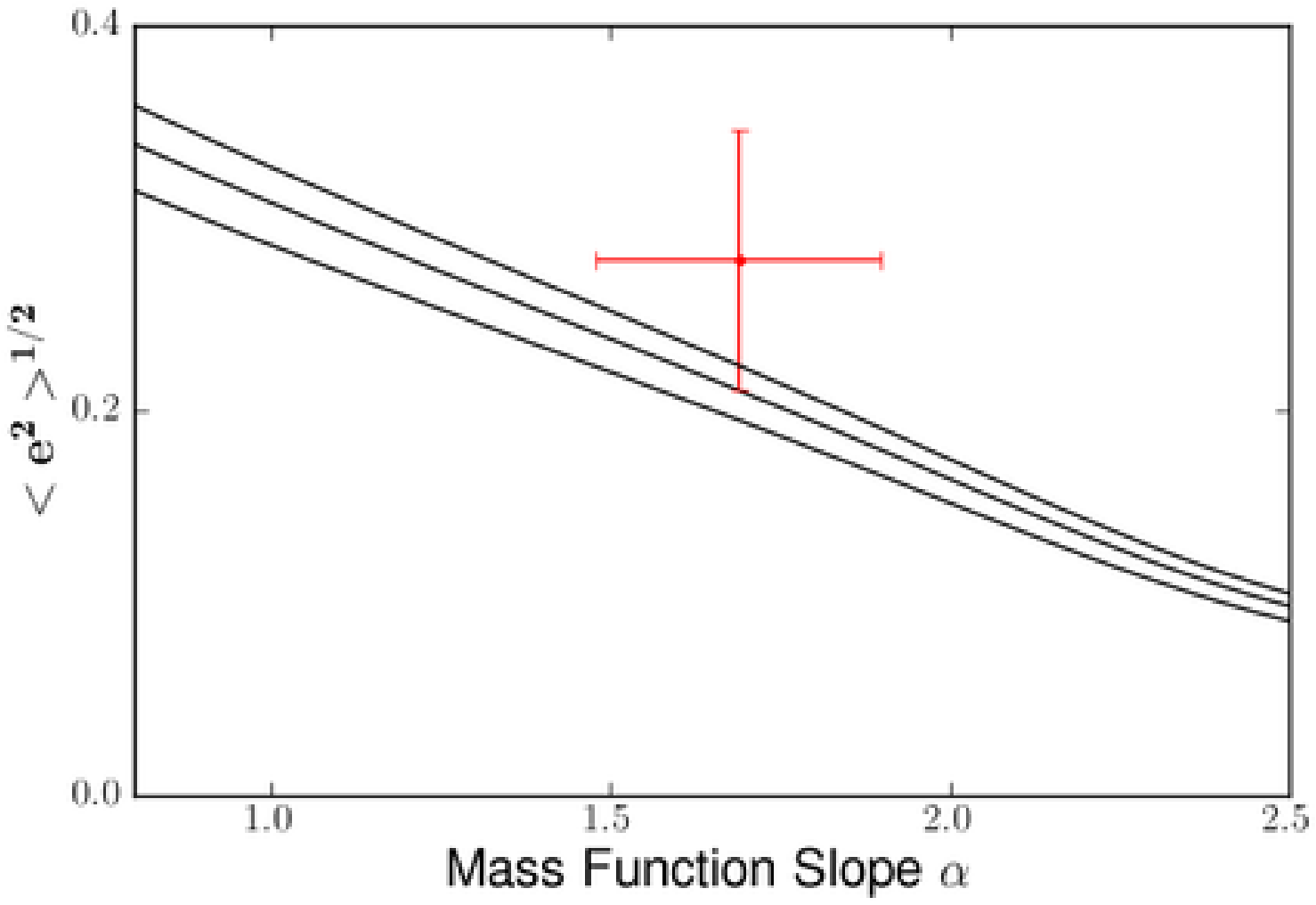}
\figcaption{Predicted rms eccentricity ({\em black curves}) of 50 stars, each with mass
$M$ = 25 \msun, resulting from 2-body interactions within a stellar disk with 
various mass function slopes.  This is a version of Figure 4 in \citet{alexander07}, 
which has been updated to reflect the latest black hole mass estimates 
(4 $\times$ 10$^6$ \msun) and a recent estimate of the age of the young 
star population from \citet{lu13} of 3.65 $\pm$ 0.87 Myr.
From bottom to top, the three curves represent the rms eccentricity after 
2.78 Myr, 3.65 Myr, and 4.52 Myr, respectively.  The red point shows the rms
eccentricity and estimated uncertainty of our best measured stars 
($\langle e^2 \rangle^{1/2}$ = 0.28 $\pm$ 0.07) and the latest estimate of the
mass function slope ($\alpha$ = 1.7 $\pm$ 0.2) from \citet{lu13}.
The observed values are within 1$\sigma$ of the predicted values from the
Alexander et al. model.
}
\label{fig:ecc_evol}
\end{figure}

Several {\em in situ} formation scenarios in which the stars form in either an
initially circular or eccentric gaseous disk have been proposed in attempts to
explain the previous estimations of the eccentricity distributions and kinematic
structures \citep{nayakshin07,alexander08,bonnell08,mapelli08,yusef08,wardle08,
hobbs09,mapelli12}. In all of these scenarios, it is necessary to include interactions 
within the disk \citep{alexander07,kocsis11} and any surrounding cusp 
\citep{madigan09,lockmann09a,lockmann09b}
over the lifetime of the stars to produce an eccentricity distribution that can be 
compared with present-day values. However, the viability of these scenarios and the 
need for more extreme scenarios that invoke initially eccentric disks must be 
re-examined in light of our revised understanding of, for example,
the young stars' eccentricity distribution (this study), their mass 
function and age \citep{lu13}, and the stellar cusp
\citep{do09,buchholz09,bartko10,do13}. In particular, the apparent lack of a
stellar cusp and the lower eccentricity distribution of the stars in the disk
simplify the dynamics. Thus, we revisit a simpler origin scenario
in which stars form in a circular gaseous disk and examine the effects of
two-body interactions on the stellar orbits \citep{alexander07}. 
The degree to which the eccentricities are excited depends on both the age and the 
mass function of the population. Using the latest values of the mass function
slope of $\alpha$ $\sim$ 1.7 and age of $\sim$3.9 Myr \citep{lu13}, we estimate
the expected rms eccentricity after dynamical evolution of the stars
(Figure \ref{fig:ecc_evol}; R. Alexander, private communication). This is a 
version of Figure 4 of \citet{alexander07}\footnote{\citet{alexander07} use $\Gamma$
to denote the mass function slope, where $\Gamma$ = 2.35 is the Salpeter slope. However,
$\Gamma$ generally represents the logarithmic slope (1.35 for Salpeter), and
$\alpha$ is typically used to denote the linear slope \citep[see e.g.,][]{bastian10}.}, 
where here we show the final eccentricities expected
after 2.78 Myr, 3.65 Myr, and 4.52 Myr, which spans the $\sim$1$\sigma$ range of 
the Lu et al. estimates for the age of the population\footnote{The ages used in
Figure \ref{fig:ecc_evol} are
from an early version of \citet{lu13}, but are within 1$\sigma$ of
their published age estimate for the young stars.}. For the latest age and mass function
slope of the young stars, the expected rms eccentricity agrees with our
observed value to within 1$\sigma$.
We conclude that our results are 
consistent with formation in a circular gas disk, which has been proposed by others
\citep{nayakshin05,alexander07,lockmann09a} and which does not require 
radially-infalling gas clouds or a stellar cusp.

% Generated by python code:
% alexander_imf_ecc.go()

\subsection{Disk Structure}
\label{sec:zBias}
Our analysis of the radial structure of the disk
reveals a prominent clockwise disk between $r$ = 0$\farcs$8-3$\farcs$2 and no other 
significant features until $r >$ 6$\farcs$5, where a set of three co-orbiting stars
is evidenced.  These two CW structures are offset from 
one another by $\sim$77$\deg$, with the outer feature being much less significant
than the feature seen at small radii.  Due to the lack of structure at intermediate
radii, we do not find evidence of a warped disk, as found by 
\citet{bartko09}\footnote{In Table 3 of
\citet{bartko09}, there is a mistake in the $i$ and $\Omega$ angles reported
for their outer radial bin. While their reported angles $\phi$ and $\theta$ are
consistent with the corresponding plot in their Figure 11, the conversion
to $i$ and $\Omega$ in Table 3 is incorrect. The correct values are ($i$, $\Omega$) = 
(118$\deg$, 179$\deg$).}. Our data support the existence of a single 
clockwise disk with a radial extent of at least $r \sim$ 3$\farcs$2. The feature seen 
at large radii may be a small cluster or a filamentary structure such as those
found in simulations of cloud-cloud collisions \citep{hobbs09} and of 
single-cloud infall \citep{lucas13}, although the stellar masses ($<$2 \msun) 
in those simulations were much lower than what is observed.
We caution, however, that more uniform sampling of the outer radial bin 
is necessary to make a definitive claim about any kinematic structures.

\subsection{The Relation of the B Stars to the Disk}
\label{sec:bstars}
The original claim of a stellar disk in the Galactic center was based on a
kinematic analysis that included only O and WR stars, as these were the
only known young stars at the time \citep{levin03}. It is unclear how the more
recently-identified B-type main sequence stars \citep{allen90,krabbe91,krabbe95,
blum95,tamblyn96,najarro97,ghez03,paumard06,bartko10,do13} are associated with the
O/WR population, if at all. While the age of the O/WR stars is 
estimated at $\sim$3-8 Myr \citep{paumard06,lu13}, the B stars have main
sequence lifetimes of up to $\sim$30 Myr for the faintest stars in our sample
($K$ = 15.9) and therefore may not have originated in the most recent star 
formation event. However, recent statistical analysis of the B star's $h$-statistic,
$h = (xv_y - yv_x) / \sqrt{GM_{BH}R}$, suggests that brighter B stars 
($K$ = 14-15) may be more consistent with formation in
a disk than by binary star infall \citep{madigan13}. Our analysis provides a more direct 
disk-association test with the use of acceleration measurements or constraints and 
radial velocity information.

If the B stars are, in fact, unrelated to the disk stars, then their inclusion in 
the analysis of \S\ref{sec:diskfrac} would decrease the significance of any kinematic
features and lead to an artificially low disk fraction. After excluding B stars having
$K >$ 14 ($N$ = 18 stars), we repeat our analysis and simulations for the O/WR stars
($N$ = 98 stars). The resulting analysis gives a location of the peak density at
($i$, $\Omega$) = (130$\deg$, 96$\deg$), which is identical to that found with
the full sample with a slightly reduced significance ($S$ = 13.3 compared to 
$S$ = 20.7 for the full sample). 
The O/WR sample was also divided into the inner, middle, and outer
radial bins ($N$=29, 35, and 34 stars, respectively), and we repeated the analyses
looking for significant kinematic features. In the inner bin, we find a strong peak at 
($i$, $\Omega$) = (124$\deg$, 103$\deg$), with $S$ = 17.9. Again, there is no 
significant feature in the middle bin ($S_{max}$ = 1.1). Finally, the outer 
radial bin shows a peak at ($i$, $\Omega$) = (117$\deg$, 192$\deg$) with $S$ = 6.0,
slightly more significant than the results from the full sample.
We therefore conclude that the overall kinematic structure remains the same whether B
stars are included in our analysis or not. 

We tested the effects of the B stars on the disk fraction by repeating the disk 
fraction simulations using 98 stars and comparing the results to the density map of 
the observed O/WR sample. The value of $\xi$ (Equation \ref{eq:diskfrac}) is minimized 
at $f_{disk}$ = 0.20 as compared to $f_{disk}$ = 0.21 for the full sample. The similarity 
in these values with those from our original analysis leads us to conclude that the 
inclusion of the B stars does not impact the true disk fraction. 

\begin{figure}[!h]
\epsscale{1.0}
\plottwo{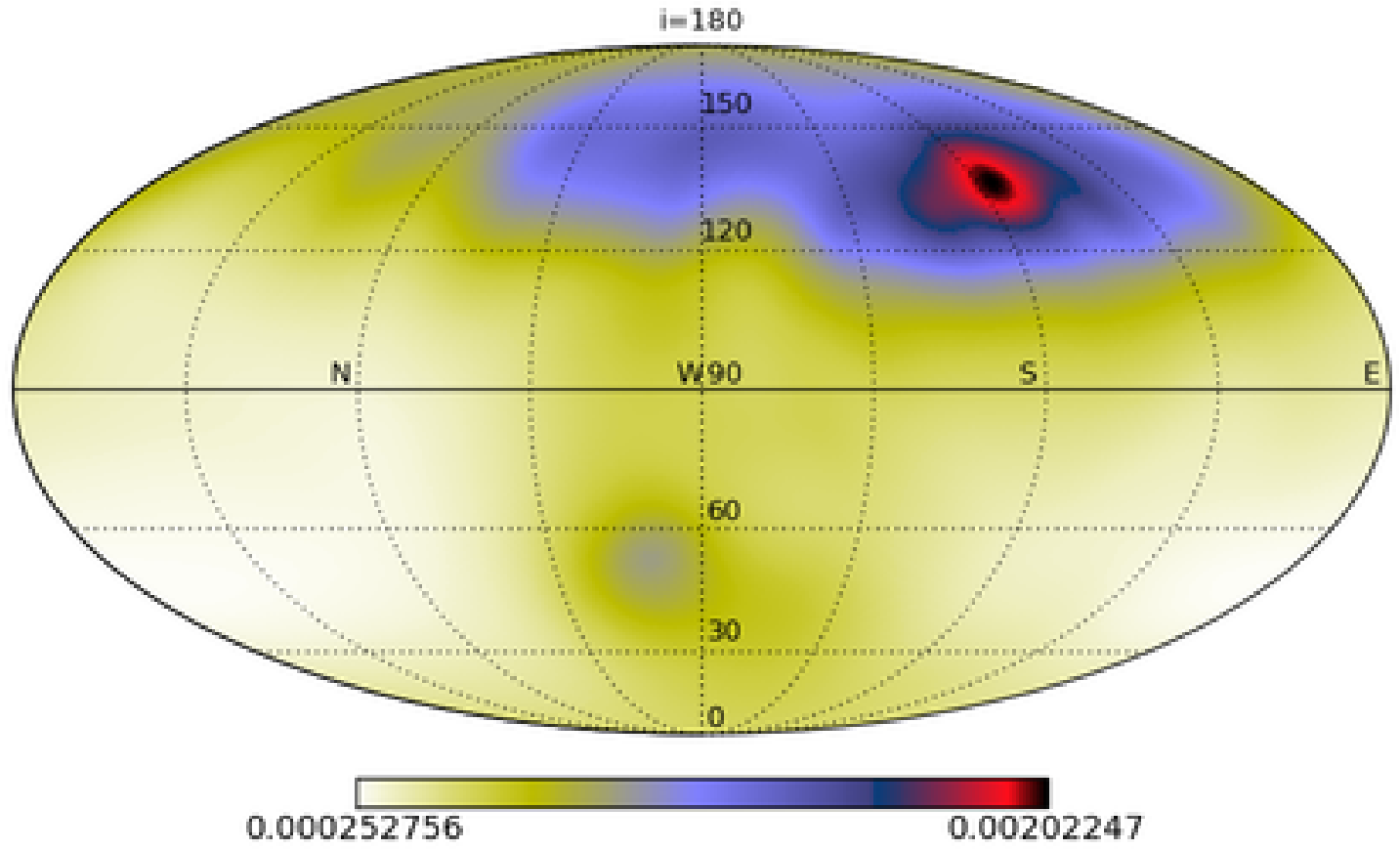}{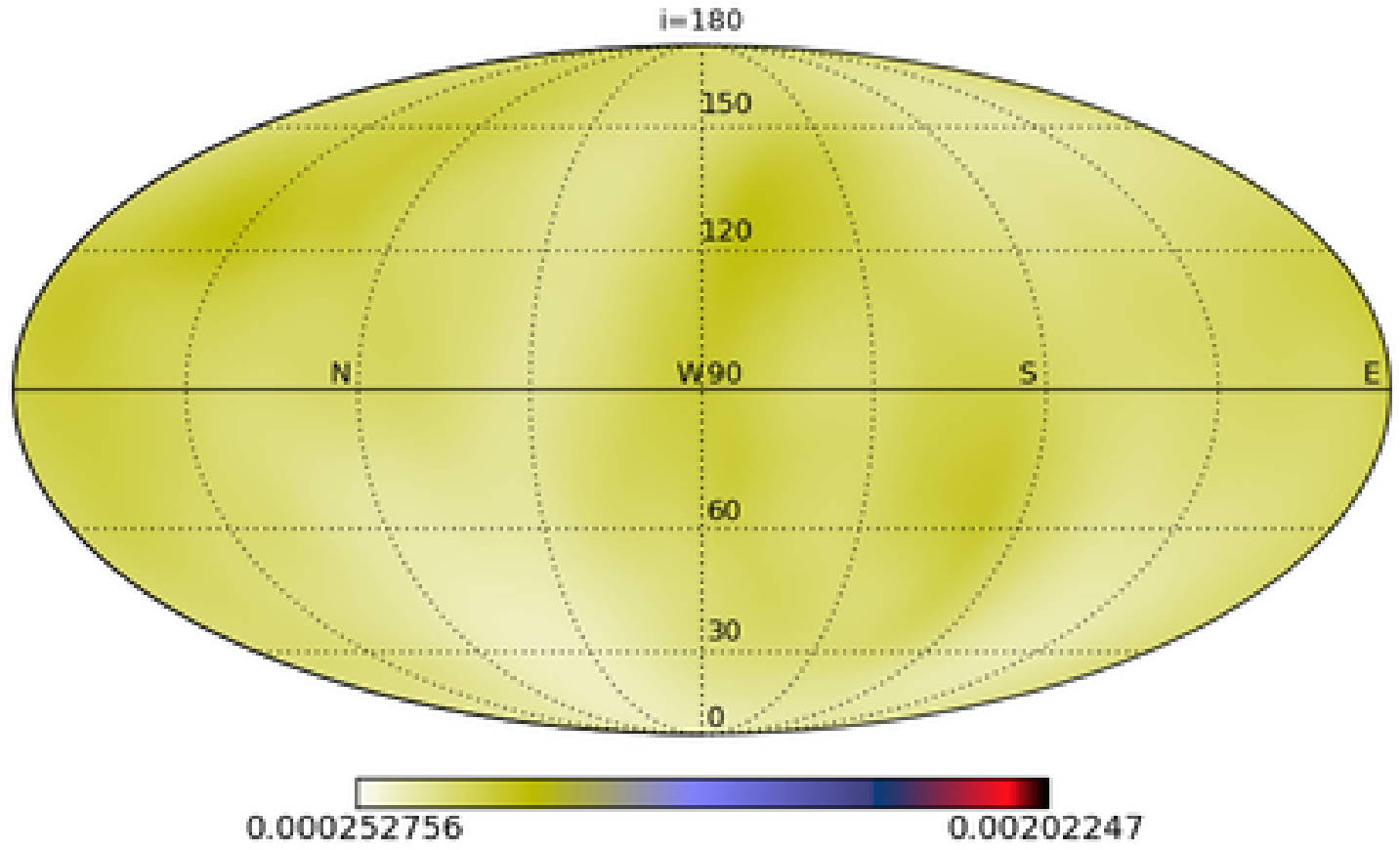}
\figcaption{{\em Left:} Density of normal vectors for the B stars (K $\ge$ 14) in our sample.
The peak density, 0.002 stars deg$^{-2}$, is located at 
($i$, $\Omega$) = (136$\deg$, 85$\deg$).
{\em Right:} Density of normal vectors for an isotropically-distributed sample of
18 stars. The same stretch is used for both density maps.
}
\label{fig:Bdensity}
\end{figure}

Finally, we created a PDF($i$, $\Omega$) density map for just the 18 B stars in
our sample (Figure \ref{fig:Bdensity}). Given the decrease in the number of stars, 
the density at each pixel in the sky was calculated using the nearest 4 neighbors 
(approximately 20\% of 18). The peak density (0.002 stars deg$^{-2}$) is found at 
($i$, $\Omega$) = (136$\deg$, 85$\deg$) $\pm$ (6$\deg$, 12$\deg$), which
is within 1$\sigma$ of the peak location using the full sample.
The significance of this feature is $S$ = 6.0 as compared to the 
expected density of normal vectors for an isotropic population of 18 stars.
Repeating the disk fraction simulations, but with a total of 18 stars, we find
a minimum $\xi$ value at $f_{disk}$ = 0.23, which is consistent with the results from
the full sample in which 20\% of the B stars were identified as disk members
(Table \ref{tab:nodes_app_table}). These stars are: S1-8 ($K$ = 14.1), 
S3-190 ($K$ = 14.0), and S10-32 ($K$ = 14.4).
This suggests that the disk fraction does not change with magnitude and therefore
that the B stars on the disk formed in the same starburst as the O/WR stars.
Furthermore, given the similarity in the observed KLFs of the disk and 
non-disk members (Figure \ref{fig:klf_topNodes}), most of the brightest B stars 
($K <$ 15) whose main sequence lifetimes are $t <$ 13 Myr, likely formed with the 
O/WR stars. The results from the analyses done on each subset of our sample
are shown in Table \ref{tab:diskResults}.

% Generated by python code:
% plot_disk_healpix.go()

\begin{deluxetable}{ccccc}
\tabletypesize{\scriptsize}
\tablewidth{0pt}
\tablecaption{Disk Properties by Sample}
\tablehead{
  \colhead{$K$} & 
  \colhead{$N_{stars}$} & 
  \colhead{($i$, $\Omega$) $\pm$ ($\sigma_i$, $\sigma_{\Omega}$)\tablenotemark{a}} & 
  \colhead{$S$} &
  \colhead{$f_{disk}$} 
}
\startdata
      all  & 116 & (130$\deg$, 96$\deg$) $\pm$ (2$\deg$, 3$\deg$)  & 20.7  & 0.21 $\pm$ 0.02  \\
    $<$14  &  98 & (130$\deg$, 96$\deg$) $\pm$ (2$\deg$, 2$\deg$)  & 13.3  & 0.20 $\pm$ 0.02 \\
  $\ge$14  &  18 & (136$\deg$, 85$\deg$) $\pm$ (6$\deg$, 12$\deg$) &  6.0  & 0.23 $\pm$ 0.11 \\
\enddata 
\tablenotetext{a}{Uncertainties in the peak location are estimated as the rms error 
from the $f_{disk}$ = 0.2 disk fraction simulations done for each sub-sample.}
\label{tab:diskResults}

\end{deluxetable}

\section{Conclusions}
\label{sec:conc}
We have analyzed the orbits of 116 young stars in the Galactic center
between projected radii $R$ = 0$\farcs$8 - 13$\arcsec$ ($\sim$0.032 pc - 0.52 pc).
Our acceleration uncertainties are, on average, 10 $\mu$as yr$^{-2}$ and are a factor
of six smaller than in our previous efforts \citep{lu09}.  We have thus been able
to make six significant acceleration measurements outside the central arcsecond 
($R$ = 1$\arcsec$) and out to
$R$ = 1$\farcs$5 ($\sim$0.06 pc), which provides the stars' line-of-sight distances and 
enables precise orbital parameter estimates.  We confirm the existence of the 
clockwise disk, which has an orbital plane oriented at ($i$, $\Omega$) = 
(130$\deg$, 96$\deg$). With simulations of mock data sets of disk stars and
an isotropically distributed population, we have shown that the true disk fraction 
of young stars is $\sim$20\%, a factor of $\sim$2.5 lower than previous estimates, 
suggesting that we are curently observing a remnant disk. The kinematic properties
of the brightest B stars are similar to those of the O/WR stars, suggesting a
common star formation event. The
opening angle of the disk is $\sim$8$\deg$ and our data do 
not reveal a change in the direction of the orbital plane as a function of radius.
The mean eccentricity of the members of the clockwise disk is
$\langle e \rangle$ = 0.27 $\pm$ 0.07. Given the recent finding by \citet{lu13} regarding 
the initial mass function and the age of the population, which is consistent with earlier 
work by \citet{paumard06}, the eccentricities of the disk stars can be explained by 
dynamical relaxation in an initially circular disk. 
The previously-claimed counterclockwise disk is not detected, despite the fact
that we use higher-precision astrometric measurements and a larger field of
view than in \citet{lu09} and \citet{bartko09}.  We confirm 
the kinematic structure seen by \citet{bartko09} at large radii, 
which may be a small cluster of stars that share similar motions but that are 
distinct from the clockwise disk seen at $r <$ 3$\farcs$2.

Constraining the stars' line-of-sight distances through precise
acceleration measurements is key for estimating stellar orbits and removes
the need for prior assumptions that may lead to significant biases.  Thus, it is
critical to increase both the precision and the time baseline of astrometric 
measurements for stars at large radii from the SMBH. Furthermore, complete
azimuthal coverage with spectroscopy at large radii will allow for the identification
of more young stars and a better characterization of the dynamics of the off-disk 
population.

\acknowledgements
Support for this work was provided by the NSF grant AST-0909218. The authors thank
the referee for useful comments that helped improve the paper.
We are also grateful to Eric Becklin for the many helpful discussions about the project 
and to Brad Hansen for suggestions  
on the disk simulations. We thank Richard Alexander for generously providing 
updated eccentricity values from dynamical evolution models of the GC young stars.
We thank the staff of the Keck Observatory for their help in obtaining the observations. 
S.Y. thanks the Aspen Center for Physics and the organizers of the 
summer 2012 workshop, as well as Ann-Marie Madigan for insightful discussions. 
The W. M. Keck Observatory is operated as a scientific 
partnership among the California Institute of Technology, the University 
of California and the National Aeronautics and Space Administration.  
The Observatory was made possible by the generous financial support of 
the W. M. Keck Foundation.  The authors also recognize and acknowledge
the very significant cultural role and reverence that the summit of Mauna
Kea has always had within the indigenous Hawaiian community. We are most
fortunate to have the opportunity to conduct observations from this mountain.

{\it Facilities:} \facility{Keck:II (NIRC2, OSIRIS), Keck:I (NIRC)}

\begin{appendix}
\section{Improved Speckle Camera (NIRC) Distortion Solution}
\label{app:nircDist}
The AO images that have new corrections for geometric optical distortion and
DAR \citep{yelda10} allow for an improvement in the determination of the geometric 
optical distortion for the speckle camera \citep[NIRC;][]{matthews96}. We use 
a similar approach to that described in \cite{lu09}, but here we map the speckle
data to the predicted star list for the 2004 July speckle epoch 
(Appendix \ref{app:localDist}) as opposed to the measured star positions. We note 
that DAR was inadvertently not corrected in the speckle images. However, over the 5"
speckle field of view, DAR amounts to $\sim$2 mas, in the extreme, and 
$\sim$1 mas on average, and is somewhat reduced when the frames are averaged 
together because the field rotates on the detector throughout the speckle
observations (in contrast to the AO observations, which are taken at a fixed
position angle). This new solution, given in Table \ref{tab:nircCoef}, results in
smaller residuals compared to our earlier solution (2 mas vs. 3 mas, on average).

% Table generated by hand
\begin{deluxetable}{lcc}
\tabletypesize{\scriptsize}
\tablewidth{0pt}
\tablecaption{Updated NIRC Reimager Distortion Coefficients}
\tablehead{
  \colhead{i} &
  \colhead{$X(a_i)$} &
  \colhead{$Y(b_i)$} 
}

\startdata
0 & 1.2972 $\times$ 10$^{-2}$ & -2.1134 $\times$ 10$^{-2}$ \\
1 & 9.9726 $\times$ 10$^{-1}$ & -1.1145 $\times$ 10$^{-3}$ \\
2 & -2.2849 $\times$ 10$^{-3}$ & 1.0034 \\
\enddata
\label{tab:nircCoef}

\end{deluxetable}

\clearpage
\section{Residual Relative Astrometric Error}
\label{app:additive}
The inaccuracies in the estimates of the PSF wings lead to an additional
source of error that is not accounted for in the estimate of the 
centroiding error.  Following the approach introduced by our group in
\citet{clarkson11}, we include an ``additive'' noise term for each
observational approach. 
For the AO data, images taken in a consistent setup 
(N=11 observations at the time of this analysis) to the 2006 June
image were aligned. Once in a common reference frame, lines were fit to the
positions as a function of time, where the positional uncertainties 
included the error on the mean from the three subset images for each epoch 
($\sigma_{rms}$; see \S\ref{sec:astrometry}) and the alignment errors
($\sigma_{aln}$), which were determined by a half-sample bootstrap 
\citep{ghez08}. Confusion was accounted for, as described in 
Section \ref{sec:align}. Only stars detected in all 11 epochs were used in 
this analysis. The velocity $\chi^2$ distribution for 1024 stars was then 
compared to the expected distribution for 9 degrees of freedom 
(11 measurements - 2 fit parameters). We determined the amount of 
error to be added to the positional uncertainties in order to 
minimize, in a least squares sense, the 
difference between the distributions. This additive
noise term for the AO data is $\sigma_{add}$ = 0.1 mas, comparable to the
centroiding error of bright stars ($K <$ 15).

The additive error for the speckle data was determined in a similar
fashion, but we aligned all speckle and LGSAO data together and used
the 2006 June image as the reference epoch. A line was
to the speckle positions as a function of time, where again, the 
positional uncertainties included $\sigma_{rms}$ and $\sigma_{aln}$. 
Only stars that were detected in all 27 speckle images and that
were not confused in any epoch were included in this analysis.
In comparing the resulting $\chi^2$ distribution for 32 stars to 
that expected for 25 degrees of freedom 
(27 speckle measurements - 2 fit parameters), 
a relatively small error (compared to $\sigma_{rms}$ for
speckle measurments, $\sim$1 mas for $K <$ 15) of 0.18 mas is necessary to fully
account for the positional scatter over time.

\section{Local Distortion Correction}
\label{app:localDist}
Of the 19 Galactic center adaptive optics data sets taken at Keck since
2004, all but three have had identical observational setups (e.g., PA=0
in the K' band).
We began observing with a consistent setup (PA=0$\deg$ and same telescope pointings) 
in 2006 May and therefore refer to this as the ``2006-setup''.  
The 2004 July image was taken at PA=200, while the 2005 July image 
was observed at PA=190.  In 2005 June, we observed the GC at PA=0 but at a
different starting position than the 2006-setup \citep[see ][]{ghez08,lu09}.  
In \citet{yelda10}, we found that a data set observed at a non-zero
PA can be transformed to the PA=0 (2006-setup) image to $\sim$0.1 pix.
To minimize the impact of this residual distortion when aligning the full
GC data set, we applied a local distortion correction to the three
images taken in different setups.  

The local distortion correction was found by comparing the positions of
the stars from the non-2006 epochs to their positions as predicted
by their best-fit proper motions.  This was done through the following 
series of steps.  First, the 2006-setup star lists (taken through 2010) 
were transformed to the 2006 June epoch using a 2nd order polynomial. 
This epoch was chosen as
the reference epoch as it is one of our highest quality images and is also
the reference frame used in our main analysis (\S\ref{sec:astrometry}).
The additive error term of 0.1 mas for AO data derived in Appendix 
\ref{app:additive} was included in the error measurements in these lists.
Once the positions were placed in a common reference frame, proper motions
were estimated by fitting a line to the positions as a function of time.
Stars with proper motion errors $>$1.5 mas yr$^{-1}$ or proper motions 
$>$10 mas yr$^{-1}$ were excluded from this analysis, as they may be 
mis-matched sources. Based on these proper motions, we created 
``predicted'' star lists for each of the three non-2006-setup epochs.

We next transformed all of the AO data (through 2010), including the 
three epochs that were taken with a different setup, to the 2006 
June image.  The transformed stellar 
positions for the three non-2006-setups were then compared to their
predicted positions based on the previous step.  The differences in
these positions represents the residual distortion in the images.
The positional differences measured over the detector for each non-2006
epoch were smoothed into a local distortion map in the following way.  
For each pixel on the
detector, the median positional difference of the 5 nearest stars was
taken as the correction for that pixel. We note that the two data sets
taken at a non-zero PA did not overlap completely with the 2006 June
field, and we assigned the pixels with no overlap a value of zero.
Similarly, we made a local distortion error map by taking the standard
deviation of the positional differences for the 5 nearest neighbors to
each pixel.

We verified that this method reduced the residuals in the the transformation
of the PA=200 to PA=0 images from 2004 July. We applied this local distortion
correction to the positions in the star lists created by 
{\em StarFinder} \citep{stf}, and added the local distortion error in 
quadrature to the centroiding errors for the three non-2006 epochs.

\section{Updated Sgr A*-Radio Rest Frame}
\label{app:new_refFrame}
Updated astrometry for the secondary standards originally presented 
in \citet{yelda10} is shown in Table \ref{tab:secondary_short}.  
The updates reflect a new mosaicking procedure, which is done on the 
individual star lists as opposed to the images (see \S\ref{sec:astrometry}). 
As compared to the previous measurements reported in
\citet{yelda10}, we find several stars with $>$3$\sigma$ difference in either
the $X$ ($N$ = 6 stars) or $Y$ ($N$ = 19 stars) velocity coordinate.  However, the
$\chi^2$ value of the velocity fits improved in almost all cases with our new
analysis, and we therefore use these updated values when constructing our
reference frame.  We present 1210 astrometric standards
here, slightly fewer than \citet{yelda10}, which had 1279 stars. 
This discrepancy is a result of the higher signal to noise in the overlapping
regions of mosaicked images as compared to mosaicked star lists.

\section{Line of Nodes Bias}
\label{app:nodes}
\begin{figure}[!h]
\epsscale{1.0}
\plottwo{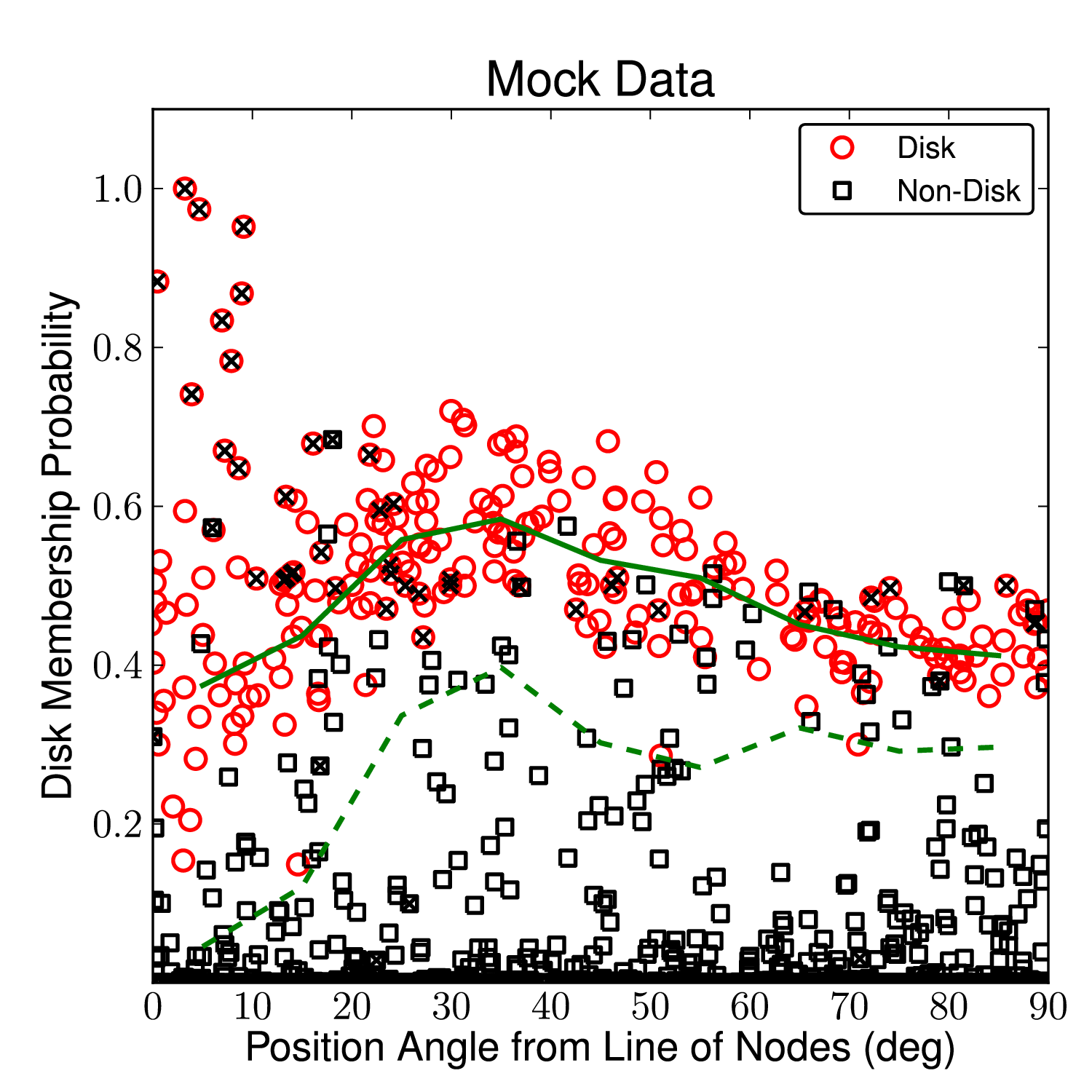}
{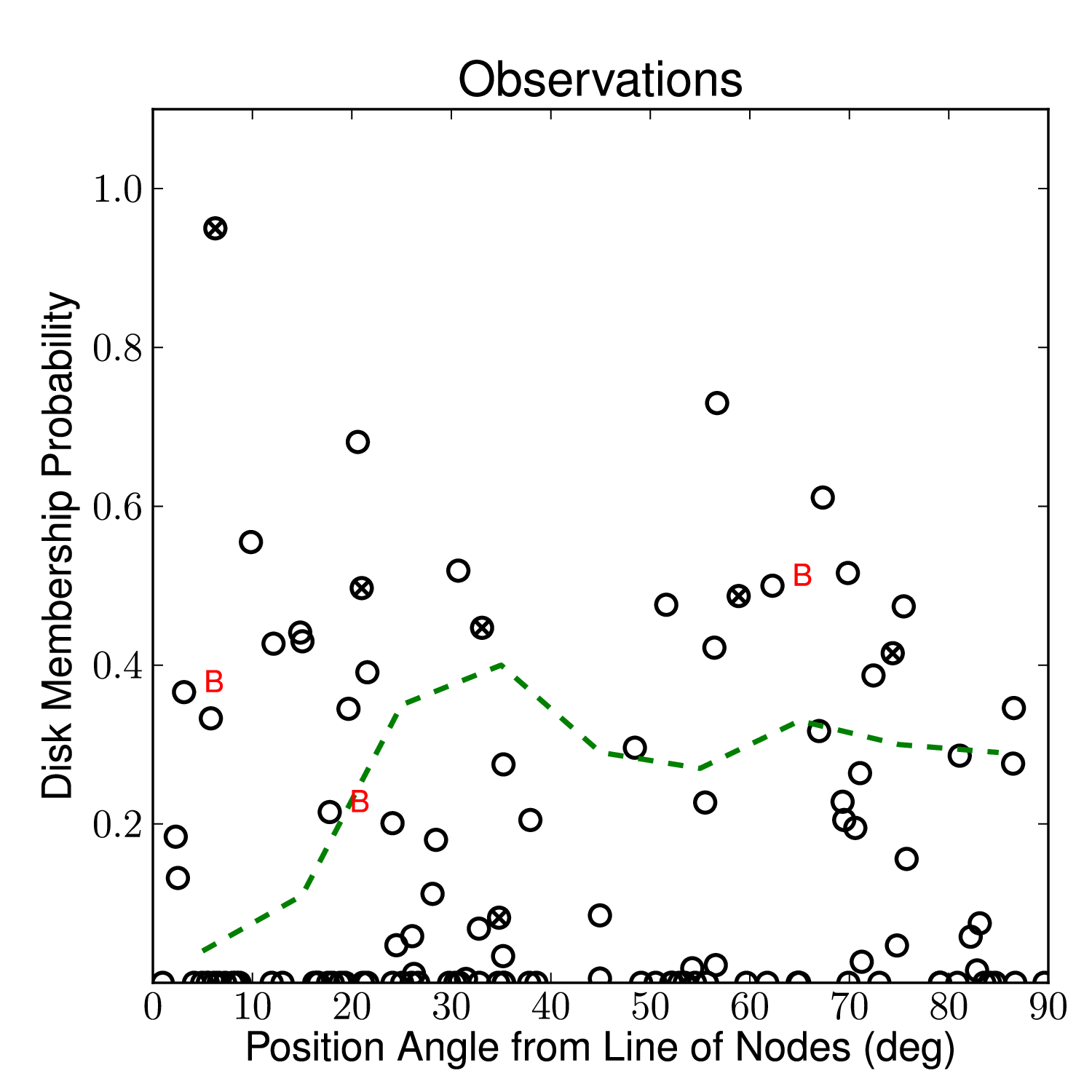}
\figcaption{{\em Left:} Disk membership probability plotted against position angle from the 
disk line of nodes ($\Omega$ = 96.3$\deg$) for the 10 simulations with $f_{disk}$ = 0.2. 
True disk members are shown as red circles (N = 240), while stars from the isotropic
population are marked as black squares (N = 960).  Stars with significant 
acceleration detections are marked with an 'x'.  For disk stars, the probability of
disk membership is a function of the angular offset from the disk's line of nodes. 
Non-accelerating disk stars with small angular offsets ($<$ 20$\deg$) show relatively 
large scatter in disk membership probability. We compute the median and standard deviation
of the probability in angular offset bins of 10$\deg$ for all non-accelerating disk stars. 
The median and 3$\sigma$ lower-limit values are shown as the green solid and dashed curves, 
respectively. 
{\em Right:} Same plot, but for the observed sample of N = 116 stars. The 3$\sigma$
lower-limit curve is overplotted.  Points above this curve represent the most likely
members of the clockwise disk. The three B stars that are on the disk are marked with a
red 'B'.
}
\label{fig:probNodes}
\end{figure}

Disk stars located near the disk's line of nodes ($\Omega$ = 96.3$\deg$) have a
small line-of-sight distance ($|z|\sim$0). For such stars, the acceleration prior 
may lead to biased orbital solutions. Unless the star has a detectable acceleration 
or an upper limit constraining the line-of-sight distance to $|z| >$ 0, the
line-of-sight distance is determined by randomly sampling from a uniform
distribution of accelerations, bounded by the minimum and maximum allowed
accelerations. With such a prior, a wide range of $z$'s is allowed, most of
which will be non-zero, thus leading to orbital solutions biased away from
the nominal disk solution.  The disk membership probability described in \S\ref{sec:2dPDF}
will therefore be a function of the disk stars' locations along their orbit. To
see this, we plot the disk membership probability against the position angle
relative to the disk line of nodes ($\Omega$ = 96.3$\deg$) for all stars in
the 10 mock data sets with $f_{disk}$ = 0.2 (Figure \ref{fig:probNodes}, left panel).
Note that the accelerating disk stars have high disk membership probability for
small angles (these stars are the innermost stars in radius).
We therefore compute the median and standard deviation
of the probability in angular offset bins of 10$\deg$ for all non-accelerating disk stars
and use this as a metric for identifying the most likely members of the disk.

% Generated by python code:
% sythesis.disk_fraction_results(getAcc=True,multiSims=True,fraction=0.2)

The observed data are shown in the right panel of Figure \ref{fig:probNodes}.  We find
that 28 stars have disk membership probabilities above the 3$\sigma$ lower-limits for
their respective position angle bin.  These stars are therefore the most likely
members of the clockwise disk. Assuming a true disk fraction of 20\%, we estimate
that 3-4 of these stars may not be true disk members.
The position angle and the significance level above which the star falls (either
1$\sigma$, 2$\sigma$, or 3$\sigma$) are shown in Table \ref{tab:nodes_app_table}.  

\clearpage
% Table generated by sythesis.nodes_bias_observed()
\begin{deluxetable}{lrr}
\tabletypesize{\scriptsize}
\tablewidth{0pt}
\tablecaption{Disk Membership Sample\label{tab:nodes_app_table}}
\tablehead{
  \colhead{Name} &
  \colhead{$PA_{nodes}$\tablenotemark{a}} &
  \colhead{Sample\tablenotemark{b}} \\ 
  \colhead{} &
  \colhead{(deg)} &
  \colhead{} 
}
\startdata
    S8-15 & 72.4 & 1$\sigma$ \\ 
   S4-169 &  9.9 & 1$\sigma$ \\ 
     S3-5 & 15.0 & 1$\sigma$ \\ 
    S7-10 & 75.5 & 1$\sigma$ \\ 
   irs34W & 14.8 & 1$\sigma$ \\ 
   S7-161 &  5.8 & 1$\sigma$ \\ 
    S6-63 & 67.3 & 1$\sigma$ \\ 
    S3-10 & 12.1 & 1$\sigma$ \\ 
   S7-236 & 20.6 & 1$\sigma$ \\ 
    S2-21 & 51.6 & 1$\sigma$ \\ 
    S2-16 & 56.7 & 1$\sigma$ \\ 
     S1-8 & 62.3 & 1$\sigma$ \\ 
    S4-36 & 19.7 & 1$\sigma$ \\ 
     S1-3 & 74.4 & 1$\sigma$ \\ 
    S0-15 &  6.2 & 1$\sigma$ \\ 
    S1-12 & 58.9 & 1$\sigma$ \\ 
    S1-14 & 21.0 & 1$\sigma$ \\ 
   S10-32 &  3.1 & 1$\sigma$ \\ 
     S8-7 & 69.9 & 1$\sigma$ \\ 
    S6-82 &  2.3 & 2$\sigma$ \\ 
  irs34NW & 30.7 & 2$\sigma$ \\ 
    irs9W & 56.4 & 2$\sigma$ \\ 
    S2-19 & 86.5 & 2$\sigma$ \\ 
  irs16CC & 21.6 & 3$\sigma$ \\ 
    S2-17 & 48.4 & 3$\sigma$ \\ 
   S3-190 & 17.8 & 3$\sigma$ \\ 
   irs16C & 33.1 & 3$\sigma$ \\ 
    S1-21 &  2.5 & 3$\sigma$ \\ 
   S10-34 & 26.1 &     other \\ 
   S10-48 & 80.9 &     other \\ 
   S5-237 & 16.6 &     other \\ 
   S5-236 & 19.3 &     other \\ 
   S5-235 & 64.9 &     other \\ 
   S5-231 &  7.3 &     other \\ 
    S10-4 & 84.1 &     other \\ 
    S10-5 & 74.8 &     other \\ 
   S6-100 & 82.8 &     other \\ 
   S10-50 & 11.9 &     other \\ 
   S4-364 & 69.9 &     other \\ 
    S13-3 & 32.8 &     other \\ 
    S10-7 & 30.8 &     other \\ 
    S11-5 & 89.6 &     other \\ 
   S11-21 & 83.1 &     other \\ 
  S10-136 & 37.8 &     other \\ 
    S0-14 & 26.2 &     other \\ 
    S6-90 & 44.9 &     other \\ 
    S9-23 & 75.8 &     other \\ 
    S9-20 & 55.5 &     other \\ 
   S9-143 & 28.1 &     other \\ 
    S9-13 & 64.8 &     other \\ 
   S9-114 & 52.9 &     other \\ 
     S9-1 &  8.0 &     other \\ 
     S8-4 & 83.6 &     other \\ 
   S8-196 & 26.0 &     other \\ 
   S8-181 & 31.5 &     other \\ 
    S7-36 & 28.5 &     other \\ 
    S7-30 & 16.2 &     other \\ 
   S7-228 &  6.1 &     other \\ 
   S7-216 &  4.1 &     other \\ 
    S7-20 & 55.7 &     other \\ 
    S7-19 & 53.5 &     other \\ 
   S7-180 & 18.8 &     other \\ 
    S6-96 & 24.1 &     other \\ 
    S6-95 & 61.7 &     other \\ 
    S6-93 & 54.5 &     other \\ 
    S6-81 &  8.7 &     other \\ 
    S7-16 & 71.1 &     other \\ 
  irs16NW & 86.7 &     other \\ 
    irs3E & 52.2 &     other \\ 
     S3-3 &  5.5 &     other \\ 
    S3-26 & 44.9 &     other \\ 
    S3-25 & 70.6 &     other \\ 
     S3-2 & 16.5 &     other \\ 
    S3-19 & 67.0 &     other \\ 
    S2-76 & 79.1 &     other \\ 
    S2-74 & 86.5 &     other \\ 
     S2-7 & 69.3 &     other \\ 
     S2-6 & 32.8 &     other \\ 
    S3-30 & 54.2 &     other \\ 
    S2-58 & 21.6 &     other \\ 
     S2-4 & 37.9 &     other \\ 
    S2-22 &  1.0 &     other \\ 
    S1-33 &  6.6 &     other \\ 
    S1-24 & 59.7 &     other \\ 
    S1-22 & 24.1 &     other \\ 
     S1-2 & 81.1 &     other \\ 
    S1-19 & 69.5 &     other \\ 
    S1-18 & 56.6 &     other \\ 
     S1-1 &  8.3 &     other \\ 
    S2-50 & 35.2 &     other \\ 
   S3-314 &  5.0 &     other \\ 
   S3-331 & 65.0 &     other \\ 
   S3-374 & 52.1 &     other \\ 
   irs33N & 84.7 &     other \\ 
   irs33E & 71.2 &     other \\ 
   irs29N & 35.2 &     other \\ 
    irs1W & 13.0 &     other \\ 
irs16SW-E & 24.5 &     other \\ 
  irs16SW & 34.8 &     other \\ 
   S9-283 & 21.1 &     other \\ 
  irs16NE & 25.1 &     other \\ 
  irs13E4 & 29.8 &     other \\ 
  irs13E2 & 34.7 &     other \\ 
  irs13E1 & 35.3 &     other \\ 
    S6-89 & 35.3 &     other \\ 
    S5-34 & 38.6 &     other \\ 
   S5-191 & 50.5 &     other \\ 
   S5-187 & 79.1 &     other \\ 
   S5-183 & 30.4 &     other \\ 
    S4-71 & 73.0 &     other \\ 
   S4-287 & 82.2 &     other \\ 
   S4-262 & 18.1 &     other \\ 
   S4-258 & 26.7 &     other \\ 
    S3-96 & 17.6 &     other \\ 
   irs7SE & 55.7 &     other \\ 
     S9-9 & 49.1 &     other \\ 
\enddata 
\tablenotetext{a}{Position angle offset from the line of nodes
of the clockwise disk ($\Omega$=96.3$\deg$) with a range of 0$\deg$-90$\deg$.}
\tablenotetext{b}{The level above which the star's disk membership probability 
falls for its respective angular offset bin (see Figure \ref{fig:probNodes}).}
\end{deluxetable}

\end{appendix}

\begin{deluxetable}{lcccccccccc}
\tabletypesize{\scriptsize}
\tablewidth{0pt}
\tablecaption{Galactic Center Secondary IR Astrometric Standards}
\tablehead{
  \colhead{Name} & 
  \colhead{K'} & 
  \colhead{$T_{0,IR}$} & 
  \colhead{Radius} & 
  \colhead{$\Delta$ R.A.} & 
  \colhead{$\sigma_{R.A.}$\tablenotemark{a}} & 
  \colhead{$\Delta$ Dec.} & 
  \colhead{$\sigma_{Dec}$\tablenotemark{a}} & 
  \colhead{v$_{RA}$\tablenotemark{b}} & 
  \colhead{v$_{Dec}$\tablenotemark{b}} \\ 
  \colhead{} & 
  \colhead{(mag)} & 
  \colhead{(year)} & 
  \colhead{(arcsec)} & 
  \colhead{(arcsec)} & 
  \colhead{(mas)} & 
  \colhead{(arcsec)} & 
  \colhead{(mas)} & 
  \colhead{(mas yr$^{-1}$)} & 
  \colhead{(mas yr$^{-1}$)} 
}
\startdata
      S0-3  & 14.8  & 2008.39  & 0.36  &   0.3351  & 1.1  &   0.1195  & 1.4  &      9.4 $\pm$      0.4     &     -1.2 $\pm$      0.6\\ 
      S0-6  & 14.2  & 2008.30  & 0.36  &   0.0292  & 1.1  &  -0.3624  & 1.2  &     -5.2 $\pm$      0.3     &      3.7 $\pm$      0.4\\ 
      S0-5  & 15.3  & 2007.99  & 0.41  &   0.1790  & 1.1  &  -0.3664  & 1.3  &     -2.1 $\pm$      0.3     &      0.4 $\pm$      0.5\\ 
\enddata 
\tablecomments{Table \ref{tab:secondary_short} is published in its entirety in the electronic version of this paper.}
\tablenotetext{a}{Positional errors include centroiding, alignment, and residual distortion (1 mas) errors, but do not include error in position of Sgr A*.}
\tablenotetext{b}{Velocity errors do not include error in velocity of Sgr A* (0.09 mas yr$^{-1}$, 0.14 mas yr$^{-1}$ in RA and Dec, respectively).}
\label{tab:secondary_short}

\end{deluxetable}
\clearpage

\end{document}